\documentclass{article}
\usepackage{graphicx}
\usepackage{amssymb,amsmath,amsthm}
\usepackage{fullpage}
\usepackage[disable]{todonotes}
\numberwithin{equation}{section}
\numberwithin{figure}{section}

\newcommand{\vu}{\mathbf{u}}
\newcommand{\abs}[1]{\left\lvert#1\right\rvert}
\newcommand{\cN}{\mathcal{N}}
\newcommand{\cF}{\mathcal{F}}
\newcommand{\cL}{\mathcal{L}}
\newcommand{\cS}{\mathcal{S}}
\newcommand{\cR}{\mathcal{R}}
\newcommand{\RR}{\mathbb{R}}
\newcommand{\ZZ}{\mathbb{Z}}

\newcommand{\VV}{\mathcal{V}}
\newcommand{\EE}{\mathcal{E}}
\newcommand{\tv}{\mathtt{v}}
\newcommand{\te}{\mathtt{e}}
\newcommand{\vv}{\mathbf{v}}
\newcommand{\vq}{\mathbf{q}}
\newcommand{\vp}{\mathbf{p}}
\newcommand{\zbar}{\bar{z}}
\newcommand{\ybar}{\bar{y}}
\DeclareMathOperator{\spann}{span}

\DeclareMathOperator{\cn}{cn}
\DeclareMathOperator{\dn}{dn}
\newcommand{\inner}[2]{\left\langle #1, #2 \right\rangle}

\newcommand{\Phiconst}{\Phi_{\rm const}}
\newcommand{\Phileft}{\Phi_{\rm left}}
\newcommand{\Phiright}{\Phi_{\rm right}}
\newcommand{\Phiodd}{\Phi_{\rm odd}}
\newcommand{\Phieven}{\Phi_{\rm even}}
\newcommand{\w}{\omega}
\newcommand{\W}{\Omega}
\newcommand{\Tcn}{T_{\rm cn}}
\newcommand{\Tdn}{T_{\rm dn}}
\newcommand{\Tcnmax}{T_{\rm cn}^{\rm max}}
\newcommand{\Tdnmin}{T_{\rm dn}^{\rm min}}
\newcommand{\tparen}[1]{\textbf{(#1)}}

\newcommand{\tPhi}{\tilde{\Phi}}
\newcommand{\intG}{\int_{\Gamma}}

\newtheorem{thm}{Theorem}
\newtheorem{rem}{Remark}[section]

\newcommand{\CC}{\mathbb{C}}
\newcommand{\cE}{\mathcal{E}}

\newcommand{\norm}[2]{\left\lVert #1 \right\rVert_{#2}}

\newcommand{\bD}{\mathbf{D}}
\newcommand{\bE}{\mathbf{E}}
\newcommand{\bI}{\mathbf{I}}
\newcommand{\bL}{\mathbf{L}}

\newcommand{\bS}{\mathbf{S}}
\newcommand{\bW}{\mathbf{W}}

\graphicspath{{Figures/}}

\title{NLS Bifurcations on the bowtie combinatorial graph and the dumbbell metric graph}
\author{Roy H. Goodman}
\date{\today}

\begin{document}
 
 \maketitle
\begin{abstract}
We consider the bifurcations of standing wave solutions to the nonlinear Schr\"odinger equation (NLS) posed on a quantum graph consisting of two loops connected by a single edge, the so-called dumbbell, recently studied in~\cite{Marzuola:2016bl}. The authors of that study found the ground state undergoes two bifurcations, first a symmetry-breaking, and the second which they call a symmetry-preserving bifurcation. We clarify the type of the symmetry-preserving bifurcation, showing it to be transcritical. We then reduce the question, and show that the phenomena described in that paper can be reproduced in a simple discrete self-trapping equation on a combinatorial graph of bowtie shape. This allows for complete analysis both by geometric methods and by parameterizing the full solution space. We then expand the question, and describe the bifurcations of \emph{all} the standing waves of this system, which can be classified into three families, and of which there exists a countably infinite set.
\end{abstract}

\section{Introduction}
Linear and nonlinear waves on quantum graphs have recently gained a lot of attention in   mathematical literature as a nontrivial generalization of well-known systems and because they possess novel mathematical properties not possible in simpler topology~\cite{Berkolaiko:2013,Berkolaiko:2017}. In physics they may arise in optical systems, Bose-Einstein condensates, and in the study of large molecules such as carbon nanotubes~\cite{Delourme:2017jd,Kuchment:2007,Niikuni:2017fu,Bolte:2014}. In the nonlinear regime they lead to  bifurcations and, in the case that one or more edges extends to the point at infinity, to interesting questions about the existence of minimizers~\cite{Pelinovsky:2016fm,Noja:2017vd,Cacciapuoti:2015fe,Adami:2015di,Adami:2017te,Adami:2016hx,Adami:2014hn,Adami:2016do, Cacciapuoti:2017hr}.

While there is an infinite variety of graphs, a lot of recent work has gone into a few simple ones. The tadpole graph consists of a half-line joined to a loop at a single vertex~\cite{Noja:2015hx}. The lollipop graph is similar, with the half-line replaced by a finite line segment~\cite{Berkolaiko:2017}. Star graphs consist of a finite number of half lines or line segments that meet at a single vertex~\cite{Adami:2012jb}.

In a recent paper, Marzuola and Pelinovsky study the bifurcations of stationary solutions to the cubic nonlinear Schr\"odinger equation posed on a dumbbell-shaped quantum graph~\cite{Marzuola:2016bl}, in which two loops are connected via a line segment. At small $L^2$ norm, the ground state, i.e.\ the lowest energy solution with that norm, is constant in the spatial coordinate. They show that this state undergoes two bifurcations as the solution amplitude is increased. The first is a symmetry-breaking or pitchfork bifurcation and the second they call a ``symmetry preserving'' bifurcation, producing a solution localized on the central edge of the graph. At the symmetry-breaking solution, the asymmetric solution becomes the ground state. 

We show that all of these phenomena are present in a simpler system, a discrete NLS equation on a graph with five vertices.  We show that in this simpler system the ``symmetry preserving'' bifurcation is actually a transcritical bifurcation, and, thus, contains a branch of solutions not present in the bifurcation diagram shown in~\cite{Marzuola:2016bl}. 
%
This is interesting, because Yang, in Ref.~\cite{Yang:2012ux}, only reported transcritical bifurcations by constructing an asymmetric potential, whereas the dumbbell graph is symmetric; see Remark~\ref{rem:generic} for more details.
We then show, via a perturbation calculation, and via numerical computation that the same is true for the problem on the quantum graph, which demonstrates the existence of the branch not found in~\cite{Marzuola:2016bl}. Further, we  perform an extensive numerical study, enumerating \emph{all} the branches of solutions that exist in a certain part of parameter/solution space.

\subsection{Organization}
The paper is organized as follows. 
The rest of this section contains a brief introduction to NLS on combinatorial and quantum graphs and a summary of recent and related results, as well as a discussion of Yang's framework for bifurcations in generalized NLS equations.	
Section~\ref{sec:bowtie} introduces a simple combinatorial graph model that reproduces the two bifurcations seen in~\cite{Marzuola:2016bl} for the dumbbell quantum graph. 
Section~\ref{sec:missing} describes arguments using perturbation theory and numerical continuation that show that the symmetry preserving bifurcation is indeed transcritical and to recover the half-branch of solutions not discussed in~\cite{Marzuola:2016bl}. 
Section~\ref{sec:singularPert} describes a singular limit of the dumbbell graph in which the length of edges $\te_1$ and $\te_3$ approaches zero. This demonstrates an interesting feature of the loop-shaped edges. 
Section~\ref{sec:more} contains a full classification and computational enumeration of all types of standing waves possible on the dumbbell graph. This is based on the observation that there are two types of behavior possible at the two vertices, and therefore three types of standing waves, as the two vertices can both have the first behavior, both have the second, or be different. After making this classification, we use a combination of exact solution formulae and numerical continuation to draw bifurcation diagrams for the three types of solutions. In fact, each of the three families are countably infinite.
Section~\ref{sec:conclusion} wraps up and discusses some possible future directions.

\subsection{NLS on combinatorial and metric graphs}

The focusing cubic nonlinear Schr\"odinger equation (NLS)
\begin{equation}
i \partial_t \Psi = \Delta \Psi  +2\abs{\Psi}^2 \Psi,
\label{NLS}
\end{equation}
subject to appropriate boundary conditions or conditions at infinity, is perhaps the most-studied system in nonlinear waves, arising as an envelope equation in systems where  dispersion is roughly balanced by nonlinearity. It is the simplest PDE with both these features, and thus is widely used as a model problem. Many variants of this equation have been studied in order to explore a wide range of phenomena.

One of the simplest variants is the discrete nonlinear Schr\"odinger equation (DNLS) in which the spatial derivative is replaced by a discrete second difference operator
\begin{equation}
i \dot{u}_n = u_{n-1} - 2 u_n + u_{n+1}  +\abs{u_n}^2 u_n.
\label{DNLS}
\end{equation}
If the total number of sites is small this system is also known as the discrete self trapping (DST) system, which has a long history~\cite{Eilbeck:1985tu,Eilbeck:2003vl,Kevrekidis:2009vv}. A DST system describing the evolution of a vector $\vu$ may be written, more generally, as 
\begin{equation}
i \dot{\vu} = \bL \vu + \cN(\vu)
\label{DST}
\end{equation}
where $\bL$ is a symmetric matrix and $\cN(\vu)$ is a nonlinearity we take to be local and cubic, ${\cN(\vu)}_n = \abs{u_n}^2 u_n$. This system may be interpreted as a Hamiltonian, with position vector $\vq=\vu \in \CC^N$ and momentum vector $\vp = i \bar{\vu}$ (such overbars will denote complex conjugates in the remainder). It is then easy to check that the Hamiltonian for this system is simply
 \begin{equation}
\label{Hu}
 H = \bar{\vu}^{\top} \bL \vu -\frac{1}{2} \sum_{n=1}^N \abs{u_n}^4.
\end{equation} 
The DST system~\eqref{DST} is then equivalent to Hamilton's equations:
\begin{equation}
i \dot{u}_n = \frac{\partial H}{\partial\bar{u}_n}.
\label{iundot}
\end{equation}

NLS may be generalized to graphs in a number of ways. Before doing so, we first set notation, briefly summarizing that of Berkolaiko~\cite{Berkolaiko:2017}. Let $\Gamma=(\VV,\EE)$ be a combinatorial graph, i.e.\ a finite collection of vertices $\VV=\{\tv_n, \, n= 1,\ldots,N\}$ connected pairwise by a collection 
of edges $\EE=\{\te_m=(\tv_i,\tv_j),\, m= 1,\ldots,M\}$. The graph is called \emph{directed} if $(\tv_i,\tv_j)$ is distinguished from $(\tv_j,\tv_i)$. Otherwise it is \emph{undirected}. 

If $\Gamma$ has no self-directed edges, i.e.\ no edges of the form $(\tv_i,\tv_i)$, the incidence matrix is defined to be the $N\times M$ matrix $\bE$ in which
$$
E_{nm} = 
\begin{cases}
1 & \text{if edge $m$ points to vertex $n$}, \\
-1 & \text{if edge $m$ points from vertex $n$}, \\
0 & \text{otherwise}.
\end{cases}
$$
Let there be a positive weight $w_m$ associated to edge $\te_m$ and let $\bW$ be a diagonal matrix with $\W_{mm} = w_m >0$. If unstated, then $\bW$ is assumed to be the identity matrix. Then the (weighted) Laplacian
$$
\Delta_\Gamma = -\bE \bW \bE^{\top}
$$ 
is a negative definite matrix, defining a DST system~\eqref{DST} with $\bL = -\Delta_\Gamma$.  While a directed graph is necessary for forming $E$ and thus $\Delta_\Gamma$, the matrix $\Delta_\Gamma$ is unchanged if the direction of any edge is interchanged.

A \emph{metric} graph is a directed graph in which a length $\ell_m$ has been assigned to each edge $\te_m$, as well as a coordinate $x_m$ with $0 < x_m < \ell_m$ which increases in the specified direction of the edge. A function $f(x)$ defined on the graph is simply a collection of functions $f_m(x_m)$ defined on the individual edges $\te_m$ such that $\left.f(x)\right\rvert_{\te_m} = f_m(x_m)$. A Schr\"odinger operator $\cL$ defined on such functions,
\begin{subequations}
\label{metricGraphLaplacian}
\begin{equation}
\left.\cL \right\rvert_{\te_m} = -\partial_{x_m}^2 + V_m(x_m),
\label{Lgraph}
\end{equation}
although we will set the potential along each edge to $V_m(x_m)=0$.
Defining the function space on which $\cL$ acts requires a boundary condition at the vertices. The graph, operator, and boundary conditions together define a \emph{quantum graph}, and we may extend the definition of equation~\eqref{NLS} to quantum graphs. W The most common boundary conditionis the Kirchhoff boundary condition, which states that the solution is continuous at the vertices, i.e.\ letting $\left.\Psi\right\rvert_{\te_m} = \psi_m$, 
\begin{equation}
\psi_{m_1}(\tv_n) = \psi_{m_2}(\tv_n), \,  \forall m_1,m_2 \in \VV_n,
\label{continuity}
\end{equation}
where $\VV_n$ is the set of all edges incident on vertex $\tv_n$, and further, that the total flux through any vertex is zero, i.e
\begin{equation}
\label{flux}
\sum_{m \in \VV_n} {(-1)}^{\sigma_m} \partial_{x_m} \psi_m(\tv_n) =0,
\end{equation}
\end{subequations}
where $\sigma_m=0$ if $\tv_n$ is the initial point on edge $\te_m$ and $\sigma_m=1$ if it is the final point. Both ends of an edge that connects a single vertex to itself contribute to this sum. Observe that for \emph{leaf} vertices, connected to only a single edge, the Kirchoff boundary condition reduces to a Neumann condition.

The set $\VV$ may contain a point at infinity, and special edges called leads that connect the finite points with the point at infinity. Such an edge must have $\ell = \infty$.

The standard norms and function spaces are defined straightforwardly, e.g.\ the $L^p$ norms
$$
\norm{\Psi}{L^p}^p = \sum_{m=1}^{M} \norm{\psi_m}{L^p}^p,
$$
and the related Sobolev norms defined similarly. In particular, we define the power to be 
\begin{equation}
Q(\Phi) = \norm{\Phi}{2}^2.
\label{Q}
\end{equation}
The $L^2$ inner products is defined by
\begin{equation*}
\inner{\Psi}{\Phi} = \sum_{m=1}^M \int_0^{\ell_m} \psi_m^*(x) \phi_m(x) \ dx.
\end{equation*}

Cubic NLS on the quantum graph is thus 
\begin{equation}
i \partial_t \Psi = \cL \Psi  +2\abs{\Psi}^2 \Psi,
\label{qNLS}
\end{equation}
which is defined on the energy space $\cE(\cL)$ consisting of all $H^1$ functions on $\Gamma$ satisfying the boundary conditions~\eqref{continuity} and~\eqref{flux}. This evolution conserves the energy
\begin{equation*}
E(\Psi) = \norm{\partial_x \Psi}{2}^2 - \norm{\Psi}{4}^4.
\end{equation*}

This paper considers standing wave solutions to the NLS equation of the form $\Psi(x,t) = \Phi(x)e^{i \Lambda t}$. Assuming $\Phi$ to be real valued, these solve
\begin{equation}
\cL \Phi - 2 \Phi^3 - \Lambda \Phi = 0.
\label{stationary}
\end{equation}

For all $\Lambda$, $\Phi=0$ is an exact solution. For $\Lambda<0$, there are additional solutions at 
\begin{equation}
\label{constantSolution}
\Phi = \pm \Phiconst(\Lambda) = \pm \sqrt{-\Lambda/2},
\end{equation}
which we refer to as the \emph{constant solutions}.

\subsection{Bifurcation Theory}

As a principal aim of this paper is to distinguish among different bifurcations among the families of standing waves, we include a brief review of the relevant bifurcation theory. Consider a stationary solution $(\Phi,\Lambda)$ to NLS on a graph~\eqref{stationary}. The associated linearization operator is
\begin{equation}
\label{linearization}
\cL_1 = \cL - \Lambda - 6 \Phi^2.
\end{equation}
As long as this operator is invertible at a particular $(\Phi_0,\Lambda_0)$, then the implicit function theorem implies that there exists an open interval $I \ni \Lambda_0$  on which there exists a unique one-parameter family of solutions, i.e.\ a continuous map
$$
\varphi: I \to H^1(\Gamma) \text{ such that } \varphi(\Lambda_0)=\Phi_0 \text{ and } \cF(\varphi(\Lambda),\Lambda) = 0.
$$  

\begin{figure}[htbp] 
   \centering
   \includegraphics[width=4in]{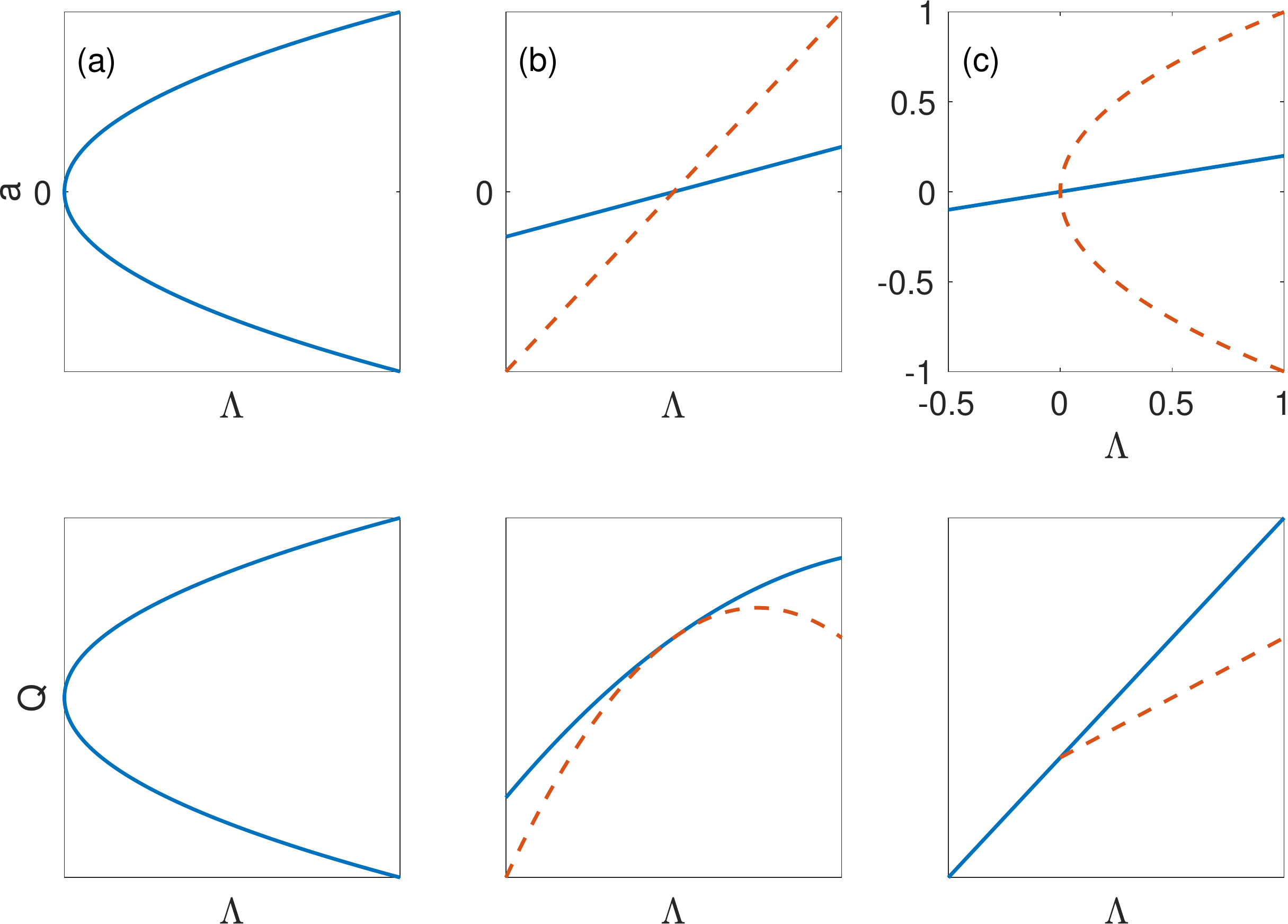} 
   \caption{The three most common bifurcations, after~\cite{Yang:2012ux}. (a)~Saddle-node, (b)~Transcritical, (c)~Pitchfork. Top row: coordinate $a$ vs.\ parameter $\Lambda$. Bottom row: power $Q$ vs.\ $\Lambda$. }
 \label{fig:bifs}
\end{figure}

Solutions $(\Phi_0,\Lambda_0)$ such that the operator~\eqref{linearization}
\begin{equation}
\cL_{10} =  {\bigl. \cL_1 \bigr\rvert}_{(\Phi_0,\Lambda_0)}
\label{Lgamma10}
\end{equation}
is singular are called bifurcation points. At such points, the implicit function theorem does not apply, and the existence or uniqueness of nearby solutions is not guaranteed. The three most commonly-seen bifurcations when $\cL_{10}$ is singular are the saddle-node, transcritical, and pitchfork bifurcation, although others exist when the conditions in the theorem below are violated. In a \emph{saddle-node}, or fold, bifurcation there exist two branches of solutions on one side of $\Lambda=\Lambda_0$, which merge smoothly at $\Lambda_0$,  and no solutions on the other; see Fig.~\ref{fig:bifs}(a). In a transcritical bifurcation, two branches cross transversely at $\Lambda_0$; Fig.~\ref{fig:bifs}(b). At a pitchfork bifurcation, three branches exist on one side of $\Lambda_0$, and only one on the other, with the branch single branch continuing smoothly through the bifurcation, and the other two branches connected smoothly at $\Lambda_0$; see Fig.~\ref{fig:bifs}(c).

Yang provides a result that classifies bifurcations into different types depending on the vanishing/nonvanishing of a sequence of nondegeneracy conditions~\cite{Yang:2012ux}. The result is given by explicitly constructing the formal asymptotic series solutions to a generalized form of the nonlinear Schr\"odinger equation. The construction goes through unaltered for the quantum graph problem. 

The results apply to a more general nonlinearity $G(\Phi,x)=F(\abs{\Phi^2},x)\Phi$ which here represents the cubic term in equation~\eqref{stationary} and any terms that would arise due to a nonzero $V(x)$ in the definition~\eqref{Lgraph}. Further, let
\begin{equation}
G_j = {\left. \partial_\Phi^j G \right\rvert}_{(\Phi_0,\Lambda_0)}
\label{Gj}
\end{equation}
Assume the linearized operator at the bifurcation $\cL_{10}$, given by equation~\eqref{Lgamma10}, has a one-dimensional nullspace with null eigenfunction $\Upsilon$. The result depends on the following quantities
\begin{equation}
\begin{gathered}
\Theta_1 = \inner{\Phi_0}{\Upsilon}; \;
\Theta_2 = \inner{G_2}{\Upsilon^3}; \;
\Theta_3 = \inner{1-G_2 \cL_{10}^{-1}\Phi_0}{\Upsilon^2}; \\
\Theta_4 = \inner{G_2{\left( \cL_{10}^{-1}\Phi_0 \right)}^2 -2 \cL_{10}^{-1}\Phi_0}{\Upsilon}; \;
\Theta_5 = \inner{G_3}{\Upsilon^4} - 3 \inner{G_2\Upsilon^2}{\cL_{10}^{-1}\left(G_2 \Upsilon^2\right)}.
\end{gathered}
\label{Thetas}
\end{equation}
While $\cL_{10}$ is by assumption not invertible, all the above terms involving $\cL_{10}^{-1}$ are well-defined.

Yang's main result is the following.
\begin{thm}
\label{YangTheorem}
Consider a system satisfying the above hypotheses. Then: 
\begin{enumerate}
\item
If $\Theta_1 \neq 0$ and $\Theta_2 \neq 0$, then a saddle-node bifurcation occurs at $\Lambda=\Lambda_0$. If $\Theta_1 \cdot \Theta_2 >0$, then the two branches exist for $\Lambda \le \Lambda_0$. If $\Theta_1 \cdot \Theta_2 < 0$, then the two branches exist for $\Lambda \ge \Lambda_0$. 
\item
If $\Theta_1 = 0$, $\Theta_2 \neq 0$, $\Theta_3 \neq 0 $ and $\Theta_3^2 > \Theta_2 \cdot \Theta_4$, then a transcritical bifurcation occurs at $\Lambda=\Lambda_0$.
\item
If $\Theta_1 = 0$, $\Theta_2 = 0$, $\Theta_3 \neq 0$, and $\Theta_5 \neq 0$, then a pitchfork bifurcation occurs at $\Lambda= \Lambda_0$. If $\Theta_3 \cdot \Theta_4 >0$, then there are three branches  for $\Lambda \le \Lambda_0$.  If $\Theta_3 \cdot \Theta_4 < 0$, then there are three branches for $\Lambda \ge \Lambda_0$. 
\end{enumerate}
\end{thm}

These bifurcations are all essentially one-dimensional. The Lyapunov-Schmidt theorem, or center manifold theory~\cite{Golubitsky:2013aa} essentially allow the bifurcation problem to be reduced to an equation involving one variable $a$ (due to the one-dimensional null space of $\cL_{10}$). Following~\cite{Glendinning:1994ub}, bifurcations in a one-dimensional problem depend on a normal form equation with an expansion of the form
\begin{equation*}
f(a,\Lambda) \sim \Theta_1 \Lambda + \frac{1}{2} 
\left(
\Theta_2 a^2 + 2 \Theta_3 a \Lambda  + \Theta_4 \Lambda^2 
\right)
+ \frac{1}{6} \Theta_5 a^3 + \ldots,
\end{equation*}
with the constants $\Theta_j$ determining the type of the bifurcation.

\begin{rem}
\label{rem:generic}
From this, it is apparent that the transcritical bifurcation is \emph{more generic} than the pitchfork bifurcation, in that it requires the vanishing of fewer terms in this expansion. 
Yang notes that if $F(\abs{\Phi}^2,x)=F(\abs{\Phi}^2,-x)$, and if $\Phi_0(x)$ is symmetric (even or odd) and $\Upsilon$ has the opposite symmetry, then $\Theta_1=\Theta_2=0$, and pitchfork bifurcations are to be expected~\cite[Remark 2]{Yang:2012ux}. He constructed transcritical bifurcations in a numerical example by tuning the system to satisfy $\Theta_1=0$, but only for potentials without this symmetry~\cite{Yang:2018}. By contrast, the dumbbell quantum graph is symmetric (see the next section), yet certain symmetric solutions undergo transcritical bifurcations. We discuss the conditions giving rise to transcritical bifurcations further in Section~\ref{sec:singularPert}.

Further, when a system has such a symmetry, then the pitchfork bifurcation is a symmetry-breaking bifurcation. That is, the system has a solution $\Phi$ that is invariant under symmetry transformation $R(\Phi)=\Phi$, whereas at the bifurcation, the new pair of solutions arises with $\Phi_1 = R(\Phi_2)$ and $\Phi_2 = R(\Phi_1)$ instead.
\end{rem}

\begin{rem}
A second theorem of Yang is useful for identifying bifurcation types from a graph of a standing wave's power $Q$ versus its frequency $\Lambda$~\cite{Yang:2012ux}. It shows that in a transcritical bifurcation, the two branches of solutions appear tangent to each other, while in a symmetry-breaking pitchfork bifurcation, the two branches overlap in the $(\Lambda,Q)$ plot, and branch out from the point $(\Lambda_0,Q_0)$ at a nonzero angle. See the second row of Fig.~\ref{fig:bifs}. This has been seen in pitchfork bifurcations in NLS systems before, e.g.~\cite{Marzuola:2010fl,Pelinovsky:2012hb}.
\end{rem}

\begin{figure}[htbp] 
   \centering
   \includegraphics[width=4in]{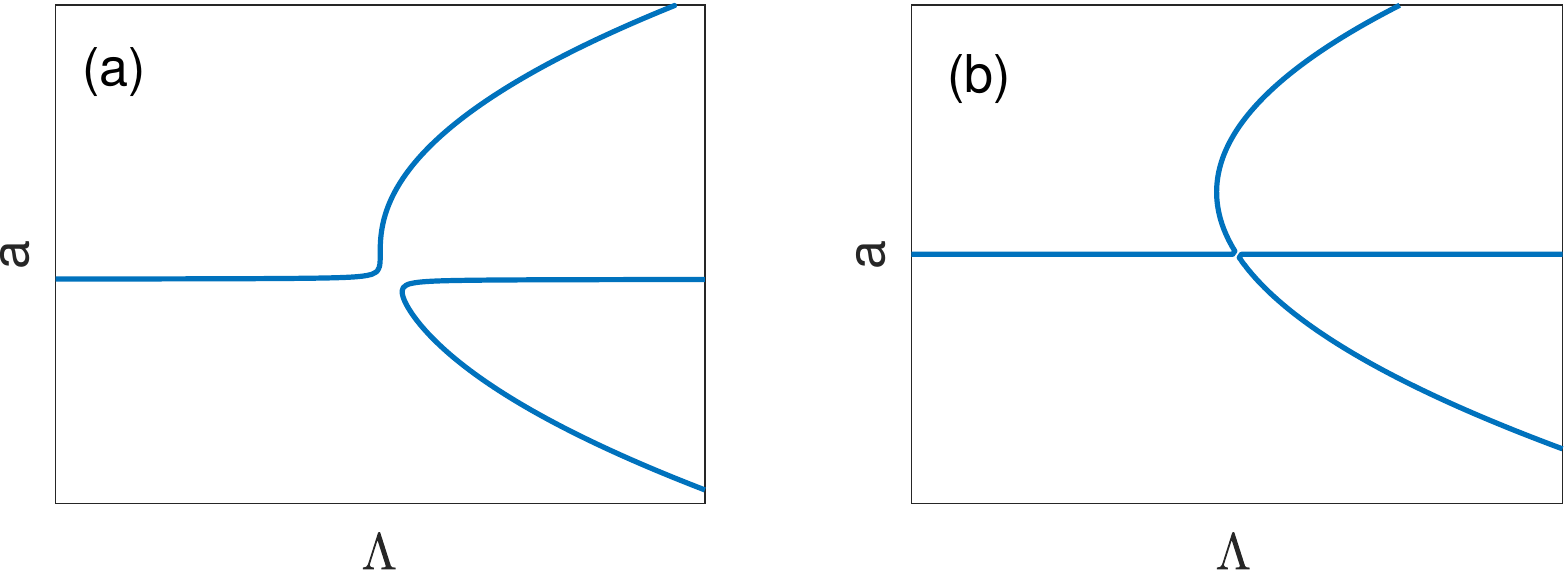} 
   \caption{A pitchfork bifurcation may split into either (a) one branch with no bifurcations and one branch with a saddle node (b) a saddle-node and a transcritical bifurcation. }
 \label{fig:imperfect}
\end{figure}

\begin{rem}
\label{rem:imperfect}
The pitchfork bifurcation is generic in systems possessing certain symmetries. When those symmetries are broken by a small general perturbation, the coefficient $\Theta_1$ will become nonzero. Thus, a saddle-node bifurcation must occur, as is shown in Fig.~\ref{fig:imperfect}(a). If the system is such that perturbations leave $\Theta_1 \equiv 0$, but results in nonzero $\Theta_2$, then the pitchfork bifurcation splits into a saddle-node and a transcritical as shown in Fig.~\ref{fig:imperfect}(b), with the two bifurcation points merging to a pitchfork in the limit $\Theta_2\to 0$. 
\end{rem}

\subsection{Previous work on the dumbbell quantum graph}

\begin{figure}[htbp] 
   \centering
   \includegraphics[width=3in]{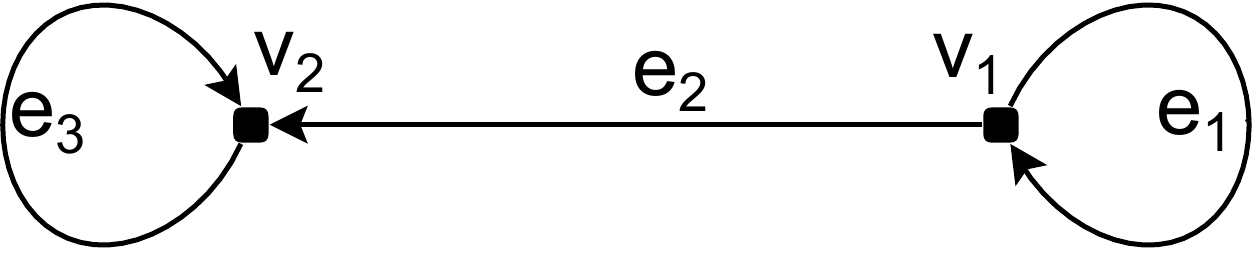} 
   \caption{The dumbbell graph with its vertices and edges labeled.}
\label{fig:dumbbell}
\end{figure}

Ref.~\cite{Marzuola:2016bl} considers stationary solutions to equation~\eqref{stationary}
on a dumbbell-shaped graph with two vertices and three edges, as displayed in Fig.~\ref{fig:dumbbell}. Two of this graph's edges are loops that connect the two vertices to themselves, and the third connects them. It makes the symmetry assumption $\ell_1 = \ell_3$ which, without loss of generality they take to be $\ell_1=\ell_3=2\pi$, and let $\ell_2 = 2 L$, parameterized by  $-\pi< x_{1,3}<\pi$ and $-L<x_2 <L$.

We note that symmetries of the graph play a role in the solutions to a quantum graph NLS system. The underlying dumbbell metric graph has the following three reflection symmetries, which we state somewhat informally:

\begin{equation}
\label{dumbbellSymmetries}
\begin{split}
\cR_1:& \;\text{Reverse the direction of edge $\te_1$.} \\
\cR_2: &\;\text{\parbox{.75\textwidth}{Exchange the labels of edges $\te_1$ and $\te_3$ and of vertices $\tv_1$ and $\tv_2$ and reverse the direction of edge $\te_2$.}} \\
\cR_3: &\;\text{Reverse the direction of edge $\te_3$.}
\end{split}
\end{equation}

A first step is to  enumerate all the eigenfunctions and eigenvalues on the dumbbell obtained by ignoring the nonlinear term in Equation~\eqref{stationary},
\begin{equation}
\cL \Phi  = \lambda \Phi.
\label{eigenproblem} 
\end{equation}
On each edge, the eigenfunction satisfies $\phi_m''(x_m) - \lambda \phi_m(x_m) = 0$, and thus the restriction of each eigenfunction to a single edge is the linear combination of a sine and a cosine function, and the eigenfunctions are obtained by enforcing the boundary conditions~\eqref{continuity} and~\eqref{flux}.

This observation is made systematic using the secular determinant, which is well-described by Berkolaiko~\cite{Berkolaiko:2017}. The nonzero eigenvalues  are of the form $\lambda = k^2$ where $k$ solves
\begin{equation}
\Sigma(k) =   \det{\left(\bI - \bS \bD(k) \right)} = 0,
\label{secular}
\end{equation}
and the dimension of the square matrices $\bS$ and $\bD(k)$ is twice the number of edges. This equation is derived by assuming that the solution is given as a linear combination of $e^{i k x_j}$ and $e^{-i k x_j}$ on each edge $\te_j$, and enforcing that the vertex conditions be satisfied. For the dumbbell graph,
$$
\bS=
\frac{1}{3}
\left(
\begin{array}{rrrrrr}
 2 & -1 & 0 & 2 & 0 & 0 \\
 -1 & 2 & 0 & 2 & 0 & 0 \\
 2 & 2 & 0 & -1 & 0 & 0 \\
 0 & 0 & -1 & 0 & 2 & 2 \\
 0 & 0 & 2 & 0 & 2 & -1 \\
 0 & 0 & 2 & 0 & -1 & 2 \\
\end{array}
\right)
\text{ and }
\bD(k) = 
\begin{pmatrix}
e^{i k L} &0 & 0 & 0 & 0 & 0 \\
0 & e^{i k L}& 0 & 0 & 0 & 0 \\
0 & 0 & e^{i k \pi} & 0 & 0 & 0 \\
0 & 0 & 0 & e^{i k \pi} & 0 & 0 \\
0 & 0 & 0 & 0 & e^{i k L} & 0 \\
0 & 0 & 0 & 0 & 0 & e^{i k L}
\end{pmatrix}
.
$$
After some manipulations, equation~\eqref{secular} is equivalent to
\begin{equation}
\left(
\sin{ k(L -\pi ) }-3 \sin{k(L +\pi ) }
\right) 
\left(
\cos{k (L-\pi) }-3 \cos{k (L+\pi ) }
\right)
\left(
\sin^2{k \pi}
\right)
= 0.
\label{secularDet}
\end{equation}

\begin{figure}[htbp] 
   \centering
   \includegraphics[width=.31\textwidth]{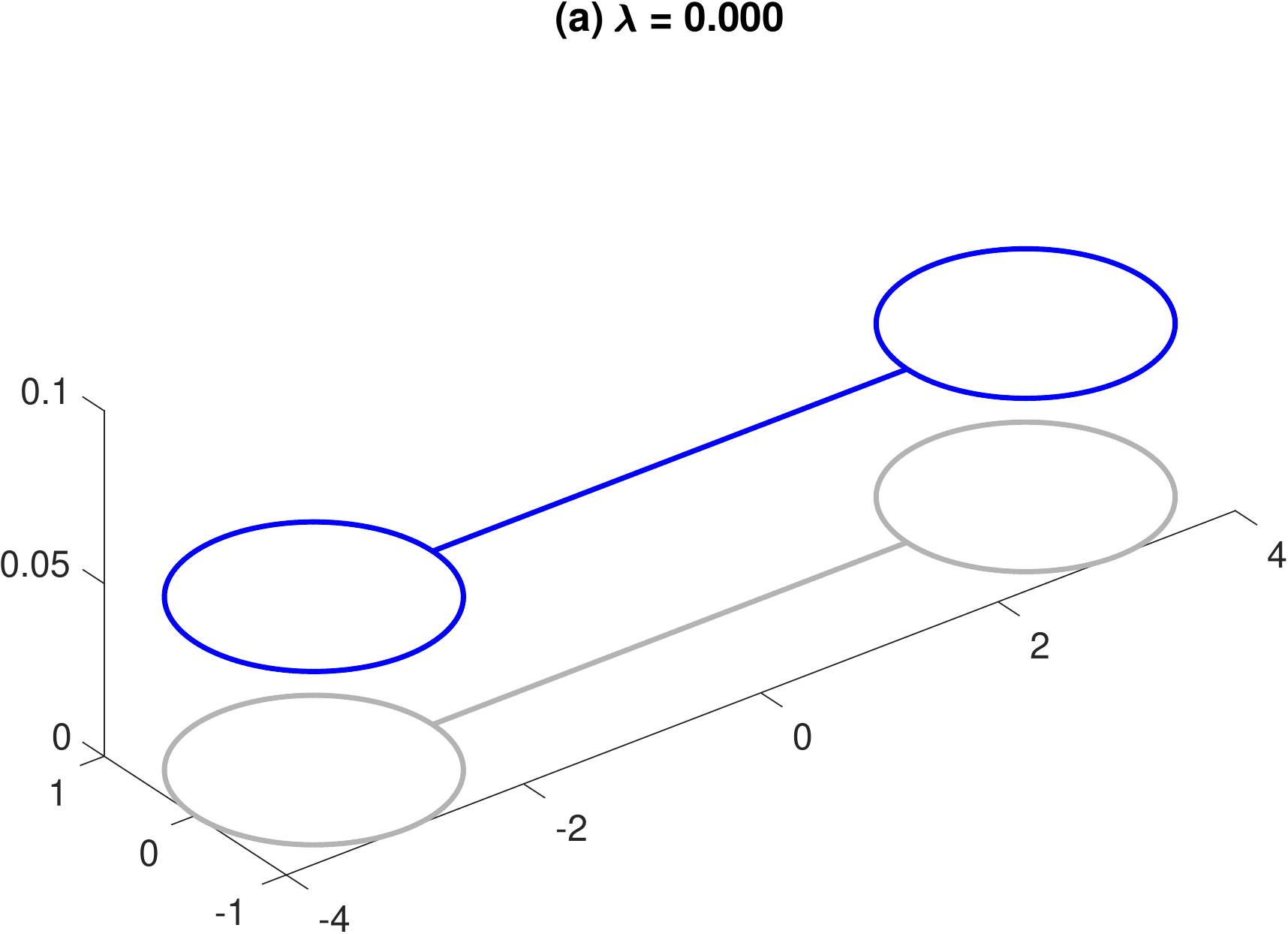} 
   \includegraphics[width=.31\textwidth]{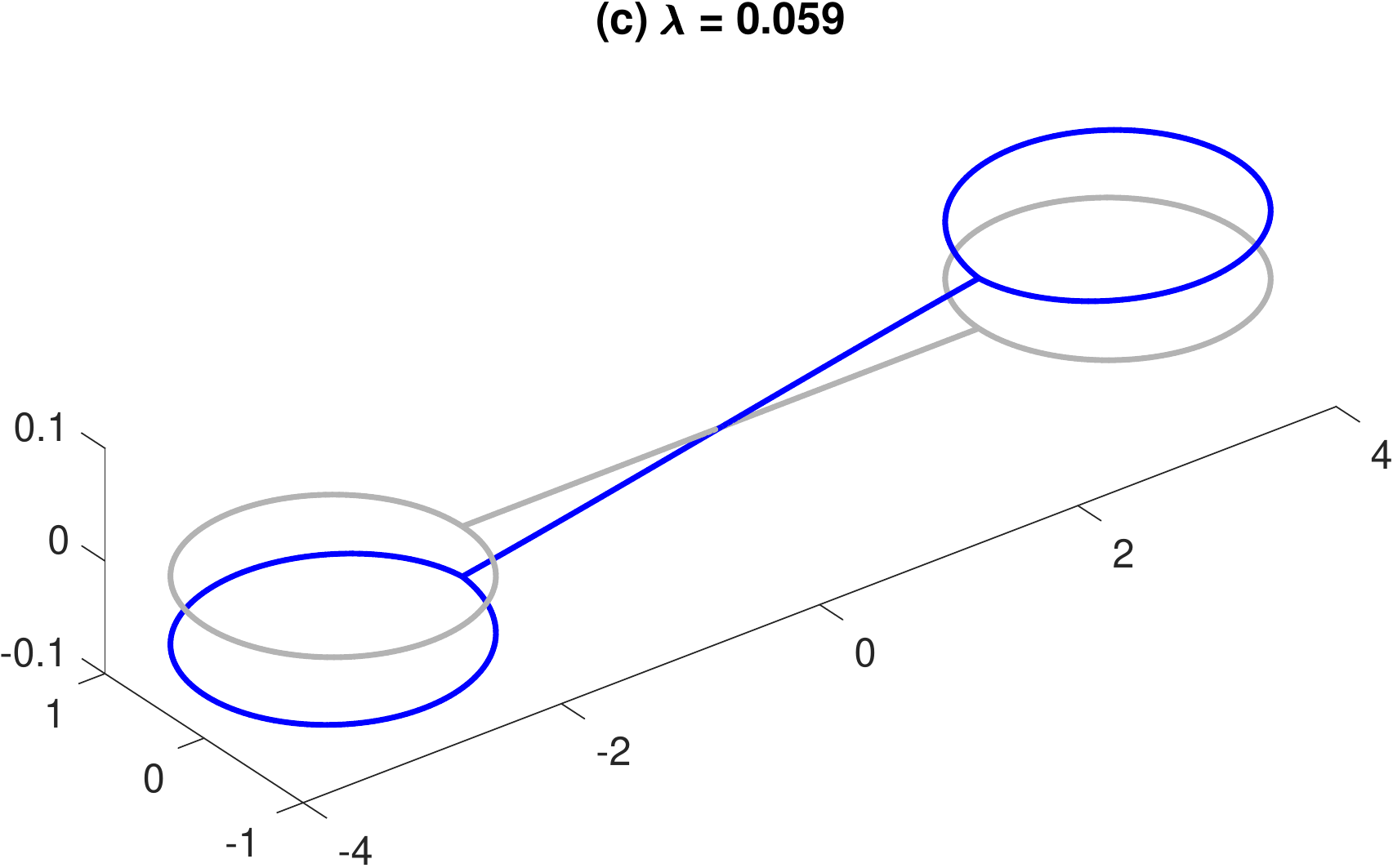} 
   \includegraphics[width=.31\textwidth]{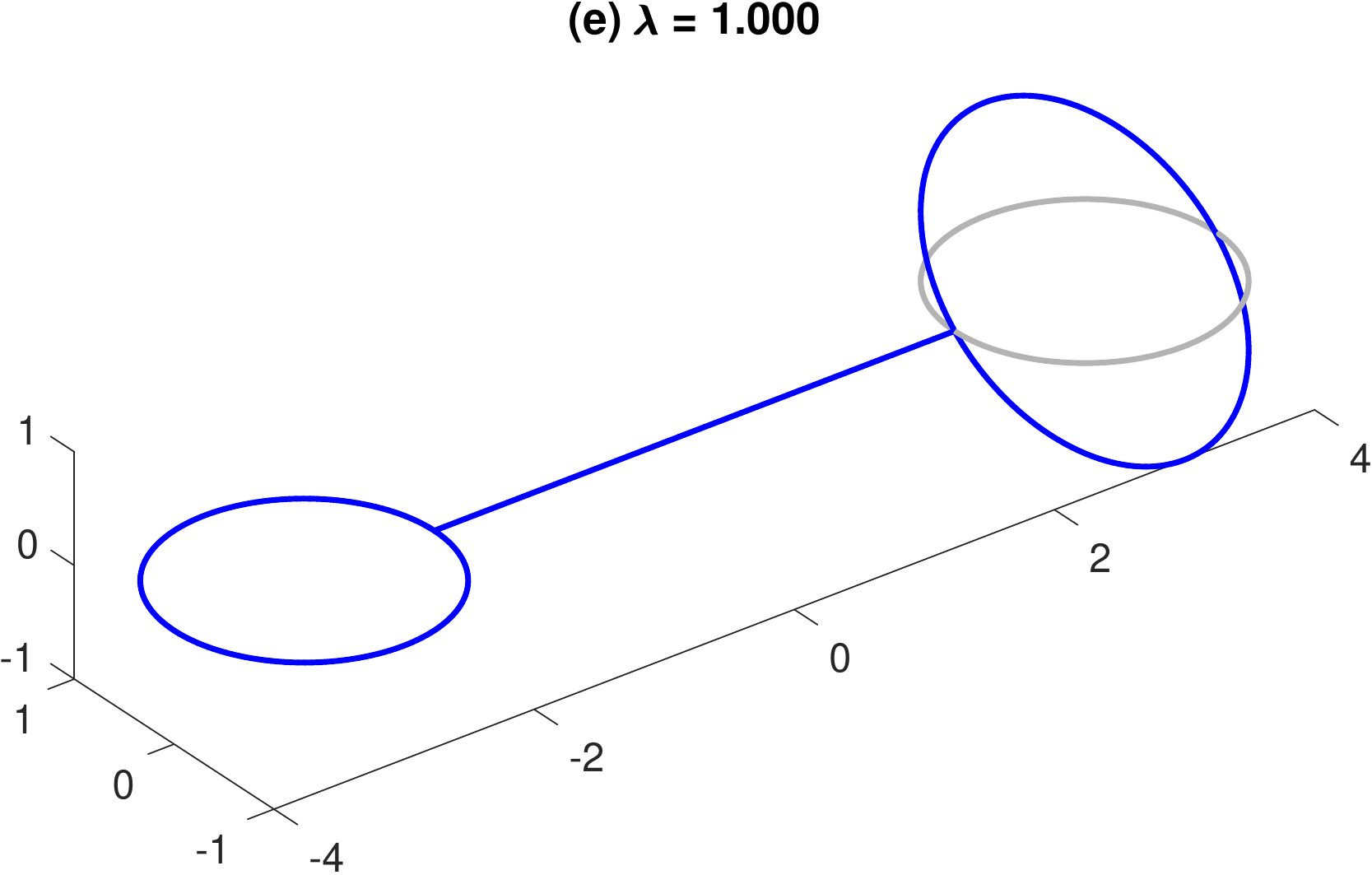} 
   \\
   \includegraphics[width=.31\textwidth]{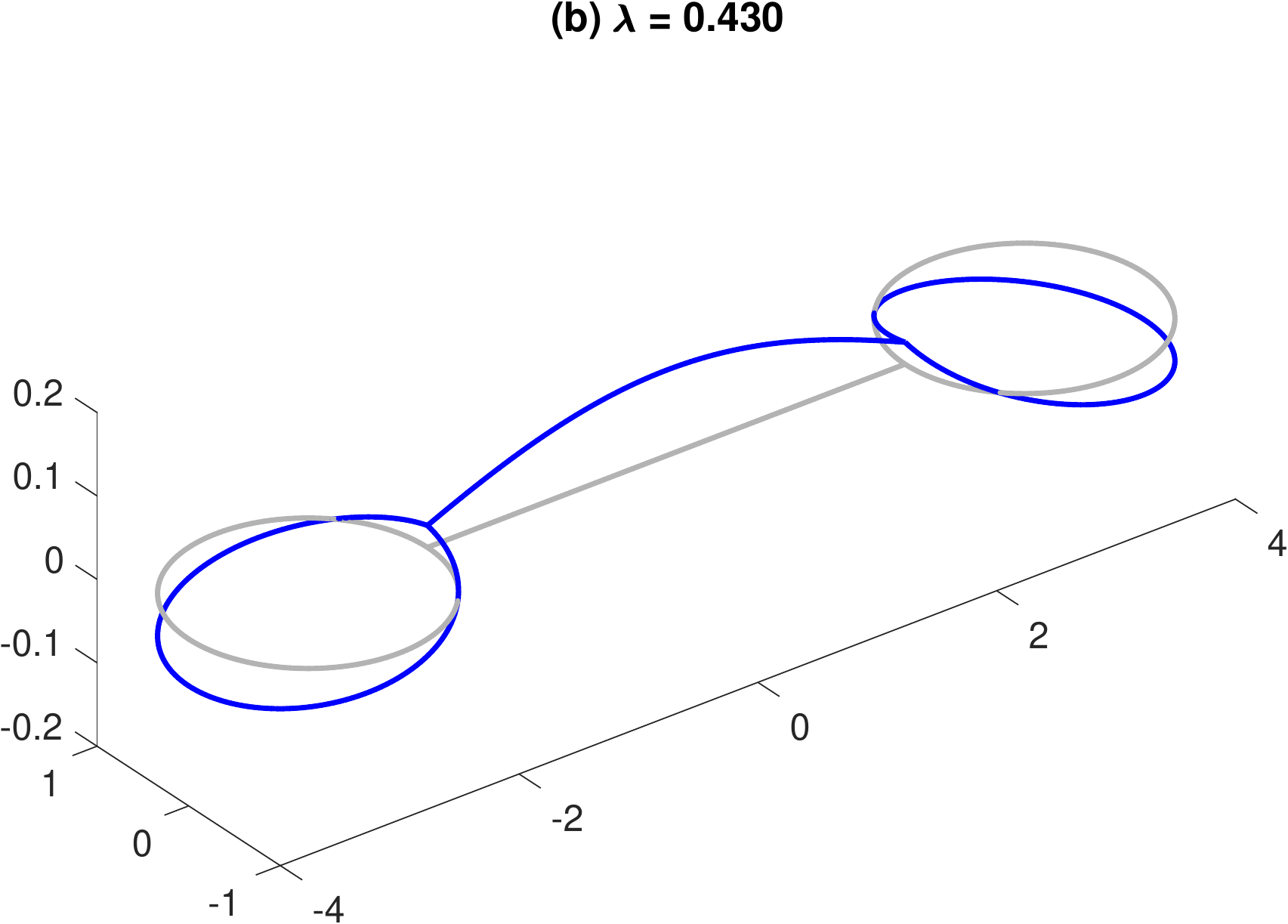} 
   \includegraphics[width=.31\textwidth]{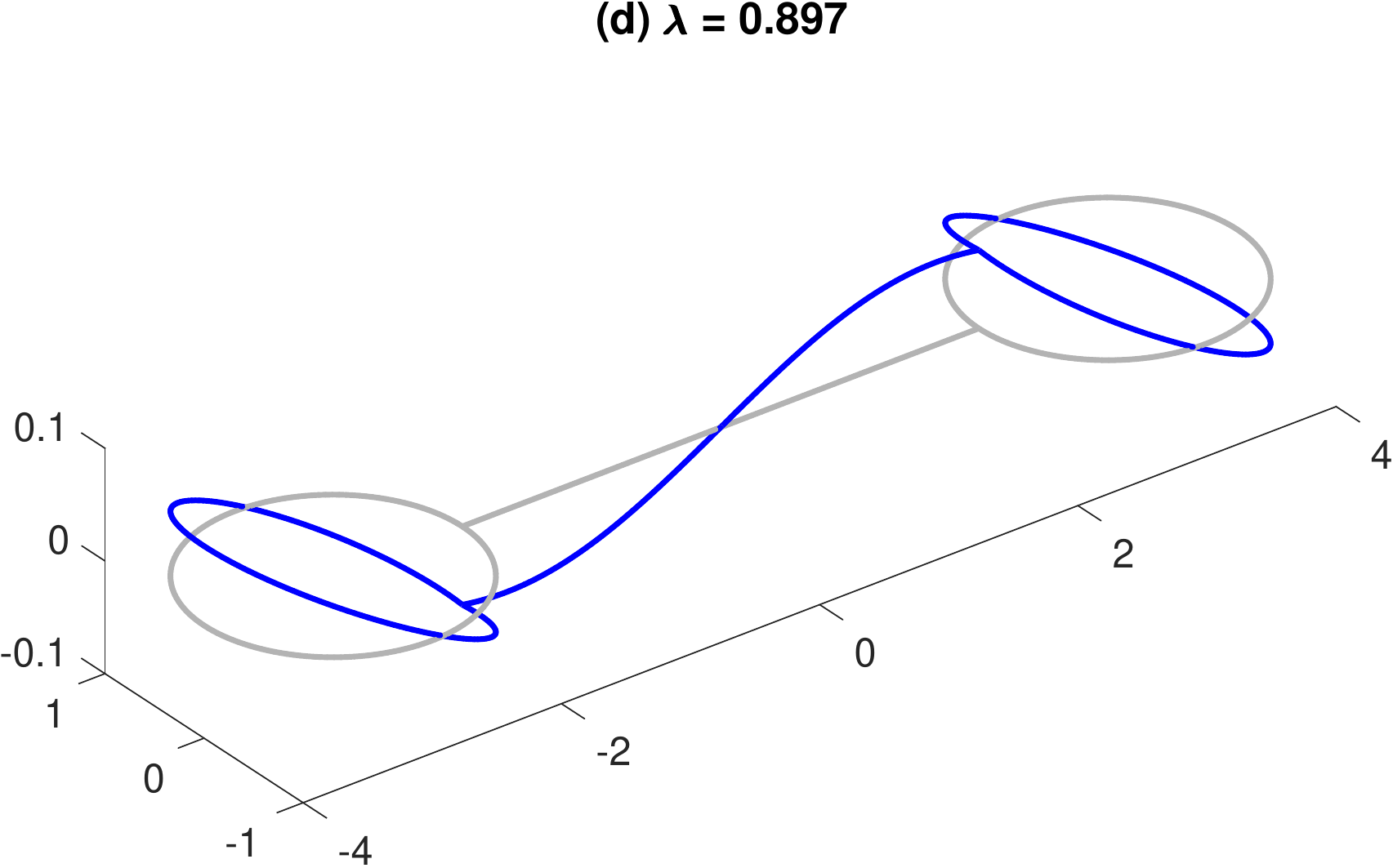} 
   \includegraphics[width=.31\textwidth]{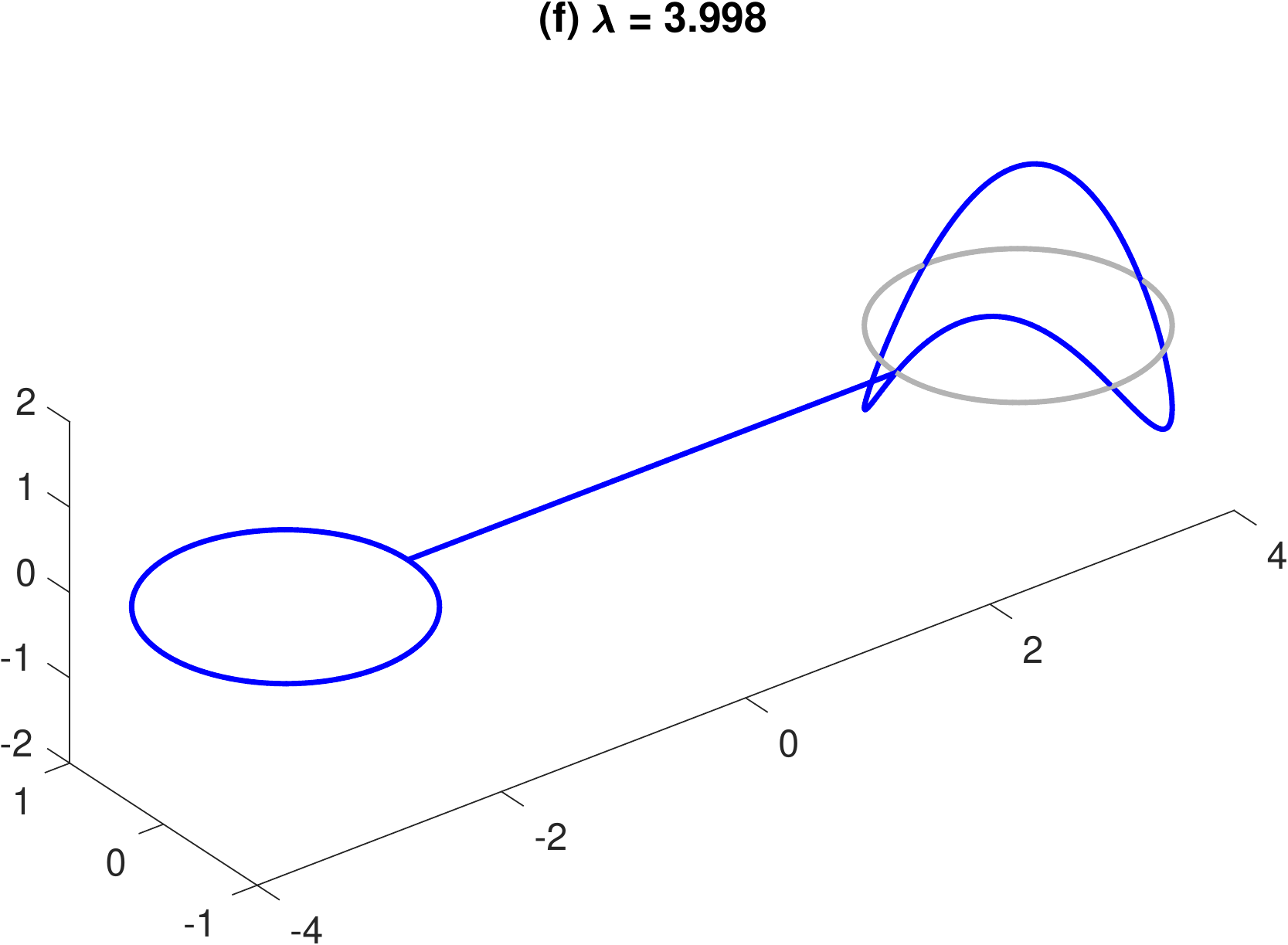} 
   \caption{The first two members of the even family of eigenfunctions (a-b), odd family (c-d), and loop-localized family (e-f) of the linear eigenvalue problem~\eqref{eigenproblem} on the dumbbell graph, computed numerically, along with the associated eigenvaluess. In subfigure (f) the analytical value is obviously $\lambda=4$, giving an indication of the accuracy of this computation.}
\label{fig:dumbbellEigs}
\end{figure}

The three factors of equation~\eqref{secularDet} correspond to three families of eigenfunctions:
\begin{description}
\item[Even Eigenfunctions]
Solutions in this family are nonzero on all three edges and satisfy $\cR_2(\Phi) = \Phi$. The eigenvalues in this family are labeled as $\w_j^2, \ j= 0,1,\ldots$ in~\cite{Marzuola:2016bl}. The linear ground state $\Phi_0=1$ is a member of this family, although the secular determinant equation holds only for $k\neq0$.
\item[Odd Eigenfunctions] Solutions in this family are nonzero on all three edges and satisfy $\cR_2(\Phi) = -\Phi$. The eigenvalues in this family are labeled as $\W_j^2, \ j= \in \ZZ^+$ in~\cite{Marzuola:2016bl}.
\item[Loop-localized Eigenfunctions] Solutions in this family have multiplicity two and can be normalized such that their support lies entirely on one or the other of the  edges  $\te_1$ and $\te_3$ and satisfy $\cR_1(\Phi)=-\Phi$ and $\cR_3(\Phi)=-\Phi$. These solutions are given by $\Phi(x) = \sin{(j x_m)}, \ j\in \ZZ^+$ on the edge $\te_m$ where they are supported and have eigenvalues $j^2$.
\end{description} 
 If $L$ is chosen as an integer multiple of $\pi/2$, eigenvalues from the loops may coincide with those of the even and odd eigenfunctions. We will assume no such resonances exist. Numerical examples, discretized using second-order finite differences, with ghost-points to implement the boundary conditions, are shown in Fig.~\ref{fig:dumbbellEigs}.

Ref.~\cite{Marzuola:2016bl} then considers the continuations of these standing waves into the nonlinear regime, where the power $Q(\Phi)$, defined in equation~\eqref{Q}, can be used as a bifurcation parameter, and where the linear problem applies in the limit of vanishing norm. The restriction of a standing wave to an individual edge can here be determined in terms of Jacobi elliptic functions; see Appendix~\ref{sec:cnoidal}.

Ref.~\cite{Marzuola:2016bl} shows that the constant solution $\Phi_0$ undergoes a pair of bifurcations with increasing $Q$, equivalently with decreasing $\Lambda$. The first is a symmetry-breaking bifurcation, at which the newly-created states are not invariant to $\cR_2$ but which together form a group orbit for that symmetry.  More concretely, two nonlinear eigenfunctions $\Phileft$ and $\Phiright$ emerge, each asymmetric, but with $\Phileft = \cR_2(\Phiright)$. The second, which they call a ``symmetry-preserving'' bifurcation, gives rise to a symmetric solution that is concentrated along the central edge $\te_2$ of the dumbbell as $\Lambda \to -\infty$. 

\begin{figure}[htbp] 
   \centering
   \includegraphics[width=3in]{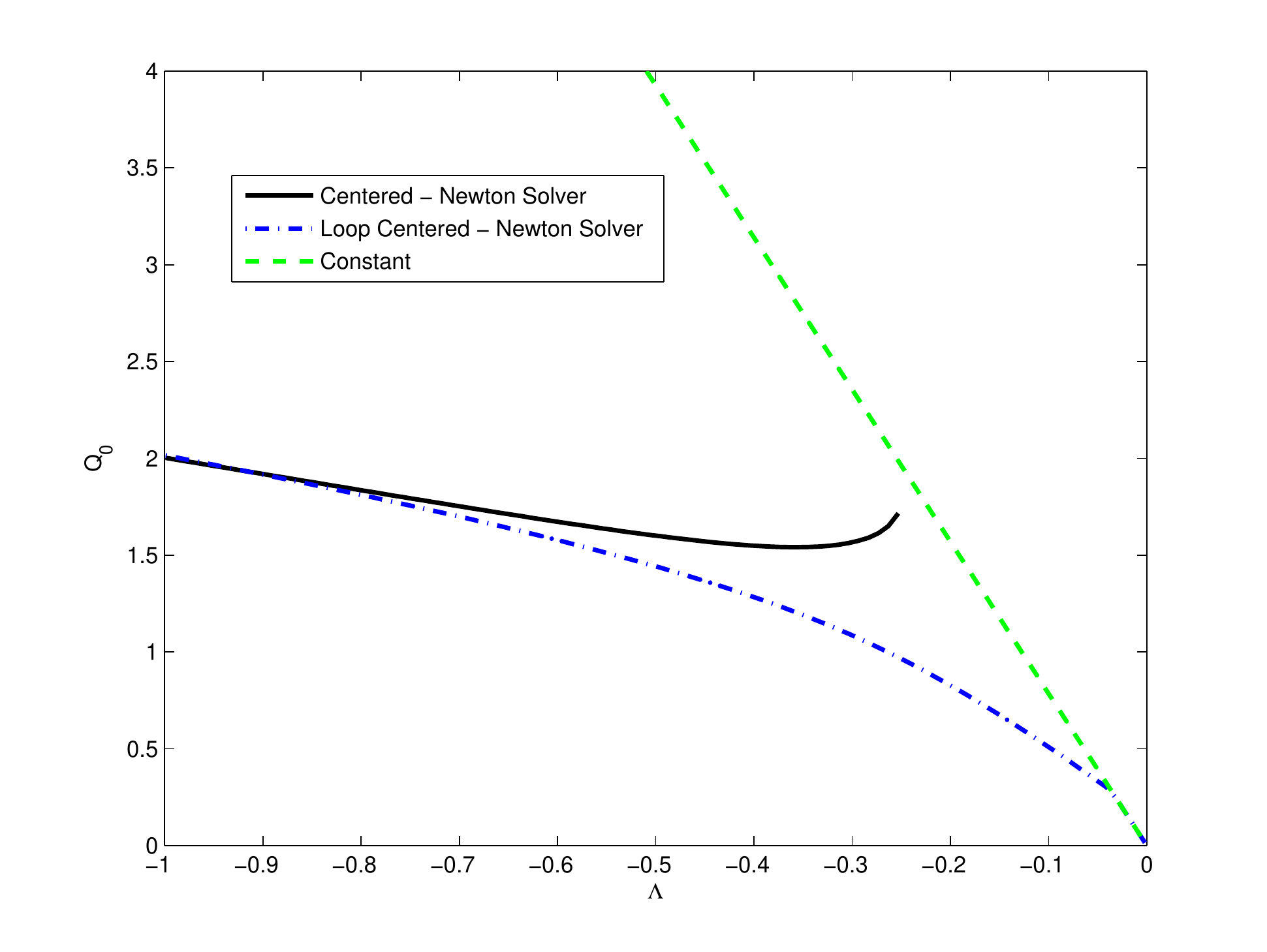} 
   \caption{A numerically computed bifurcation diagram from Ref.~\cite{Marzuola:2016bl}. }
\label{fig:Old}
\end{figure}

A bifurcation diagram computed in Ref.~\cite{Marzuola:2016bl} for $L= \pi/2$ is shown in Fig.~\ref{fig:Old}, showing the two bifurcations discussed in that paper. Observe that this diagram is incomplete: the branch labeled ``Centered'' does not meet the branch labeled ``Constant.'' The present paper was initially motivated by two questions: to understand what exactly occurs at the symmetry preserving bifurcation and to determine whether this effect is novel to the dumbbell quantum graph or if, instead, it occurs in simpler systems. The answers are simple: the symmetry-preserving bifurcation is of transcritical type, and the same phenomenon can be found in a significantly simpler system, the bowtie combinatorial graph. It appears that the closed loops in both the dumbbell and the bowtie contribute to the transcritical nature of the second bifurcation. In a simpler related problem, the Schr\"odinger equation on an interval with Neumann boundary conditions, all the bifurcations are pitchforks. This distinction is due to a subtle breaking of symmetry by the loops, as we shall discuss. Moreover, on the lollipop graph, featuring one loop and one straight edge, all of these bifurcations are transcritical.

The ground state of the quantum graph is defined to be the lowest energy state with a given $L^2$ norm, i.e.\ the solution of
\begin{equation}
E_0 = \inf_{\Phi \in \cE(\cL)} \left\{ E(\Phi)  \left\lvert \, Q(\Phi) = Q_0 \right. \right \}.
\label{grounddef}
\end{equation}
The ground state for the DST system is defined analogously for the DST with the Hamiltonian~\eqref{Hu} taking the role of the energy, in particular, it is the solution to
$$
H_0 = \inf_{\vu \in \RR^n} \left\{ H(\vu)  \left\lvert \, Q(\vu) = Q_0 \right. \right \}.
$$

 The central finding of Ref.~\cite{Marzuola:2016bl} concerned the ground state and was subsequently found to be mistaken. It has since been corrected by its authors online~\cite{Marzuola:2017}. Upon its creation in the symmetry-breaking bifurcation, the asymmetric \emph{loop-centered} state is the ground state of the system. The analysis of~\cite{Marzuola:2016bl} seemed to show, incorrectly, that for sufficiently large values of $L^2$, the \emph{centered} solution concentrated on the central edge of the graph has the lower energy. While this solution is a local constrained minimizer of the energy in equation~\eqref{grounddef}, while the loop-centered solution in the global minimizer.
 
\section{Three approaches to DST on the bowtie  graph}
\label{sec:bowtie}
\begin{figure}[htbp] 
   \centering
   \includegraphics[width=3in]{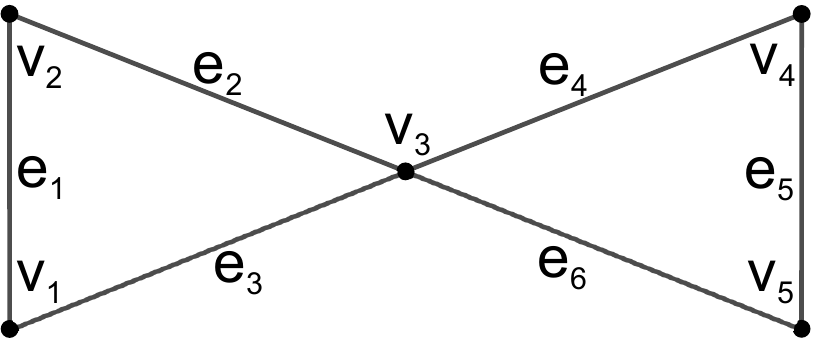} 
   \caption{The bowtie combinatorial graph.}
\label{fig:bowtie}
\end{figure}

Here we introduce a simpler combinatorial graph, on which the DST equation~\eqref{DST} undergoes the same sequence of bifurcations described in the previous section, which we consider a very coarse model of the bowtie quantum graph. The graph consists of five vertices, joined into a bowtie configuration as in Fig.~\ref{fig:bowtie}.

Setting the weights to one, the negative of the Laplacian matrix is
\begin{equation}
\label{bowtieDST}
\bL_\Gamma = \begin{pmatrix}
 2 & -1 & -1 & 0 & 0 \\
 -1 & 2 & -1 & 0 & 0 \\
 -1 & -1 & 4 & -1 & -1 \\
 0 & 0 & -1 & 2 & -1 \\
 0 & 0 & -1 & -1 & 2
 \end{pmatrix},
\end{equation}
with eigenvalues $\lambda_1=0$, $\lambda_2=5$, $\lambda_3=1$, and $\lambda_4=\lambda_5=3$
(numbered in this strange order purposely) and associated normalized eigenvectors,
$$
\vv_1 = \frac{1}{\sqrt{5}} \begin{pmatrix} 1 \\ 1 \\ 1 \\ 1 \\ 1 \end{pmatrix}, 
\vv_2 = \frac{1}{\sqrt{20}} \begin{pmatrix}1 \\ 1 \\ -4 \\ 1 \\ 1 \end{pmatrix},
\vv_3 = \frac{1}{2} \begin{pmatrix}-1 \\ -1 \\ 0 \\ 1 \\1 \end{pmatrix},
\vv_4 = \frac{1}{\sqrt{2}} \begin{pmatrix}1 \\ -1 \\ 0  \\ 0 \\ 0 \end{pmatrix},
\vv_5 = \frac{1}{\sqrt{2}} \begin{pmatrix}0  \\ 0 \\ 0  \\ 1 \\ -1 \end{pmatrix}.
$$
The DST system on this graph is invariant to left multiplication by the three permutations, in cyclic notation:
$$
R_1 = (12), \, R_2 = (14)(25), \, \text{and } R_3 = (45),
$$
which are discrete analogs of the symmetries~\eqref{dumbbellSymmetries}.%
\footnote{Recall that in cyclic notation, each set of indices grouped between parentheses represents a subset of list elements which undergoes a cyclic permutation. Fixed points, which would be represented by singleton groups are, by convention, omitted.}
The space $\cS_1 =\spann{\{\vv_1,\vv_2\}}$ is invariant under $R_1$, $R_2$, and $R_3$, while the space $\cS_2=\spann{\{\vv_1,\vv_2,\vv_3\}}$ is invariant under just $R_1$ and $R_3$. The space $\cS_3 =\spann{\{\vv_4,\vv_5\}}$ is invariant under $R_2$. Put more simply, on $\cS_1$, $u_1 = u_2 = u_4 = u_5$, on $\cS_2$, $u_1 = u_2$ and $u_4 = u_5$, and on $\cS_3$, $u_1=-u_2$, $u_3=0$, and $u_4=-u_5$.

The change of variables to diagonalize the linear part, 
$$
\vu = \sum_{j=1}^5 z_j \vv_j,
$$
is canonical (i.e.\ it preserves Hamiltonian structure) and results in the Hamiltonian
\begin{equation}
\label{Hdiag}
\begin{split}
H  = & 5 \abs{z_2}^2 + \abs{z_3}^2 + 3 \abs{z_4}^2 + 3 \abs{z_5}^2  
- \frac{1}{50}\abs{z_1 - 2 z_2}^4  \\
& - \frac{1}{800} \left(
   \abs{2 z_1+z_2+\sqrt{5} z_3 +  \sqrt{10} z_4}^4
 + \abs{2 z_1+z_2+\sqrt{5} z_3 -  \sqrt{10} z_4}^4  \right.\\
& + \left.   \phantom{\frac{1}{800}\left(\right.}
   \abs{2 z_1+z_2 -\sqrt{5} z_3 +  \sqrt{10} z_5}^4 
 + \abs{2 z_1+z_2-\sqrt{5} z_3 -  \sqrt{10} z_5}^4
 \right).
\end{split}
\end{equation}
In these coordinates, the subspaces $\cS_1$--$\cS_3$ are all invariant under the dynamics of $H$, which are now written as $\cS_1 = \{(z_1,z_2,0,0,0)\}$, $\cS_2 = \{(z_1,z_2,z_3,0,0)\}$, $\cS_3  = \{(0,0,0,z_4,z_5)\}$.

\subsection{Linear Stability via Symmetry Reduction on $\cS_2$}
The two bifurcations discussed in Ref.~\cite{Marzuola:2016bl} take place on the invariant subspace $\cS_2$, obtained by setting $z_4 = z_5 = 0$ in Hamiltonian~\eqref{Hdiag}. The canonical change of variables 
\begin{equation}
\label{ztoy}
z_1 = \sqrt{R- \abs{y_2}^2 - \abs{y_3}^2} e^{i\phi}, \,
z_2 = y_2 e^{i\phi}, \,
z_3 = y_3 e^{i\phi}
\end{equation}
yields a Hamiltonian that is independent of $\phi$. Thus, $R = \abs{z_1}^2 + \abs{z_2}^2 + \abs{z_3}^2$, the conjugate variable to $\phi$, is conserved. This reduces the number of degrees of freedom to two. In these variables, the continuation of the linear ground state is represented by $y_2 = y_3 = 0$. The leading order quadratic part of the Hamiltonian, obtained from Hamiltonian~\eqref{Hdiag} by setting $z_4=z_5=0$, using change of variables~\eqref{ztoy} and ignoring quartic terms, is given by
$$
H_{\rm lead} = 5 \abs{y_2}^2 +  \abs{y_3}^2
-\frac{R}{10} {\left( y_2 + \ybar_2 \right)}^2
-\frac{R}{10} {\left(y_3 + \ybar_3 \right)}^2.
$$
Applying equation~\eqref{iundot} to this Hamiltonian yields linearized evolution equations
$$
i \dot{y}_2 = 5 y_2-\frac{R}{5} \left(y_2+\ybar_2\right), \,
i \dot{y}_3 =  y_3-\frac{R}{5} \left(y_3+\ybar_3\right), 
$$
with eigenvalues
$$
\lambda = \pm \sqrt{\tfrac{2R}{5}-1} \text{ and } 
\lambda = \pm \sqrt{2R-25}.
$$
For small values of $R$ all eigenvalues are imaginary and the ground state is stable.
The system has bifurcations at $R = \tfrac{5}{2}$ and $R = \tfrac{25}{2}$, which we will see are of pitchfork and transcritical type, respectively, and involve the directions $y_3$ and $y_2$, respectively.

\subsection{Geometric Reduction on $\cS_1$}
In order to better understand the analog of ``symmetry preserving'' bifurcation of Ref.~\cite{Marzuola:2016bl}, we consider the restriction to the invariant subspace $\cS_1$ on which solutions are invariant to all three discrete symmetries. On this subspace 
\begin{equation*}
H = 
5 \abs{z_2}^2
-\frac{1}{10}\abs{z_1}^4
-\frac{1}{10} \left( z_1^2 \zbar_2^2 + 4 \abs{z_1}^2 \abs{z_2}^2 + \zbar_1^2 z_2^2  \right)
+\frac{3}{10} \abs{z_2}^2 \left(z_1 \zbar_2 + \zbar_1 z_2 \right)
-\frac{13}{40} \abs{z_2}^4.
\end{equation*}
This Hamiltonian is $S^1$ invariant, i.e.\ $H(e^{i\phi}z_1,e^{i\phi}z_2)= H(z_1,z_2)$. This implies that 
\begin{subequations}
\label{RXYZ}
\begin{equation}
R = \abs{z_1}^2+ \abs{z_2}^2 
\label{R}
\end{equation}
is conserved, i.e.\ that trajectories are confined to three-spheres. Therefore the following change of coordinates, the Hopf fibration of $S^3$, which is standard in geometric mechanics~\cite{Holm:2011uw}, applies. We define a new change of variables given by equation~\eqref{R} and
\begin{equation}
Z = \abs{z_1}^2 - \abs{z_2}^2, \,
X+ i Y = 2 \zbar_1 z_2.
\label{XYZ}
\end{equation}
\end{subequations}
While this change of variables is non-canonical, they put the system in a \emph{Poisson} form which has many similar properties to a Hamiltonian system.

Then a simple calculation confirms that
\begin{equation}
R^2 = X^2 + Y^2 + Z^2,
\label{sphere}
\end{equation}
Thus $R$ is a Casimir for the reduced system and $(X,Y,Z)$ evolve as a Poisson system
\begin{equation}
\frac{d}{dt} \begin{pmatrix} X \\ Y \\ Z \end{pmatrix} 
= -\nabla\frac{R^2(X,Y,Z)}{2} \times H(X,Y,Z),
\label{Poisson}
\end{equation}
where the $\times$-symbol denotes the cross product in $\RR^3$ and, in these coordinates,
$$
H = 
-\frac{5 Z}{2}
+ \frac{3 R X}{20} 
+ \frac{9 R Z}{80}
- \frac{X^2}{20}
- \frac{3 X Z}{20}
+ \frac{Y^2}{20}
-\frac{Z^2}{160}
+ \left( \frac{5 R}{2}
-\frac{33 R^2}{160} \right).
$$
While the derivation and theory of such systems would take us too far afield, we note that the derivation of evolution equations~\eqref{Poisson} from the general Hamiltonian form~\eqref{iundot} and the change of variables~\eqref{RXYZ} is an exercise in the chain rule.

Then $(X,Y,Z)$ evolve according to 
\begin{subequations}
\label{XYZdot}
\begin{align}
\dot X & = \frac{1}{80}  (-9 R+12 X+9 Z+200) Y;\label{Xdot} \\
\dot Y & = \frac{1}{80} \left(-12 X^2 + 7 X Z + 12 Z^2 +(9 R -200) X-12 R Z \right);\label{Ydot} \\
\dot Z & = \frac{1}{20}  (3 R-4 X-3 Z) Y.\label{Zdot}
\end{align}
\end{subequations}
It follows from equations~\eqref{Xdot} and~\eqref{Zdot} that the fixed points  satisfy $Y=0$ (which implies that $z_1/z_2 \in \RR$) and, from equations~\eqref{sphere} and~\eqref{Ydot}
\begin{subequations}
\label{XZfix}
\begin{align}
X^2 + Z^2 & = R^2; \\
-12 X^2 + 7 X Z + 12 Z^2 +(9 R -200) X-12 R Z  & =0.
\end{align}
\end{subequations}
These are the equations for a circle and a hyperbola. The solution $(X,Z)=(0,R)$ holds for all positive values of $R$ and retracing our steps, corresponds to the continuation of the linear ground state solution $\vu \propto \vv_1$.

\begin{figure}[htbp] 
    \centering
    \includegraphics[height=0.2\textwidth]{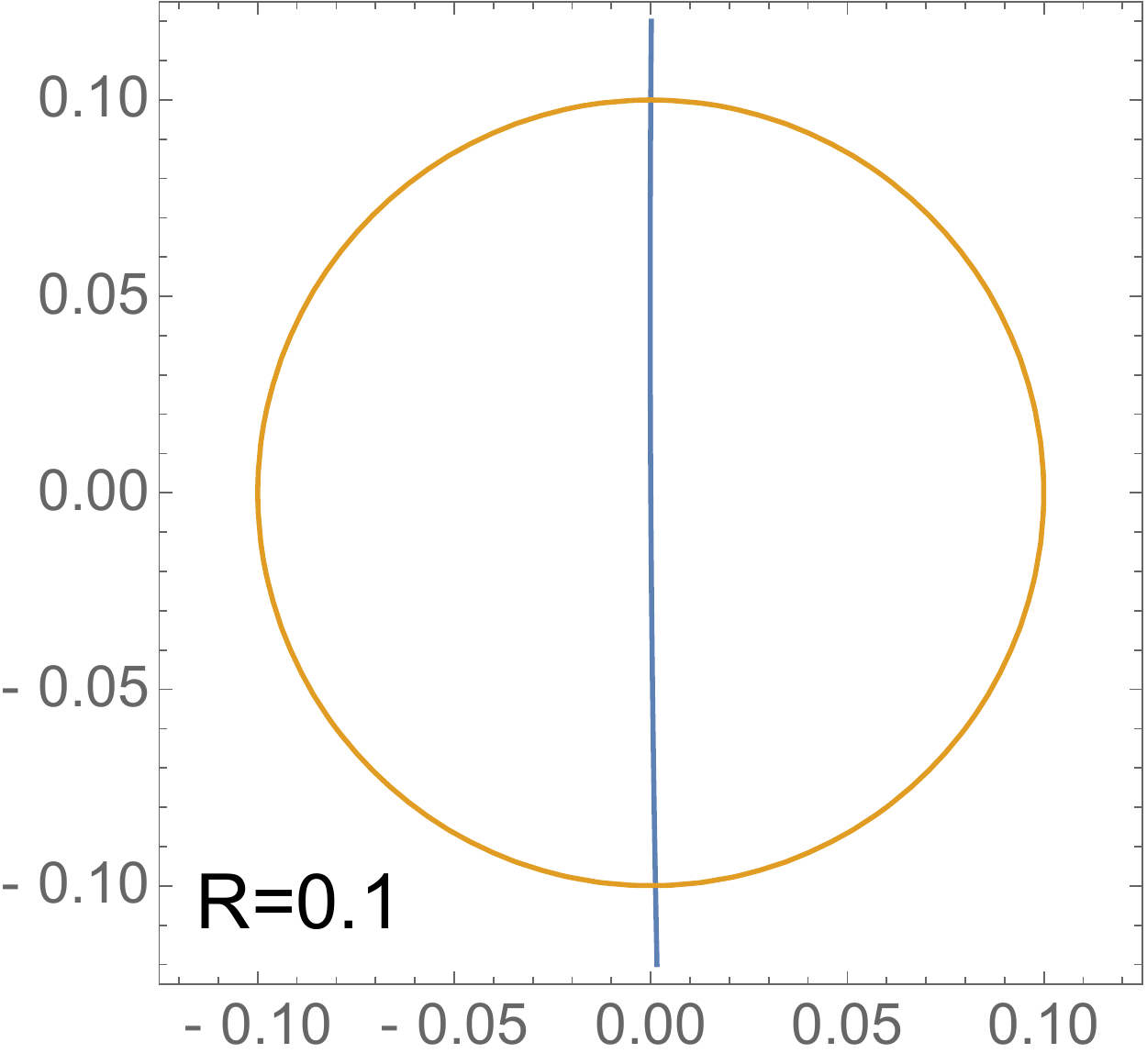}
    \includegraphics[height=0.2\textwidth]{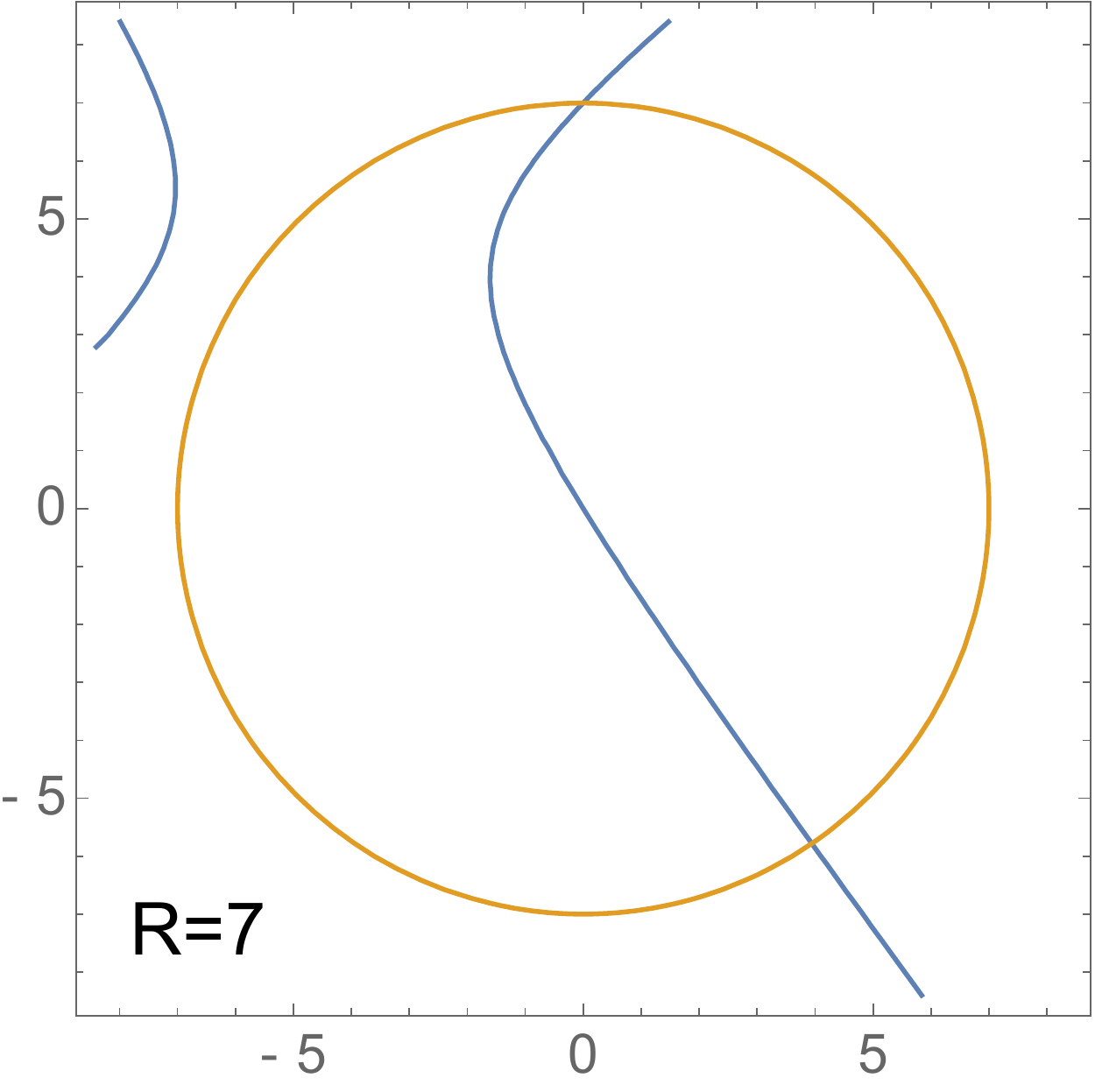}
    \includegraphics[height=0.2\textwidth]{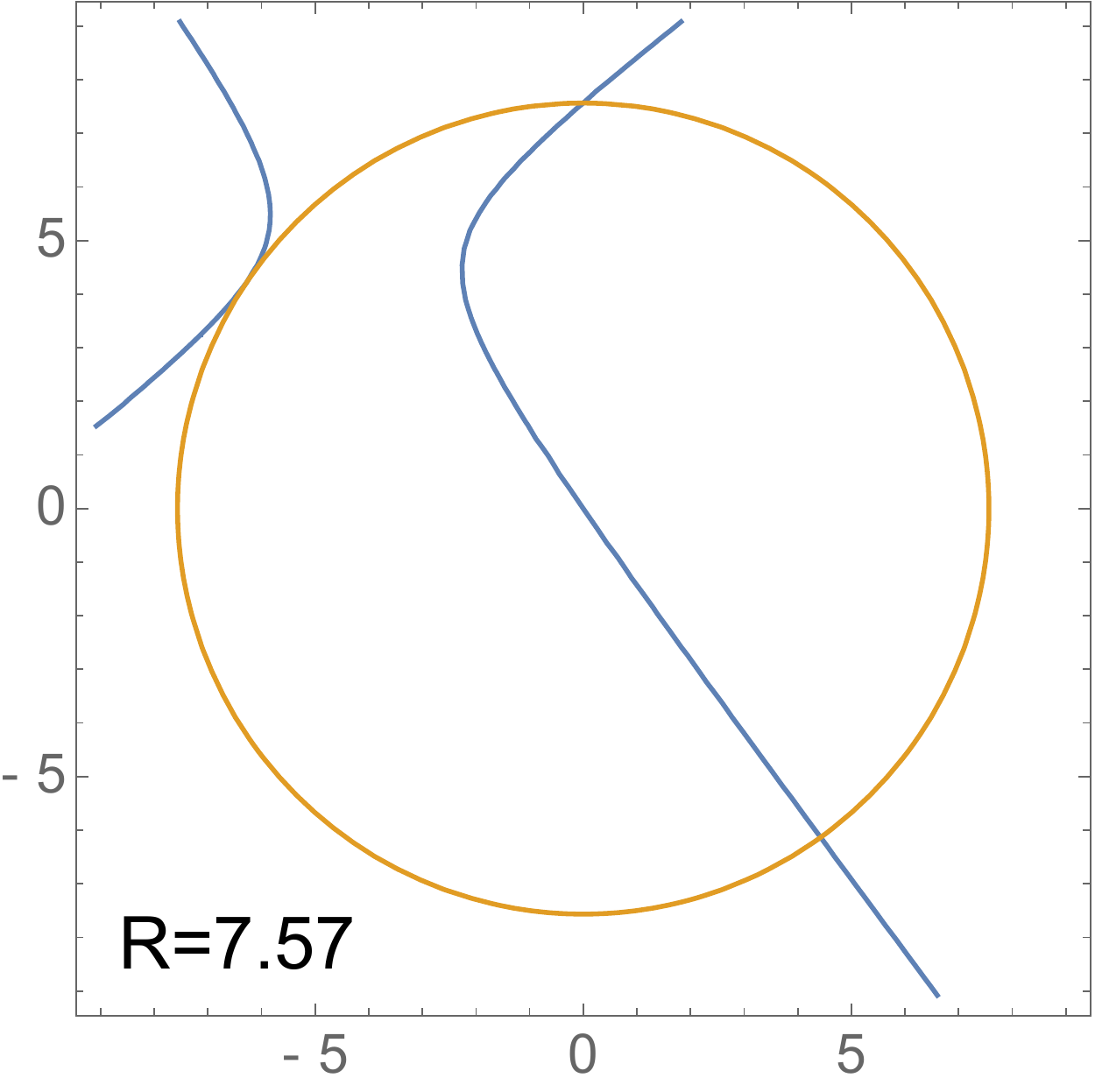}
    \includegraphics[height=0.2\textwidth]{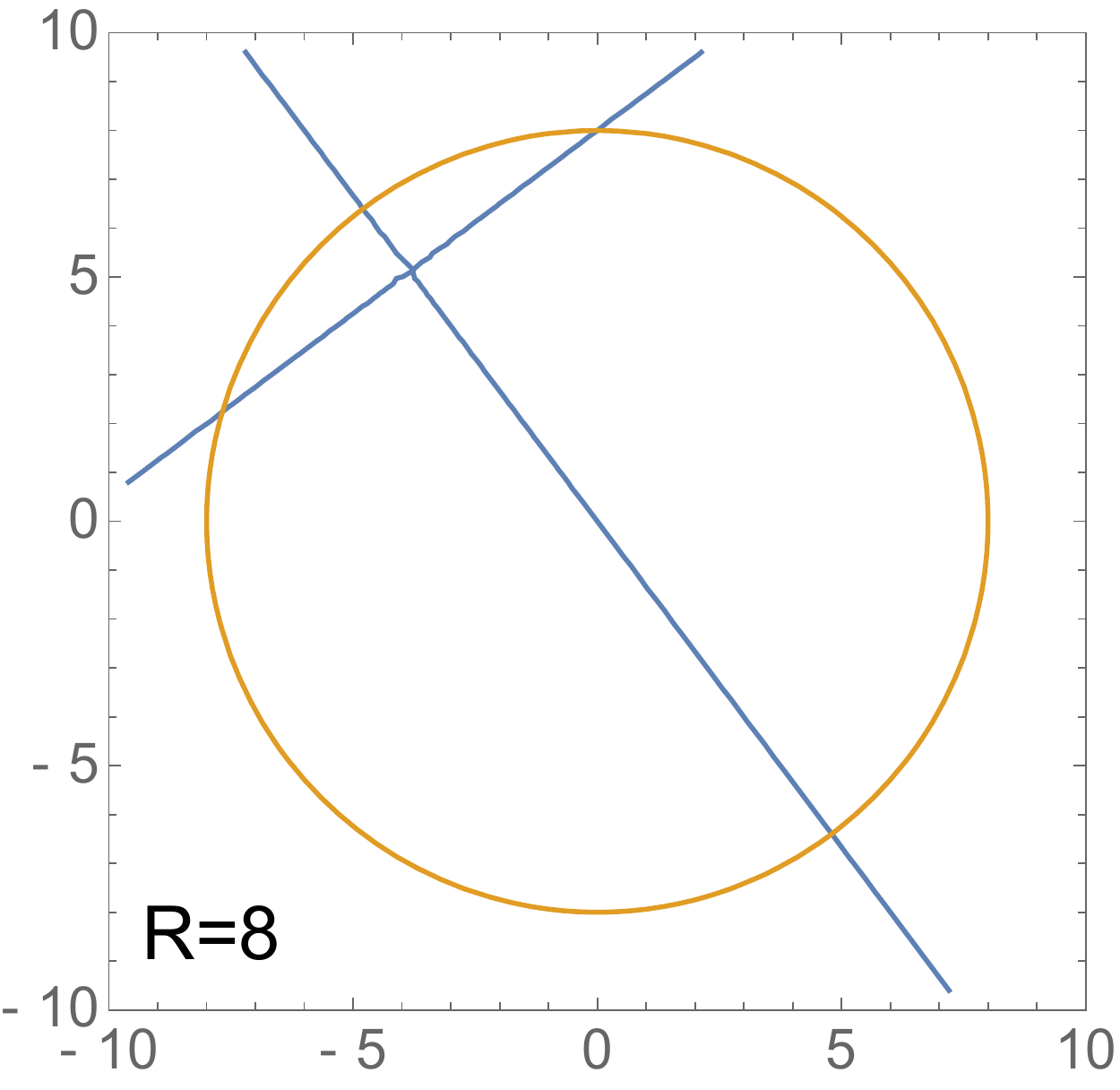}\\
    \includegraphics[height=0.2\textwidth]{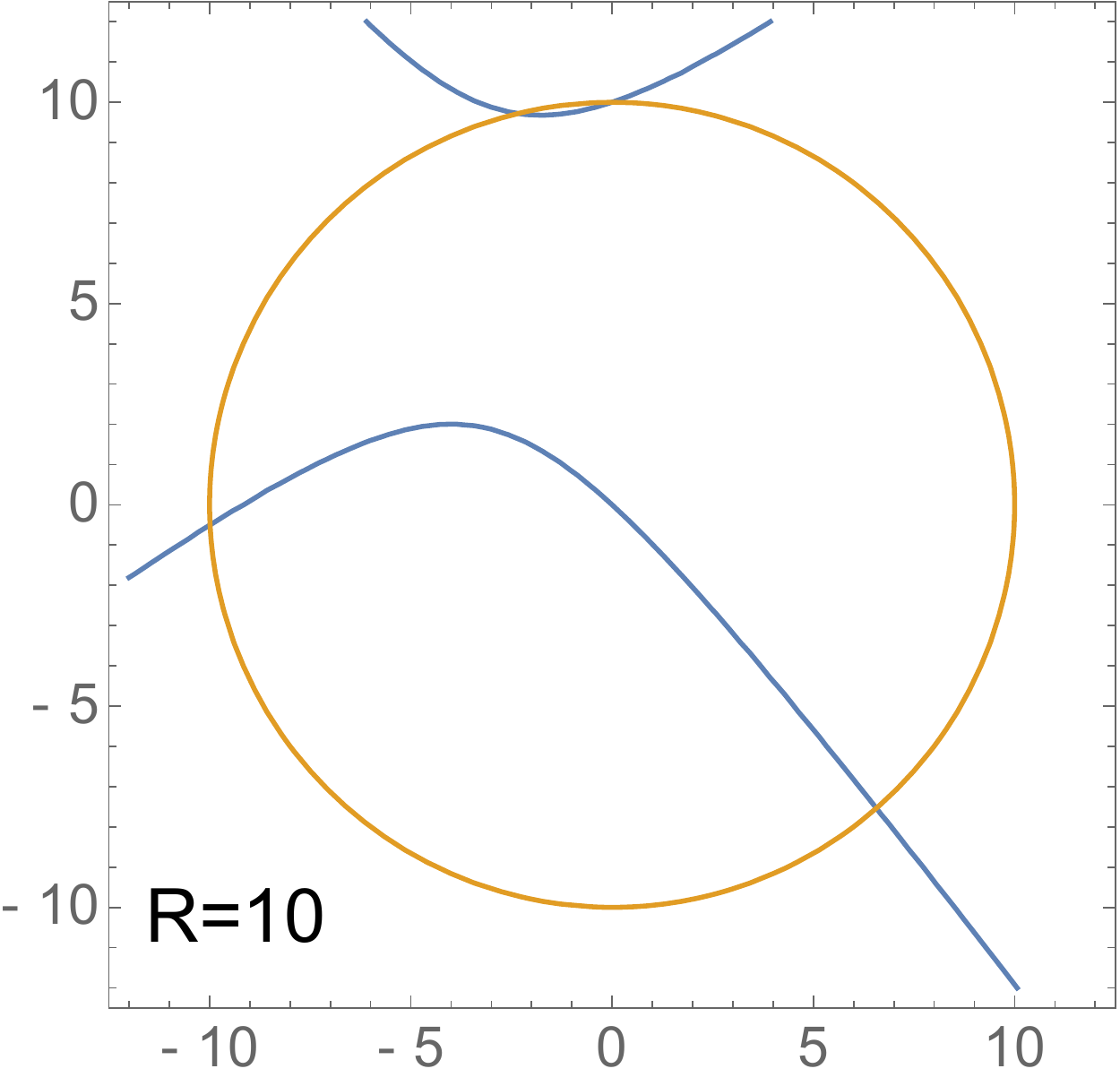}
    \includegraphics[height=0.2\textwidth]{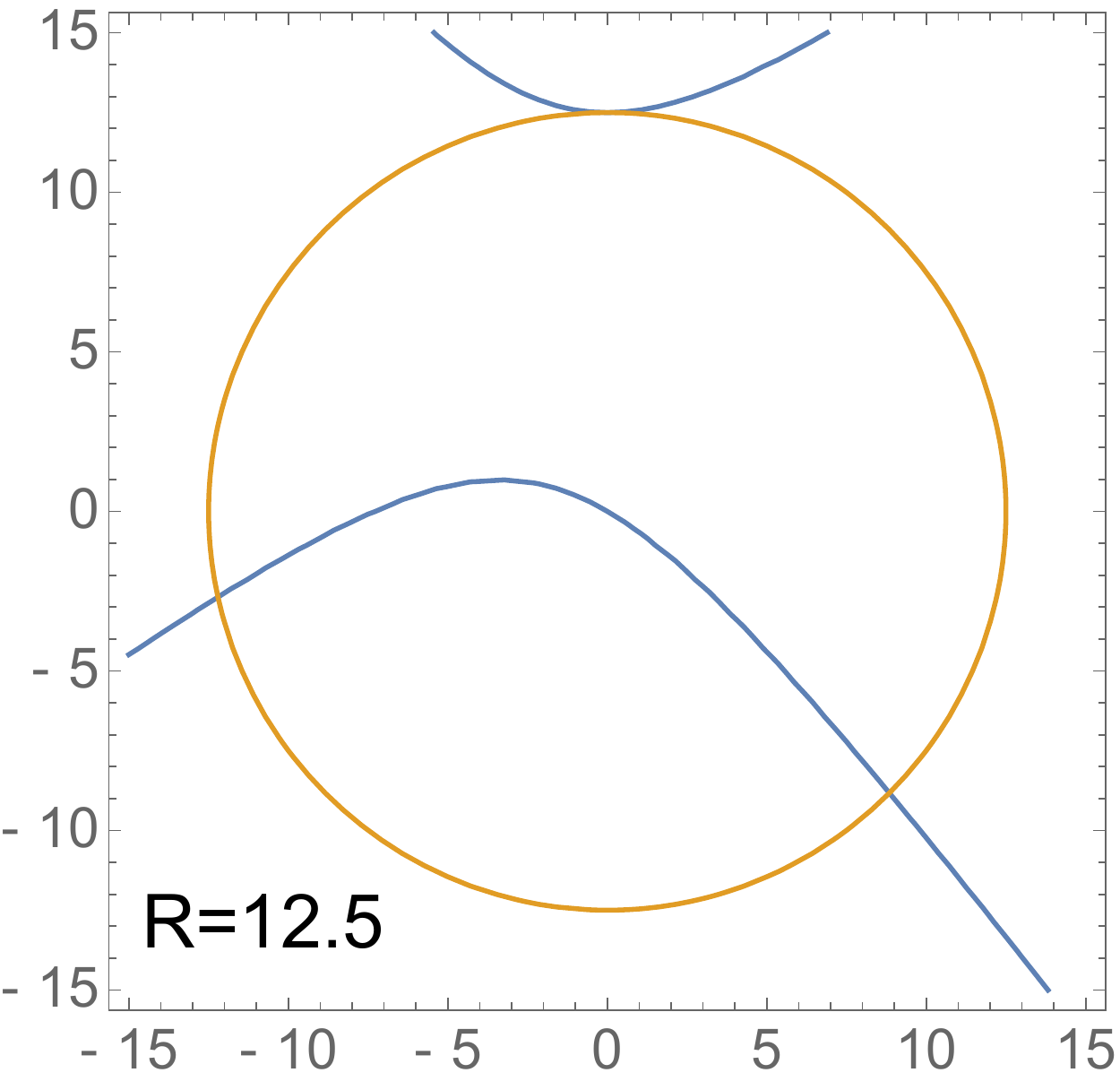}
    \includegraphics[height=0.2\textwidth]{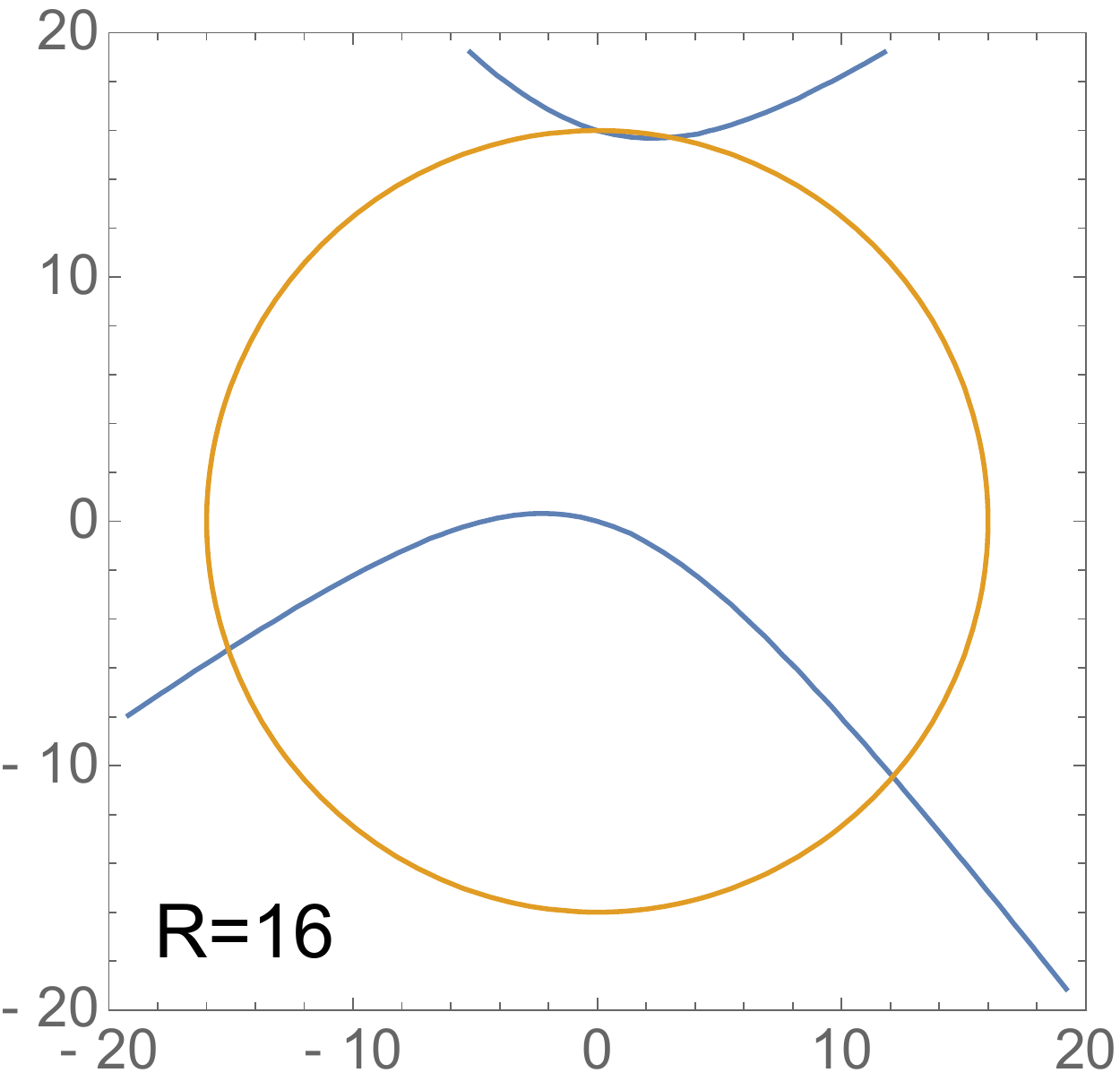}
    \caption{The solutions of system~\eqref{XZfix} in $(X,Z)$, given by the intersections of the two curves, showing the sequence of bifurcations described in the text. Note in particular that for $R=10$ there exists a fixed point just to the left of $(0,R)$, which has moved to the right for $R=16$.}
\label{fig:hypercirc}
 \end{figure}

For small values of $R$ there are two fixed points. The second corresponds to the linear solution $\vu \propto \vv_2$ which approaches $(X,Z) \approx (0,-R)$ as $R \to 0^+$. As $R$ increases there are two bifurcations. First at $R \approx 7.57$,%
\footnote{The precise value, computed by eliminating $Z$ from system~\eqref{XZfix} using the resultant, and then setting the discriminant with respect to $X$ to zero is $R=-13-\sqrt{241-12 \cdot 15^{2/3}}+\sqrt{482+12 \cdot 15^{2/3}+\tfrac{7378}{\sqrt{241-12 \cdot 15^{2/3}}}}$.}
 the other branch of the hyperbola crosses the circle, leading to a new pair of fixed points. At $R=8$, the branches of the hyperbola merge and re-connect, which is not itself a bifurcation of fixed points, but sets the stage for a transcritical bifurcation at $R=\tfrac{25}{2}$. This is the analog of Marzuola et al.'s ``symmetry-preserving bifurcation;'' see Fig.~\ref{fig:hypercirc}.

\subsection{Parameterization of solution branches on $\cS_2$}
In this section, we follow~\cite{Eilbeck:1985tu} and use simple methods from algebraic geometry to enumerate all the relative fixed points of the discrete bowtie system that lie on the invariant subspace $\cS_2$, i.e.\ we look for stationary solutions of the equation~\eqref{DST} with Laplacian matrix given by~\eqref{bowtieDST} of the form
$$
u_1=u_2= a e^{i \W t}, \, u_3 = b e^{i \W t}, \, u_4=u_5 = c e^{i \W t}; \; a,b,c,\W \in \RR.
$$
The reduced system consists of the three equations
\begin{subequations}
\label{abceqns}
\begin{align}
a-b -a^3 - \W a & = 0;\label{abca}\\
-2 a +4 b -2 c -b^3 - \W b & = 0;\label{abcb}\\
-b + c -c^3 - \W c & = 0.\label{abcc}
\end{align}
\end{subequations}

We will derive several branches of solutions. For later convenience, we will assign each branch a number. We first derive two simple solutions:

\begin{itemize}
\item \textbf{Branch 1} is the constant state $a = b = c$, the continuation of the eigenvector $\vv_1$, for which $\W = - a^2$. The squared $\ell^2$ norm for this solution is just $Q = 5 a^2$, so that $\W = -Q/5$.

\item\textbf{Branch 2} satisfies $c = -a$, $b = 0$, the continuation of eigenvector $\vv_3$, with $\W = 1-a^2$. This solution has $Q = 4 a^2$, so $\W = 1-Q/4$.
\end{itemize}

We  parameterize all the remaining solutions as follows. 
Subtracting equation~\eqref{abca} from equation~\eqref{abcc} yields 
\begin{equation}
(c-a)(a^2 + ac + c^2 + \W -1) = 0.
\label{ac_only}
\end{equation}
Letting either factor equal zero yields a solution. Setting the first factor to zero yields $a=c$, and the solution lies on the invariant subspace $\cS_1$. This reduces system~\eqref{abceqns}  to
\begin{subequations}
\begin{align}
a-b -a^3 - \W a & = 0;\label{aW}\\
-4 a +4 b-b^3 - \W b & = 0.
\end{align}
\label{abWsystem}
\end{subequations}
Subtracting the two and factoring yields
\begin{equation*}
(a-b) \left(a^2+a b+b^2 +\W -5\right) = 0.
\end{equation*}
Setting the first factor to zero yields Branch 1 already discussed. 
Setting the second factor to zero defines an ellipse when $\Omega<5$:
$$ \frac{3}{4} {\left(a+b\right)}^2 + \frac{1}{4} {\left(a-b\right)}^2 = 5 -\Omega.$$
This can be parameterized straightforwardly as $a+b = \frac{2}{\sqrt{3}}\sqrt{5-\W}\cos{\theta}$, $a-b = 2 \sqrt{5-\W} \sin{\theta}$. Basic trigonometry then yields
\begin{subequations}
\label{abWSolution}
\begin{equation}
a  = \frac{2\sqrt{5-\W}}{\sqrt3}  \sin{\left(\theta - \frac{\pi}{3}\right)}; \;
b  = \frac{2\sqrt{5-\W}}{\sqrt3}  \sin{\left(\theta + \frac{\pi}{3}\right)}.
\end{equation}
 Plugging this back into equation~\eqref{aW} to solve for $\W$ and simplifying yields
\begin{equation}
\W  = 5- \frac{3}{2} \csc{3 \theta } \left(3 \sqrt{3} \cos{\theta}-5 \sin{\theta}\right).
\end{equation}	
\end{subequations}
The solution satisfies the necessary condition that $\Omega<5$ for two subintervals of $[0,\pi]$, yielding two additional branches:
\begin{itemize}

\item \textbf{Branch 3} Solution~\eqref{abWSolution} with $0<\theta<\tan^{-1}{\left(\tfrac{3\sqrt{3}}{5}\right)}$ yields a branch with $(\W,Q) \to (-\infty,\infty)$ as $\theta \to 0^+$ and $(W,Q) \to (5,0)$ as $\theta \to \tan^{-1}{\left(\tfrac{3\sqrt{3}}{5}\right)}^-$, so this is the nonlinear continuation of the eigenvector $\vv_2$.

\item \textbf{Branch 4} Solution~\eqref{abWSolution} with $\frac{\pi}{3}<\theta<\frac{2\pi}{3}$. The branch diverges as $\theta$ approaches the endpoints of the interval. At $\theta=\frac{\pi}{2}$, this yields $a=b=\sqrt{-\W}=\sqrt{\tfrac{5}{2}}$, i.e.\ it lies on Branch 1. This is the transcritical bifurcation.
\end{itemize}

Setting the second factor of equation~\eqref{ac_only} to zero allows us to solve for $a$ and $c$ in the same manner that we solved for $a$ and $b$ in deriving Branches 3 and 4,
\begin{subequations}
\label{abcWSolution}
\begin{equation}
a  = \frac{2\sqrt{1-\W}}{\sqrt3}  \sin{\left(\theta - \frac{\pi}{3}\right)}; \;
c  = \frac{2\sqrt{1-\W}}{\sqrt3}  \sin{\left(\theta + \frac{\pi}{3}\right)}.
\end{equation}
Substituting these values back into equation~\eqref{abca} or~\eqref{abcc} and simplifying yields a formula for $b$
\begin{equation}
b = \frac{-2\left(1-\W\right)^{3/2}\sin{(3\theta)}}{3\sqrt{3}}.
\end{equation}
\end{subequations}
Finally, plugging these values into equation~\eqref{abcb} yields a quartic equation for $\W$. This quartic equation factors into linear and cubic terms, so that $\Omega$ satisfies
\begin{gather*}
 (4 \mu -3) \Omega -4 \mu +6  = 0 \text{ or} \\
4 (4 \mu -3)^2 \mu  \Omega^3 
-24 (2 \mu -1) (4 \mu -3) \mu  \Omega^2
+3 (4 \mu -3) \left(16 \mu^2-4 \mu +3\right) \Omega
-64 \mu^3 +48 \mu^2-36 \mu +81  = 0.
\end{gather*}
Here $\mu = \sin^2{\theta}$, so that $\mu = \frac{3}{4}$ corresponds to $\theta=\frac{\pi}{3}$ or $\theta=\frac{2\pi}{3}$.
\begin{itemize}
\item \textbf{Branch 5} corresponds to solutions that make the linear factor zero. This yields physically meaningful solutions only for $\frac{\pi}{3}<\theta<\frac{2\pi}{3}$. At $\theta=\frac{\pi}{2}$, the branch touches Branch 4, and in fact corresponds to a symmetry-breaking branch for that mode. 
\end{itemize}
Cardano's formula yields three solutions that make the cubic term equal zero. One of these is real for all $\theta$ and yields two branches of physically relevant solutions. 
\begin{itemize}
\item \textbf{Branch 6} arises from this formula on the interval $\tfrac{\pi}{3} < \theta < \tfrac{2\pi}{3}$. Along this branch
$$
\lim_{\theta \to \tfrac{\pi}{3}^+}(a,b,c) = (0,0,\infty), \,
\lim_{\theta \to \tfrac{2\pi}{3}^-}(a,b,c) = (\infty,0,0), \,
\text{and }
\bigl.(a,b,c)\bigr|_{\theta=\tfrac{\pi}{2}} = \frac{(1,1,1)}{\sqrt{2}}.
$$
At $\theta = \frac{\pi}{2}$, it bifurcates from Branch 1 in a symmetry-breaking bifurcation.
\item \textbf{Branch 7} arises from this formula on the interval $0<\theta<\tfrac{\pi}{3}$. $(\Omega,Q) \to (-\infty,\infty)$ as the endpoints of the interval are approached. It arises in a saddle-node bifurcation at $\W\approx -2.7$.
\end{itemize}
The other two roots of the cubic are all physically irrelevant, with $\Omega>1$ or $\Im(\Omega) \neq 0$ for all values of $\theta$.

\begin{figure}[htbp] 
   \centering
   \includegraphics[width=5in]{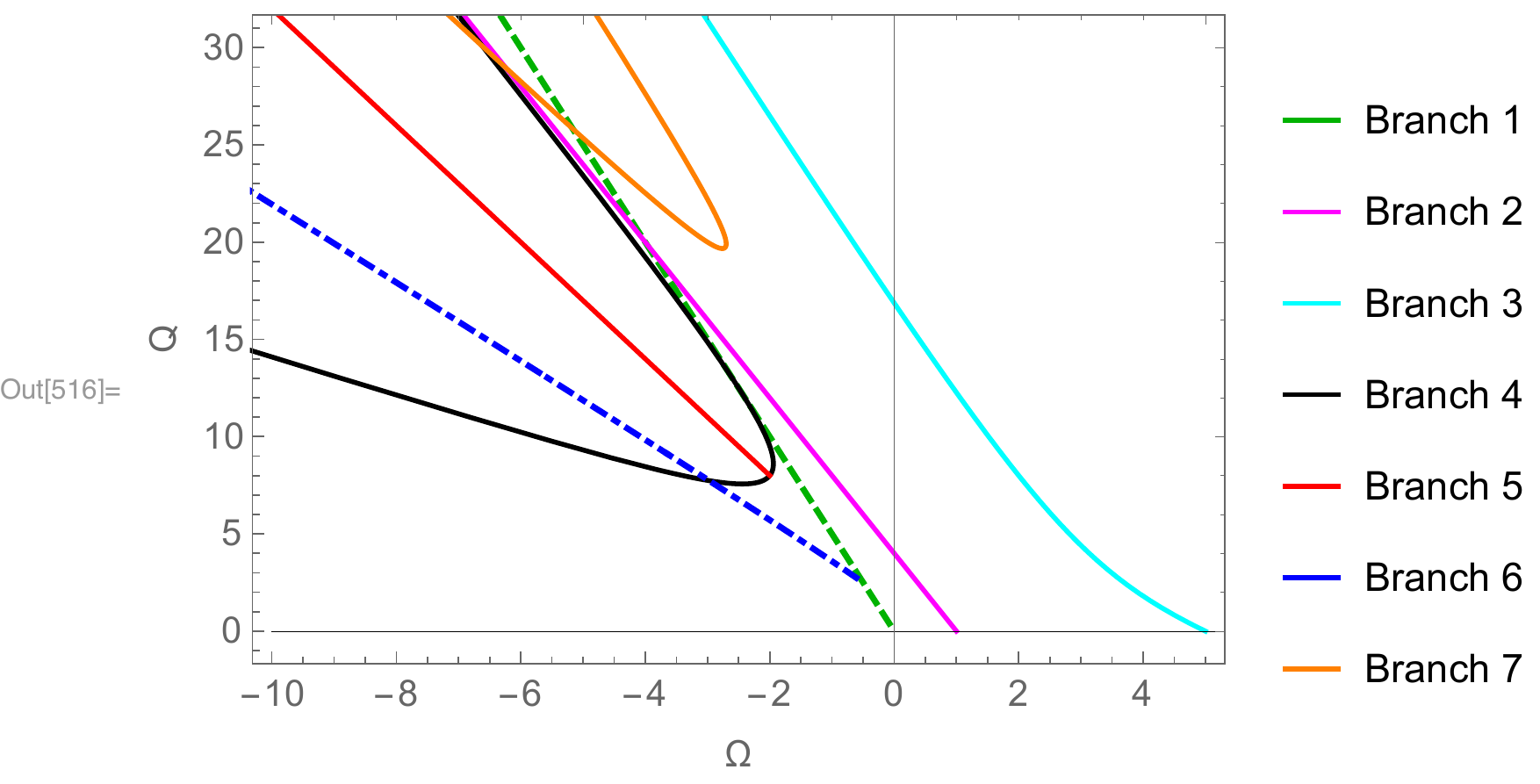} 
   \caption{Branches of stationary solutions to the bowtie-shaped DST system  on the subspace $\cS_2$.}
\label{fig:odeBif}
\end{figure}

Fig.~\ref{fig:odeBif} shows all the branches discussed above.  We summarize the conclusions of the parameterization study shown in this figure.
Branches 1, 4, and 6 correspond to the ones discussed in Ref.~\cite{Marzuola:2016bl} and are colored as in Fig.~\ref{fig:Old}.
 Branch 1, the constant solution  begins at the origin and bifurcates in the (symmetry-breaking) pitchfork bifurcation at $\W = -\tfrac{1}{2}$, creating Branch 6. Branch 1 then undergoes a transcritical bifurcation at $\W = -\tfrac{5}{2}$, creating Branch 4 of symmetric solutions. 
 Both halves of Branch 4 diverge to $(-\infty,\infty)$ in the $(\W,Q)$ plane. As $\W to -\infty$, $b\to0$ and $a=c\to +\infty$.  The half branch that extends to the right from the bifurcation turns around at a fold bifurcation at $\W \approx -1.94$ (this value is the root of a quartic polynomial). It then undergoes a symmetry-breaking bifurcation, creating Branch 5.   For large negative $\Omega$, the solution on Branch 5.concentrates at the center point $b$. 
Branch 7, which arises in a saddle-node bifurcation has no analog in Fig.~\ref{fig:Old}, but we find many such branches in our more complete numerical study in Section~\ref{sec:more}

Branch 5 appears to cross Branch 6 in the $(\Omega,Q)$-projection $\W \approx -2.97$, as the two branches appear to cross in Fig.~\ref{fig:Old}. The statement has no bearing on the conclusions of Refs.~\cite{Marzuola:2016bl,Marzuola:2017}. The asymptotics of the solutions on the quantum graph for large $L^2$ norm depend on the exponential decay of the standing waves as a function of the distance from their maxima. The slope of the curve of symmetric solutions is steeper than that of the curve of symmetric solutions because the symmetric solution concentrates at the single vertex $\tv_3$ for large $Q$, whereas the asymmetric solution concentrates at the two points $\tv_1$ and $\tv_2$ for large values of $Q$.

\section{The symmetry-preserving bifurcation in the dumbbell graph}
\label{sec:missing}

\begin{figure}[htbp] 
   \centering
   \includegraphics[width=0.4\textwidth]{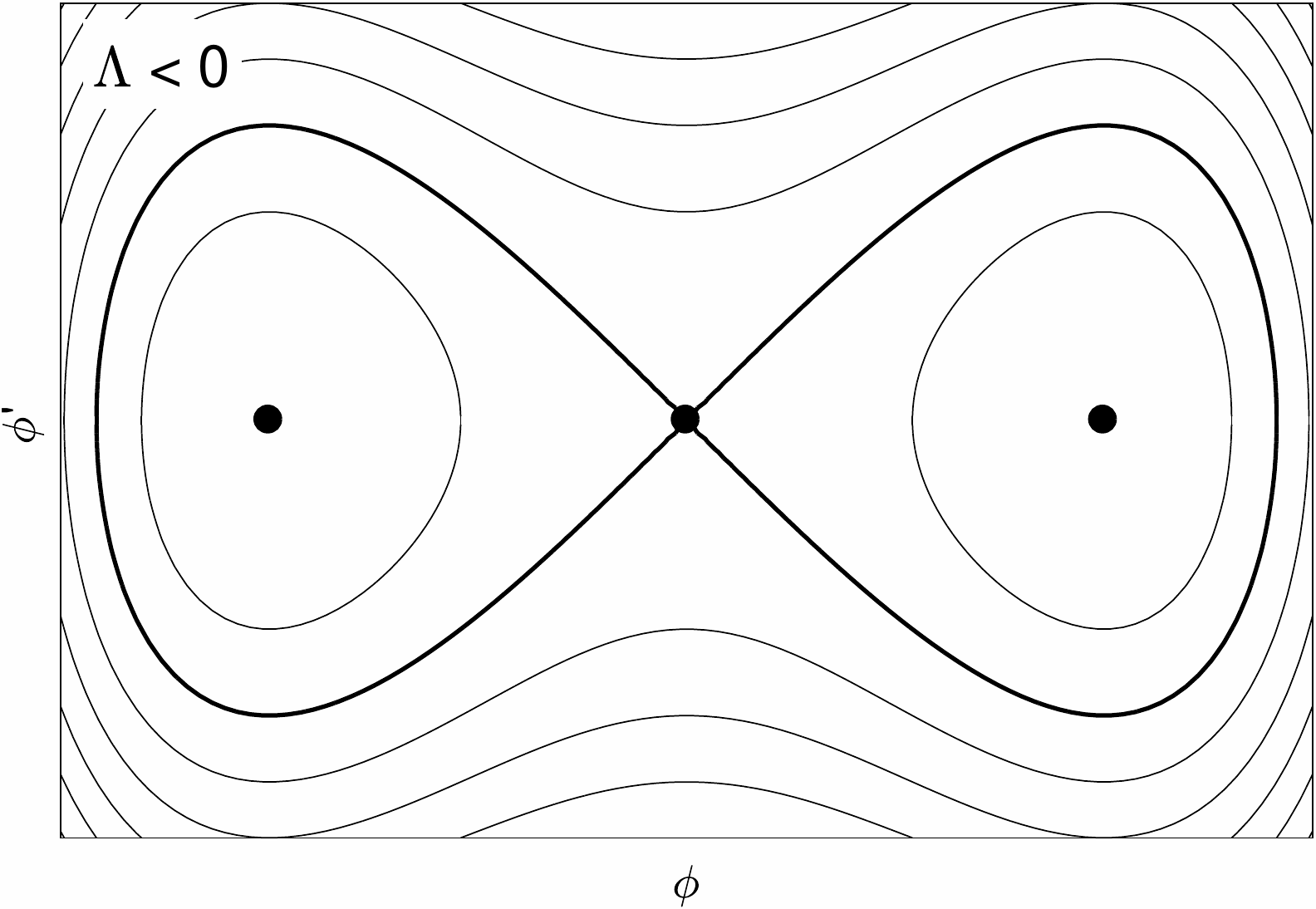}
   \includegraphics[width=0.4\textwidth]{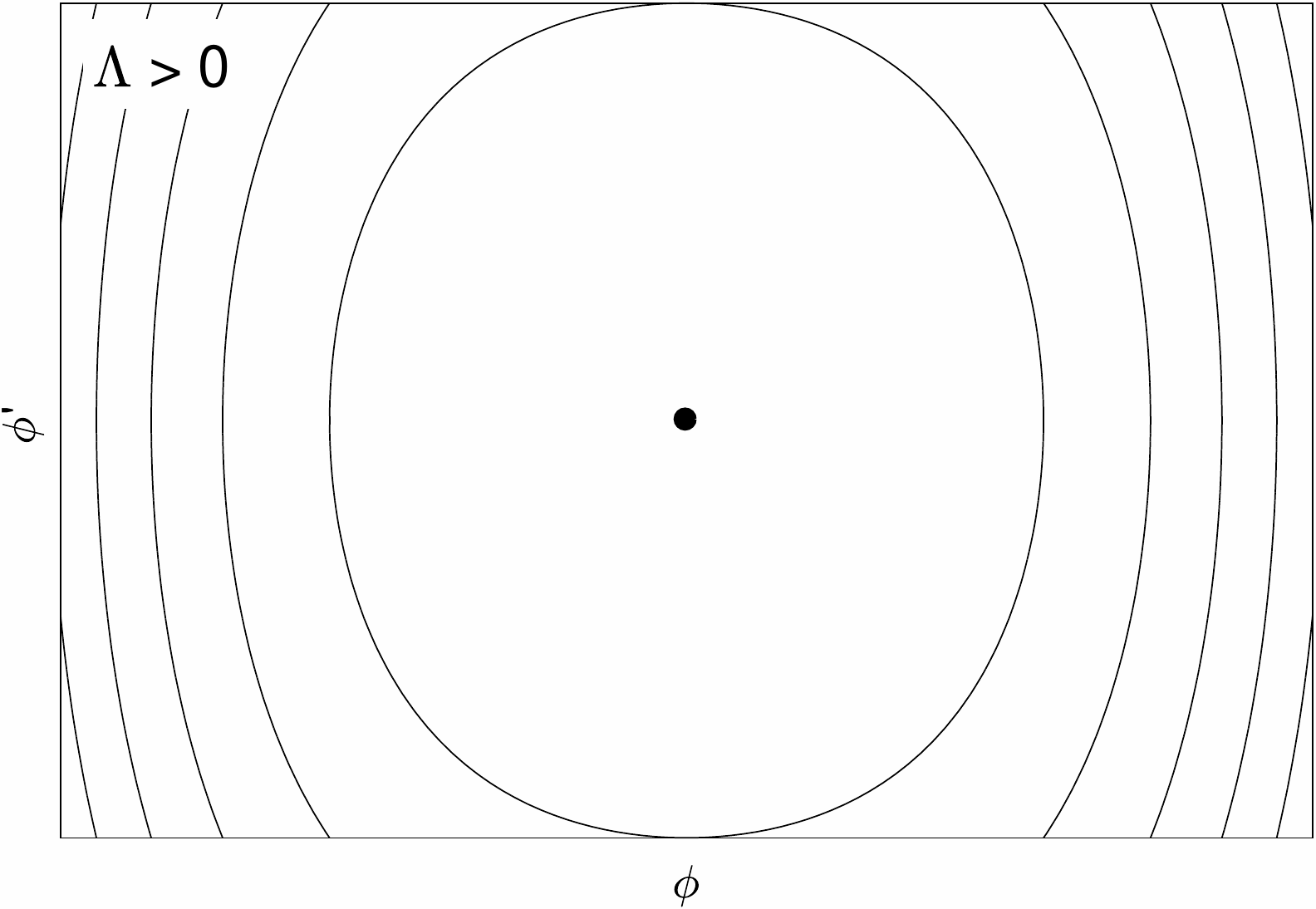}
   \caption{The phase plane of Equation~\eqref{stationary}, whose trajectories are level sets of the energy given by Equation~\eqref{energyODE}.}
\label{fig:phaseplane}
\end{figure}

We now return to the dumbbell quantum graph. The restriction of equation~\eqref{stationary} to any edge $\te_j$ is just
\begin{equation}
\phi_j'' + \Lambda \phi_j + 2 \abs{\phi_j}^2 \phi_j = 0.
\label{stationaryEdge}
\end{equation}
 whose phase plane is shown in Fig.~\ref{fig:phaseplane}, with trajectories lying along level sets of the energy
\begin{equation}
E = \frac{1}{2} \left(\phi_j'^2 + \Lambda \phi_j^2 + \phi_j^4 \right).
\label{energyODE}
\end{equation}
For $\Lambda>0$, there is one family of periodic orbits surrounding a single elliptic fixed point at the origin. For $\Lambda<0$, this fixed point has undergone a symmetry-breaking bifurcation, becoming hyperbolic and giving rise to two new fixed points. These are  defined by equation~\eqref{constantSolution}, and define the constant-valued solutions to system~\eqref{metricGraphLaplacian}. Exact formulas for the periodic orbits are provided in terms of Jacobi elliptic functions in Appendix~\ref{sec:cnoidal}.

\subsection{Numerical continuation study}
Ref.~\cite{Marzuola:2016bl} finds that the constant-valued solution~\eqref{constantSolution} to system~\eqref{metricGraphLaplacian} undergoes two bifurcations as $\Lambda$ is decreased from zero. We numerically compute both branches using pseudo-arclength continuation. We use centered differences to discretize system~\eqref{stationary}, and second-order one-sided differences to discretize the vertex condition~\eqref{flux} on the dumbbell-shaped graph depicted in Fig.~\ref{fig:dumbbell}. The system is solved using Newton's method, combined with pseudo-arclength continuation as described in the textbook of Nayfeh and Balachandran~\cite{nayfeh2008applied}. The code has two advantages over that used in Ref.~\cite{Marzuola:2016bl}: first, it is able to go around fold bifurcations, and second, it detects branching bifurcations, both pitchfork and transcritical, and is able to switch branches; following Govaerts~\cite{Govaerts:2000vv}. The nonlinear solver uses a Newton conjugate gradient-based scheme developed by Yang~\cite{Yang:2009}.

\begin{figure}[htbp] 
   \centering
   \includegraphics[width=3in]{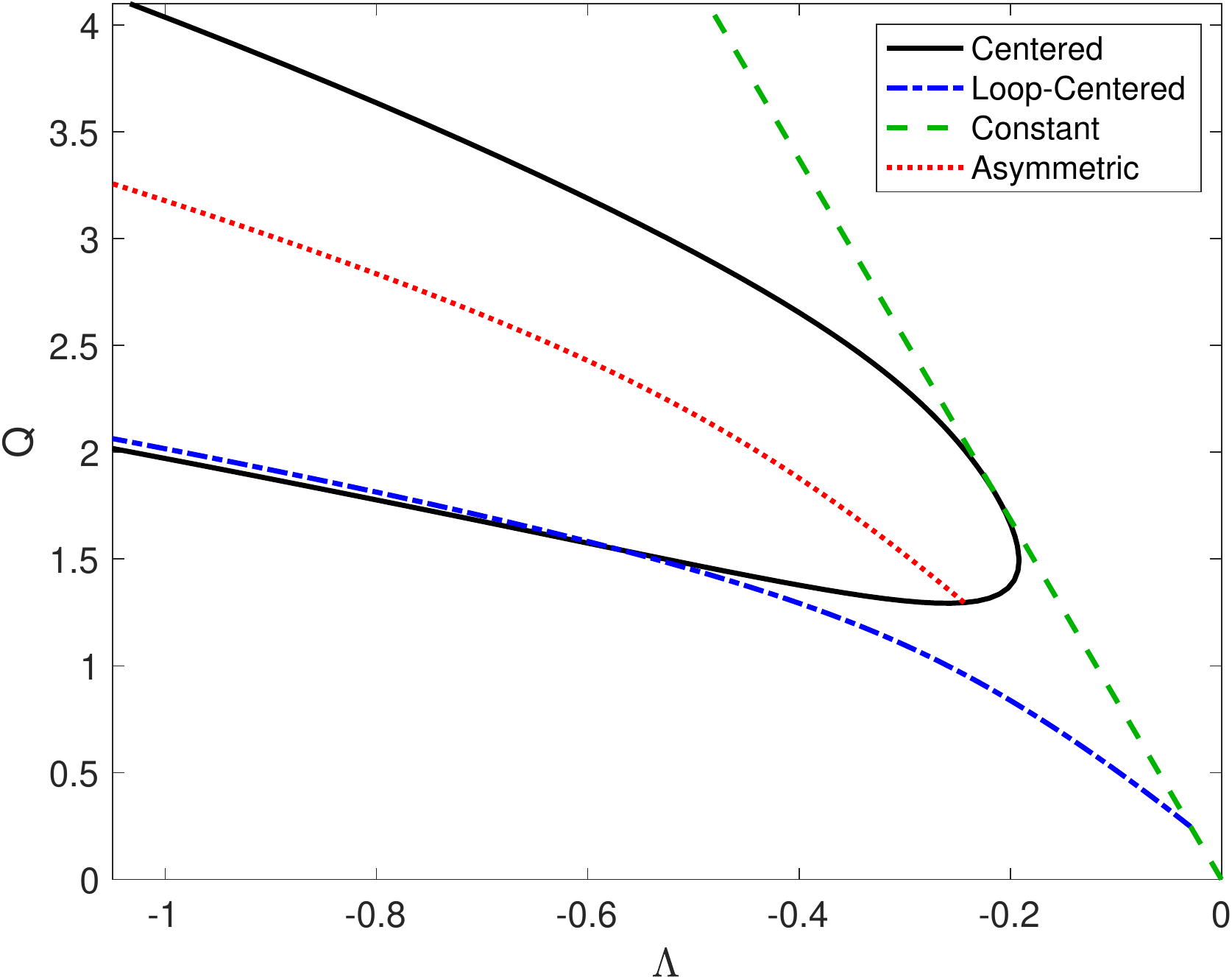} 
   \caption{Numerical continuation of the PDE on the quantum graph. Comparison with Fig.~\ref{fig:bifs} indicates that the loop-centered and constant solutions meet in a transcritical bifurcation. The computation indicates that the centered solution also undergoes saddle-node and pitchfork bifurcations.}
\label{fig:pdeBif}
\end{figure}

\begin{figure}[htbp] 
   \centering
   \includegraphics[width=.45\textwidth]{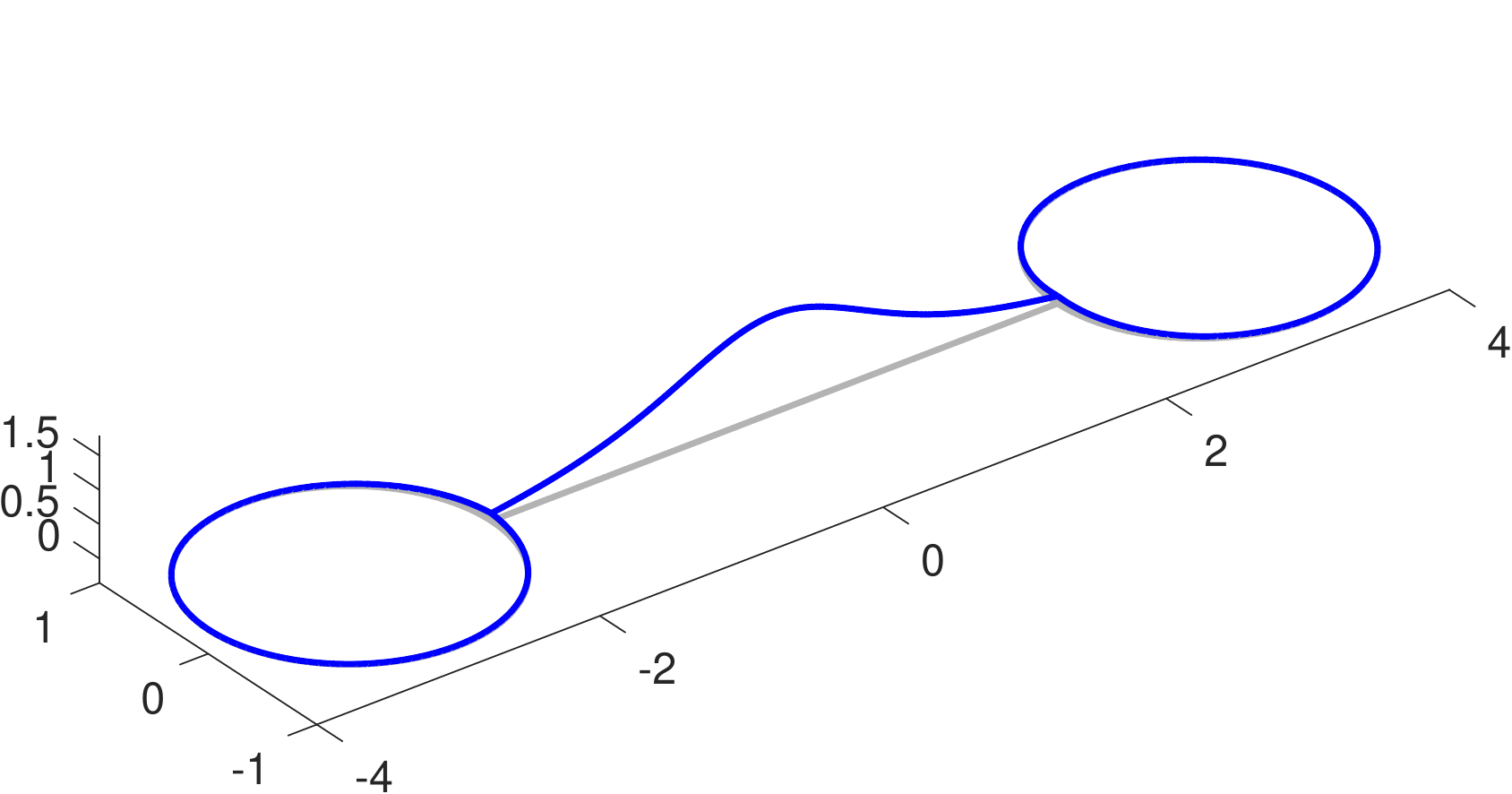}
   \includegraphics[width=.45\textwidth]{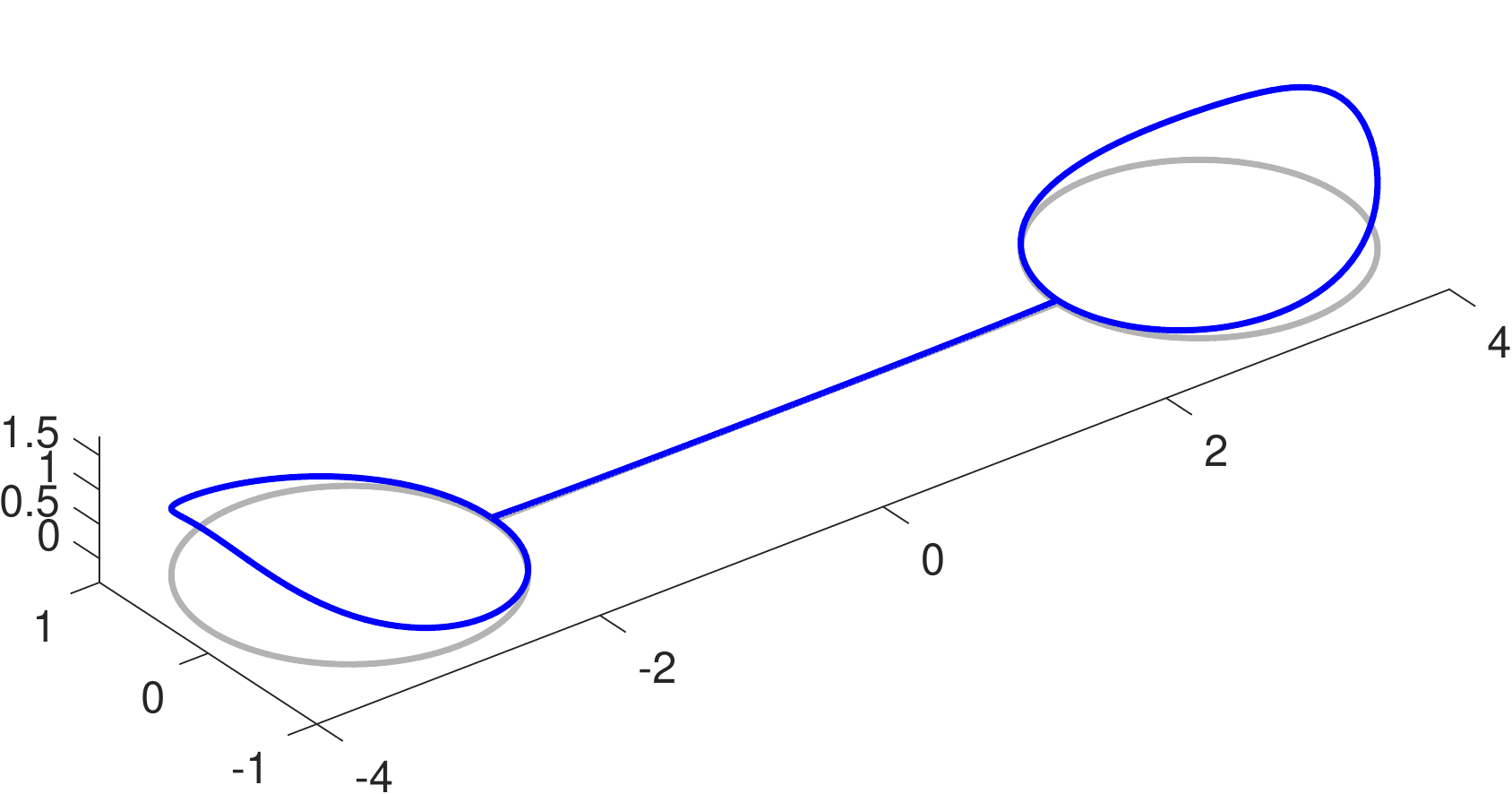}
      \includegraphics[width=.45\textwidth]{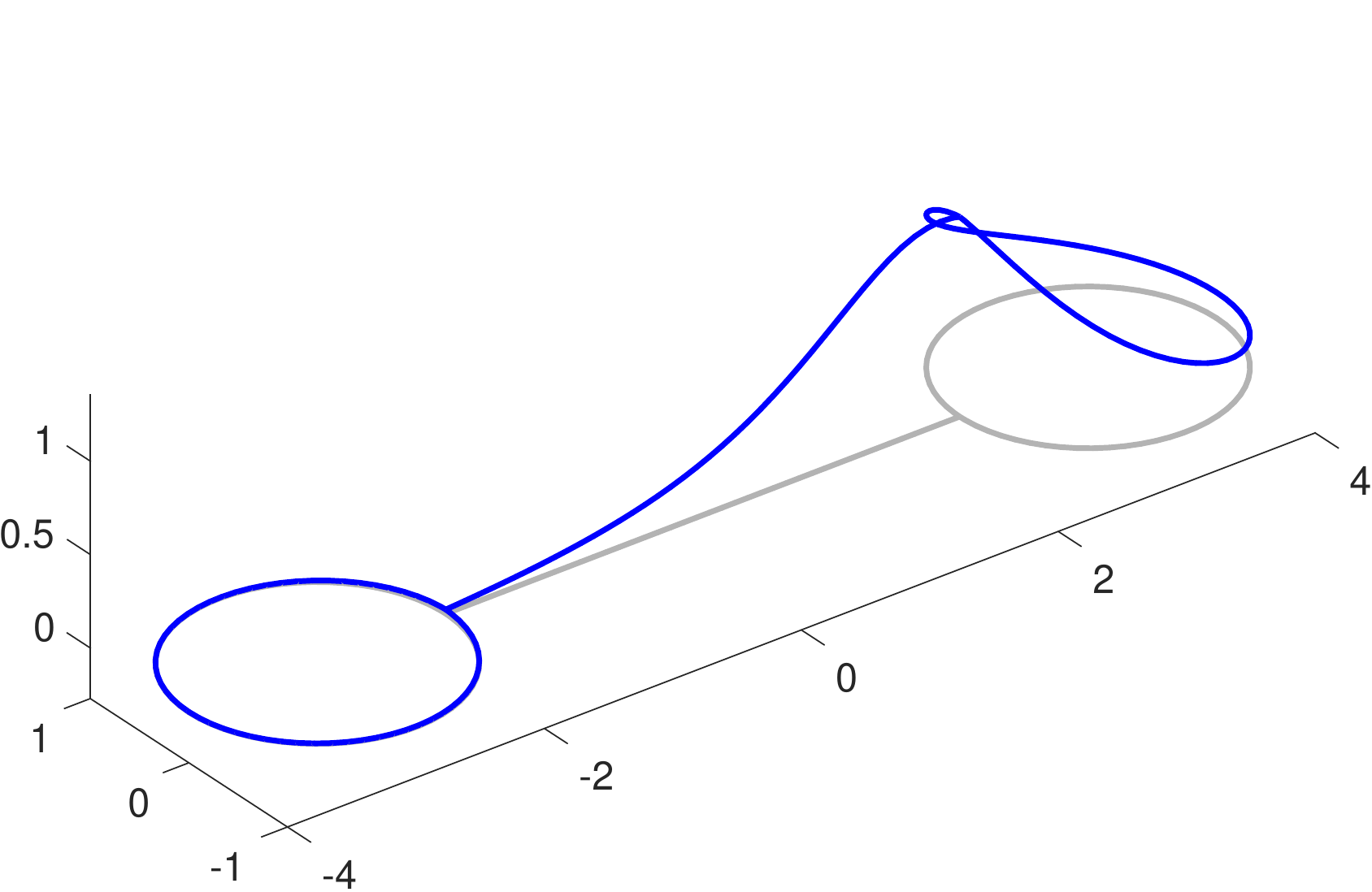}
      \includegraphics[width=.45\textwidth]{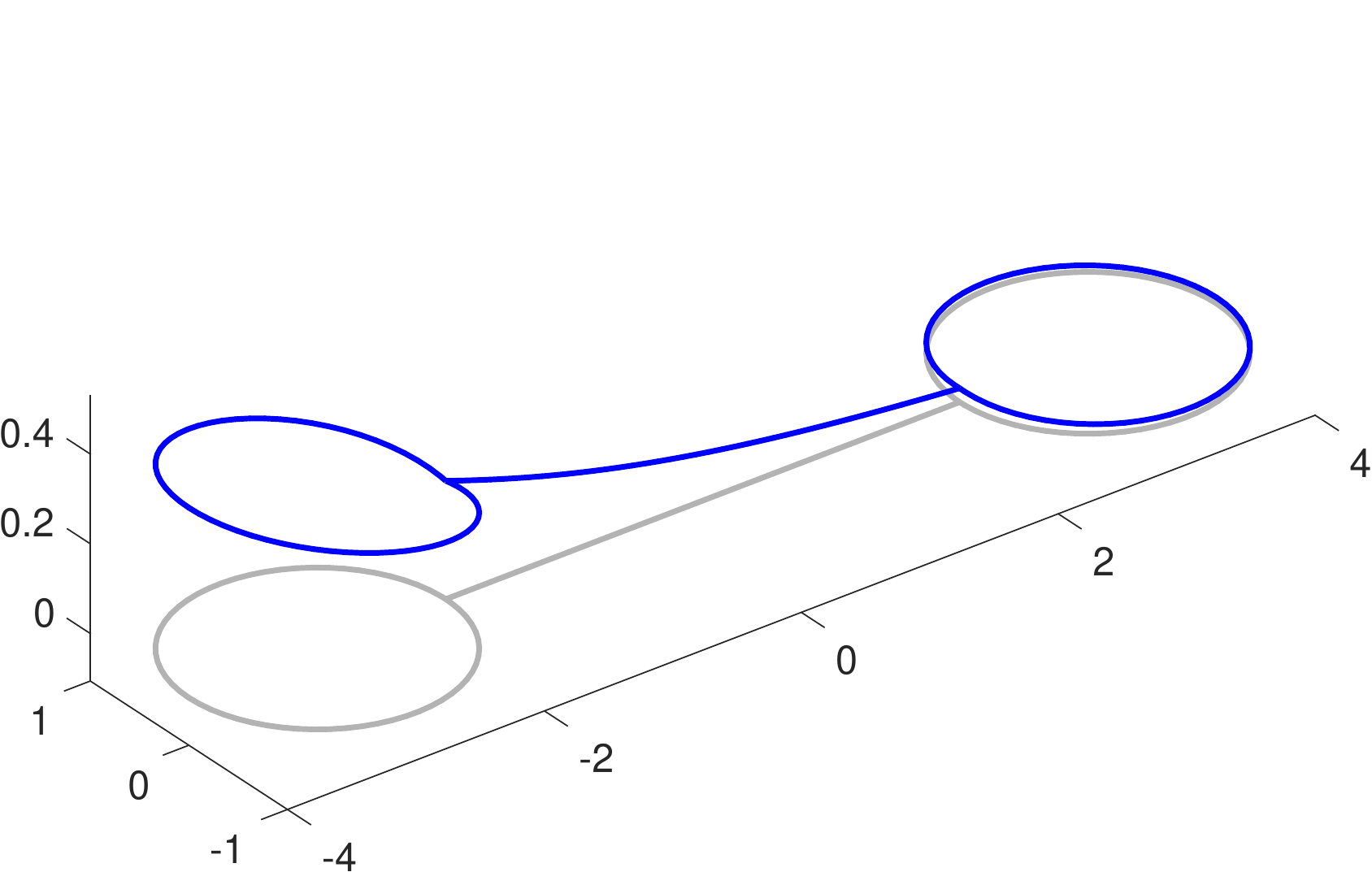}
   \caption{(Top-Left) Large-amplitude centered solution on the half-branch discovered in Ref.~\cite{Marzuola:2016bl}. (Top-Right) Large-amplitude two-soliton solution. (Bottom-Left) Solution arising from symmetry-breaking of centered state. (Bottom-Right) Solution arising from symmetry-breaking of constant state.}
\label{fig:evenSolutions}
\end{figure}

The result of a computation with $L=2$ is shown in Fig.~\ref{fig:pdeBif}. The symmetry-breaking bifurcation is exactly as in Fig.~\ref{fig:Old}. In addition, this computation captures several features not present in~\cite{Marzuola:2016bl}, but visible in the bifurcation diagram for the bowtie DST model, Fig.~\ref{fig:odeBif}. First, the branch of centered solutions of Fig.~\ref{fig:Old} undergoes a fold bifurcation at $\Lambda \approx -0.19$. The upper branch then meets the branch of constant solutions at the ``symmetry-preserving'' bifurcation Fig.~\ref{fig:Old} near $\Lambda = -\w_1^2 \approx-0.215$, in exactly the same manner found for DST on the bowtie graph. By comparison with Fig.~\ref{fig:bifs}, it appears that this should be a transcritical bifurcation, which will be confirmed below. In addition, the branch undergoes a symmetry-breaking pitchfork bifurcation very close to the saddle-node bifurcation. Solutions from both the top and bottom halves of the symmetric branch, as well as the symmetry-broken branch are displayed in Fig.~\ref{fig:evenSolutions}.

\subsection{Analysis of the bifurcations}
Ref.~\cite{Marzuola:2016bl} contains a perturbation theory parameterizing symmetry-broken solutions close to the constant-valued solution using, essentially, the Poincar\'e-Lindstedt method. We here use Theorem~\ref{YangTheorem} to classify the first bifurcation as a pitchfork and the second as transcritical.

Before classifying the individual bifurcations of the constant solution~\eqref{constantSolution}, we compute the necessary quantities~\eqref{Gj} and~\eqref{Thetas} for the application of Theorem~\ref{YangTheorem}. The linearized operator about the constant solution is 
$$
\bigl. \cL_{1} \bigr \rvert_{\Phi=\sqrt{-\frac{\Lambda}{2}}}
=
\cL+ 2 \Lambda.
$$
Therefore if $(\gamma^2,\tPhi)$ is an eigenpair of $\cL$, then along the branch of constant solutions $(\gamma^2+2\Lambda,\tPhi)$ is an eigenpair of $\cL_1$, and the system undergoes a bifurcation at $(\Lambda_0,\Phi_0)= \left(-\gamma^2/2,\gamma/2\right)$ with null vector $\Upsilon=\tPhi$.

For this problem, $G(\Phi) = 2 \Phi^3$, in equation~\eqref{Gj}, and at the bifurcation point
$$
G_2 = 12 \tPhi = 6 \gamma; \; G_3 = 12.
$$
The constants in equation~\eqref{Thetas} depend on $\cL_{10}^{-1} \tPhi$. Since $\left. \cL_{10}\Phi \right|_{\Phi=1} = -\gamma^2$, we have 
$$
\cL_{10}^{-1} \tPhi = \frac{-1}{\gamma^2} \cdot \frac{\gamma}{2} = \frac{-1}{2 \gamma}
$$ 
and

\begin{equation*}
\begin{gathered}
\Theta_1 = \frac{\gamma}{2} \intG \tPhi \ dx = 0; \qquad
\Theta_2 = \frac{3}{2}\gamma^2 \intG \tPhi^3 \ dx ; \qquad
\Theta_3 = 4 \intG \tPhi^2 \ dx > 0; \\
\Theta_4 = \frac{28}{\gamma} \int_G \tPhi \ dx = 0; \qquad
\Theta_5 = 12 \intG \tPhi^2 \ dx - 144 \gamma^2 \inner{\tPhi^2}{\cL_{10}^{-1} \tPhi^2}
\end{gathered}
\end{equation*}
All of these quantities depends on known functions and quantities, except for which $\Theta_5$ requires $\cL_{10}^{-1} \tPhi^2$. The constant $\Theta_5$ is only required at pitchfork bifurcations. 
In particular, if $\tPhi$ is an odd function, then all elements of the null space of $\cL_{10}$ are multiples of $\tPhi$. Thus, since $\tPhi^2$ is even, the operator can be inverted.

\subsubsection{The Pitchfork Bifurcation}
\label{sec:pitchfork}
Setting $\gamma=\W_1$, where $\W_1^2$ is the eigenvalue corresponding to the first odd eigenfunction, then a bifurcation occurs at $\Lambda = -\W_1^2/2$. We call this eigenfunction $\Phiodd$ and normalize it such that $\intG \Phiodd^2 dx = 1$. Then since $\Phiodd^3(x)$ is odd, $\Theta_2=0$. Since $\Theta_3$ is positive and $\Theta_5$ will in general be nonzero (though this would need to be checked numerically), the bifurcation is of pitchfork type. 

Following~\cite{Marzuola:2017}, the perturbation expansion is given by 
\begin{align*}
\Phi & = \frac{\W_1}{2} + a \Phiodd + a^2 \Phi_2 + a^3 \Phi_3 + O(a^4); \\
\Lambda &= \frac{-\W_1^2}{2} + a^2 \beta_2 + a^3 \beta_3 + O(a^4).
\end{align*}
In order to solve for terms at $O(a^2)$, we must satisfy a Fredholm condition at $O(a^3)$ but we do not otherwise need the terms at that higher order.

We find 
$$
\beta_2  = 9 \W_1^2 \intG \Phiodd^2 \tPhi_2 \ dx \text{ and }
\Phi_2  = \tPhi_2 - \frac{\beta_2}{2 \W_1},
$$
where $\tPhi_2$ is the unique even-symmetric solution to 
$$
\cL_{10} \tPhi_2 = \Phiodd^2.
$$
This is well-defined since the null space of the self-adjoint operator on the left is spanned by the odd-symmetric function $\Phiodd$ and the right-hand side has even symmetry. Ref.~\cite{Marzuola:2016bl} notes that this equation may be solved explicitly, and uses the exact solution to prove the direction in which the solution branch bends.

\subsubsection{The Transcritical Bifurcation}
\label{sec:transcritical}

Setting $\gamma=\w_1$, where $\w_1^2$ is the eigenvalue corresponding to the first even eigenfunction, then a bifurcation occurs at $\Lambda = -\w_1^2/2$. We call this eigenfunction $\Phieven$ and normalize it such that $\intG \Phieven^2 dx = 1$. Then since $\Phieven^3(x)$ is even, then in general $\Theta_2\neq0$. In this case, following~\cite{Yang:2012ux}, the perturbation solution is given by 
\begin{align*}
\Phi & = \frac{\w_1}{2} + a \Phi_1 + a^2 \Phi_2 + a^3 \Phi_3 + O(a^4); \\
\Lambda &= \frac{-\w_1^2}{2} + a \beta_1 + a^2 \beta_2 + a^3 \beta_3 + O(a^4).
\end{align*}

This leads a sequence of equations:
\begin{align*}
O(a): & & \cL_{10} \Phi_1 & = \frac{ \w_1 \beta_1}{2}; \\
O(a^2): & & \cL_{10} \Phi_2 & = 
\frac{ \w_1\beta_2 }{2}
+\beta_1 \Phi_1
+3  \w_1 \Phi_1^2
; \\
O(a^3): & & \cL_{10} \Phi_3 & = 
\frac{ \w_1\beta_3}{2}
+\beta_2 \Phi_1
+\beta_1 \Phi_2
+6  \w_1 \Phi_1 \Phi_2
+2 \Phi_1^3.
\end{align*}
At $O(a)$ we find that $\Phi_1 = \Phieven + C$ and $\beta_1 = -2 \w_1 C$, with $C$  to be determined. The equation at $O(a^2)$ then becomes
$$
\cL_{10} \Phi_2 = \w_1 \left(\tfrac{\beta_2}{2}+C^2 \right) +  \w_1 \left( 4 C \Phieven + 3 \Phieven^2 \right).
$$
The constant term is solvable, but the remaining term must be orthogonal to $\Phieven$ for solvability. This then yields a condition
\begin{equation}
C = -\frac{3}{4} \intG \Phieven^3 \ dx.
\label{C}
\end{equation}
Solving the equation at this order yields
$$
\Phi_2 = -\tfrac{1}{\w_1} \left(\tfrac{\beta_2}{2} + C^2 \right) + \tPhi_2,
$$
where $\tPhi_2$ is the unique solution to 
\begin{align*}
\cL_{10} \tPhi_2 & =  \w_1 \left( 4 C \Phieven + 3 \Phieven^2 \right);\\
\inner{\tPhi_2}{\Phieven} & = 0.
\end{align*}
A Fredholm condition at $O(a^3)$ then yields
$$
\beta_2  = -4 C^2 + \intG \Phieven^4 \ dx 
+ \frac{3\w_1}{2} \intG \tPhi_2 \Phieven^2 \ dx.
$$

\begin{rem}
The expansions described in this section are equally valid for every eigenvalue/eigenfunction pair in the even or odd families. Thus, the constant-valued solution undergoes a pitchfork bifurcation each time its value crosses a frequency $-\w_j^2/2$ and a transcritical bifurcation each time it crosses a frequency $-\W_j^2/2$. We delay discussion of the bifurcations that involve the loop-localized families of eigenfunctions to Section~\ref{sec:more}.
\end{rem}

\begin{rem}
Fig.~\ref{fig:pdeBif} shows that that a saddle node bifurcation occurs very near to this transcritical. Remark~\ref{rem:imperfect} discusses the fact that a pitchfork bifurcation perturbed via a perturbation with $\Theta_1=0$ splits into a saddle-node and a transcritical bifurcation. In the next section, we discuss a specific sense in which the dumbbell graph can be thought of as a perturbation of a simpler graph which features only pitchfork bifurcations.
\end{rem}
%

\section{Comparison with simpler quantum graphs}
\label{sec:singularPert}

We were able to gain some insight into the dumbbell graph by considering the simpler problem of the bowtie combinatorial graph. Here we consider two additional simpler problems, the cubic Schr\"odinger problem defined on an interval with Neumann boundary conditions, and the problem on the lollipop graph, obtained from the graph in Fig.~\ref{fig:dumbbell} by removing the edge $\te_3$ and imposing a Neumann condition at vertex $\tv_2$. 

The former can be thought of as a quantum graph consisting of just one edge and two vertices. This system, which can be derived as a singular limit of the dumbbell graph, has a different bifurcation structure. A rigorous approach to quantum graphs in the limit of vanishing edge length is given by Berkolaiko et al.~\cite{Berkolaiko:2018}. In what follows, we rescale the $x$ coordinates along the edges of the dumbbell graph so that $\ell_1= \ell_3=\epsilon$ and $\ell_2 = \pi$. The secular determinant equation~\eqref{secularDet} becomes:
$$
\left(
\sin{\frac{k (\pi -\epsilon )}{2} }-3 \sin{\frac{k (\epsilon +\pi )}{2} }
\right) 
\left(
\cos{\frac{k (\pi -\epsilon )}{2} }-3 \cos{\frac{k (\epsilon +\pi )}{2} }
\right)
\left(
\sin^2{\frac{k \epsilon }{2}}
\right)
= 0
$$
The solutions to these the three factors correspond, respectively in the order presented, to the even, odd, and loop-localized families of eigenfunctions. For the loop-localized families, this gives $k = 2 n \pi/ \epsilon$, a sort of boundary layer, in that the solution is concentrated into a small region near the boundary of the $\epsilon \to 0$ limiting problem. For the other two branches we may expand the wavenumber $k = \sum_{j=0}^\infty \epsilon^j k_j$. For the even-symmetric solutions we find $k_0 = 2 n$ and for the odd-symmetric, $k_0= 2 n -1$ and in both cases
$$
k = k_0 \left(
1 - \frac{2\epsilon}{\pi} + \frac{4 \epsilon^2}{\pi^2}  - \frac{8 \epsilon^3}{\pi^3}
\right) +  k_0^3 \frac{\epsilon^3}{2 \pi} + O(\epsilon^4).
$$
Letting the parameterization on the edge $\te_2$ be given by $0<x_2<\pi$, then all the eigenfunctions are of the form $\cos{k x}$ with $k \approx 2n$ on the even family and $k \approx (2n-1)$ on the odd family. 

Now, we take the singular limit $\epsilon \to 0^+$. The boundary layer solutions disappear, and the problem on the central edge $\te_2$ reduces to the Schr\"odinger equation on an interval with Neumann conditions, whose eigenfunctions are simply $\cos{nx}$. We examine the sequence of bifurcations undergone by the nonlinear standing wave with constant spatial profile $\Phi$ and frequency $\Lambda = -2 \Phi^2$ for this system. This solution undergoes a sequence of bifurcations at the discrete sequence of frequencies $\Lambda_n = -n^2/2$. Each of these bifurcations is a pitchfork, independently of the parity of $n$ and in contrast to the dumbbell system, which alternates between pitchfork and transcritical bifurcations. In terms of Theorem~\ref{YangTheorem}, as $\lim_{\epsilon\to 0^+} \Theta_2 =0$ and the transcritical becomes a pitchfork.

The difference between the problem on the interval and that on the dumbbell is that the Neumann problem on the interval has an additional symmetry: namely, the solution on the interval can be extended to the whole real line under an even periodic extension. Consider a solution in a neighborhood of the pitchfork bifurcation. In both the Neumann problem and the dumbbell problem, the solution looks to leading order like $\Phi = \text{constant} + a \Phiodd$. Reversing the sign of $a$ is equivalent to applying the symmetry $\cR_2$ to this solution. This argument fails near the transcritical bifurcation on the dumbbell: reversing the sign of $a$ is not equivalent to applying a symmetry. However, in the limit $\epsilon=0$, a new symmetry appears, as the evenly extended eigenfunctions have an odd symmetry about each of their zeros. This symmetry persists for  nonlinear standing waves. 

\begin{figure}[htbp] 
   \centering
   \includegraphics[width=2in]{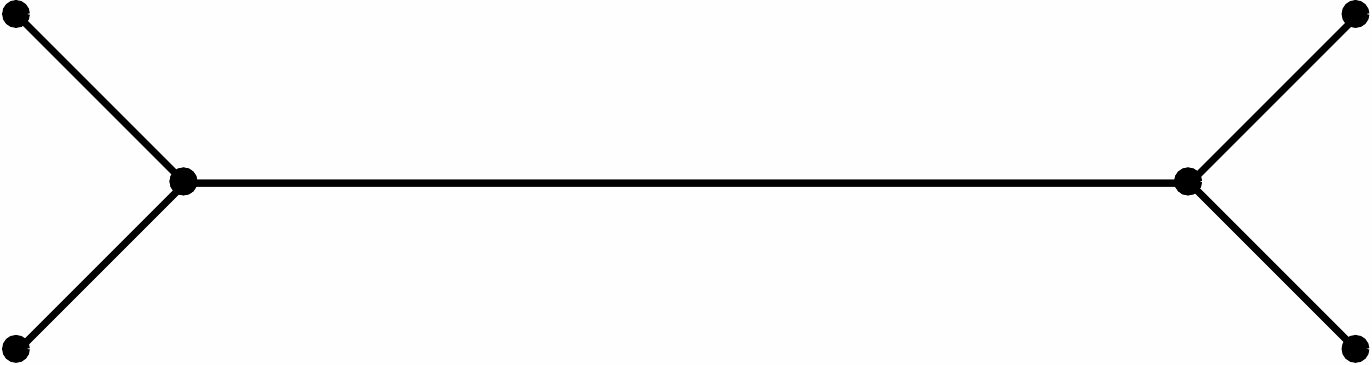} 
   \caption{A graph that should have similar bifurcations.}
\label{fig:doublefork}
\end{figure}

It is not necessary that there be loops appended to the interval. In the family of graphs  pictured in Fig.~\ref{fig:doublefork}, the extra symmetry of the Neumann interval is also destroyed, and pitchforks involving even eigenfunctions will also be converted to transcriticals. 

In general, for any compact graph with Kirchhoff conditions at all vertices, the constant-valued solution will undergo a sequence of transcritical bifurcations, since the eigenfunctions of the linearized system will be the same as the linear eigenfunction and all are orthogonal to the constant-valued solution, yielding $\Theta_1=0$. If, however, the graph is sufficiently symmetric that it supports even eigenfunctions, then these will satisfy $\Theta_2=0$ and the bifurcations will be symmetry-breaking pitchforks.

\begin{figure}[htbp] 
   \centering
   \includegraphics[width=.4\textwidth]{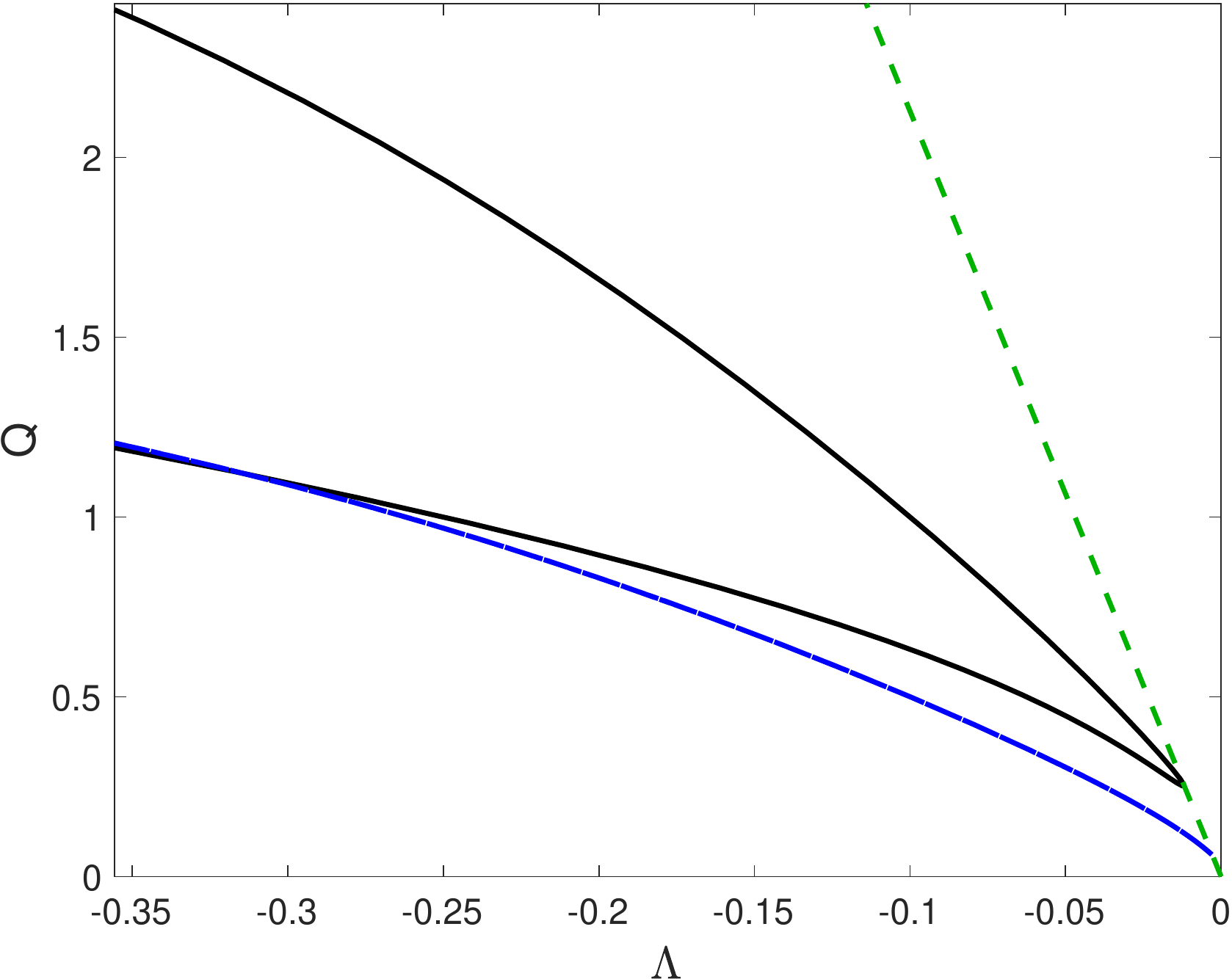} 
   \includegraphics[width=.4\textwidth]{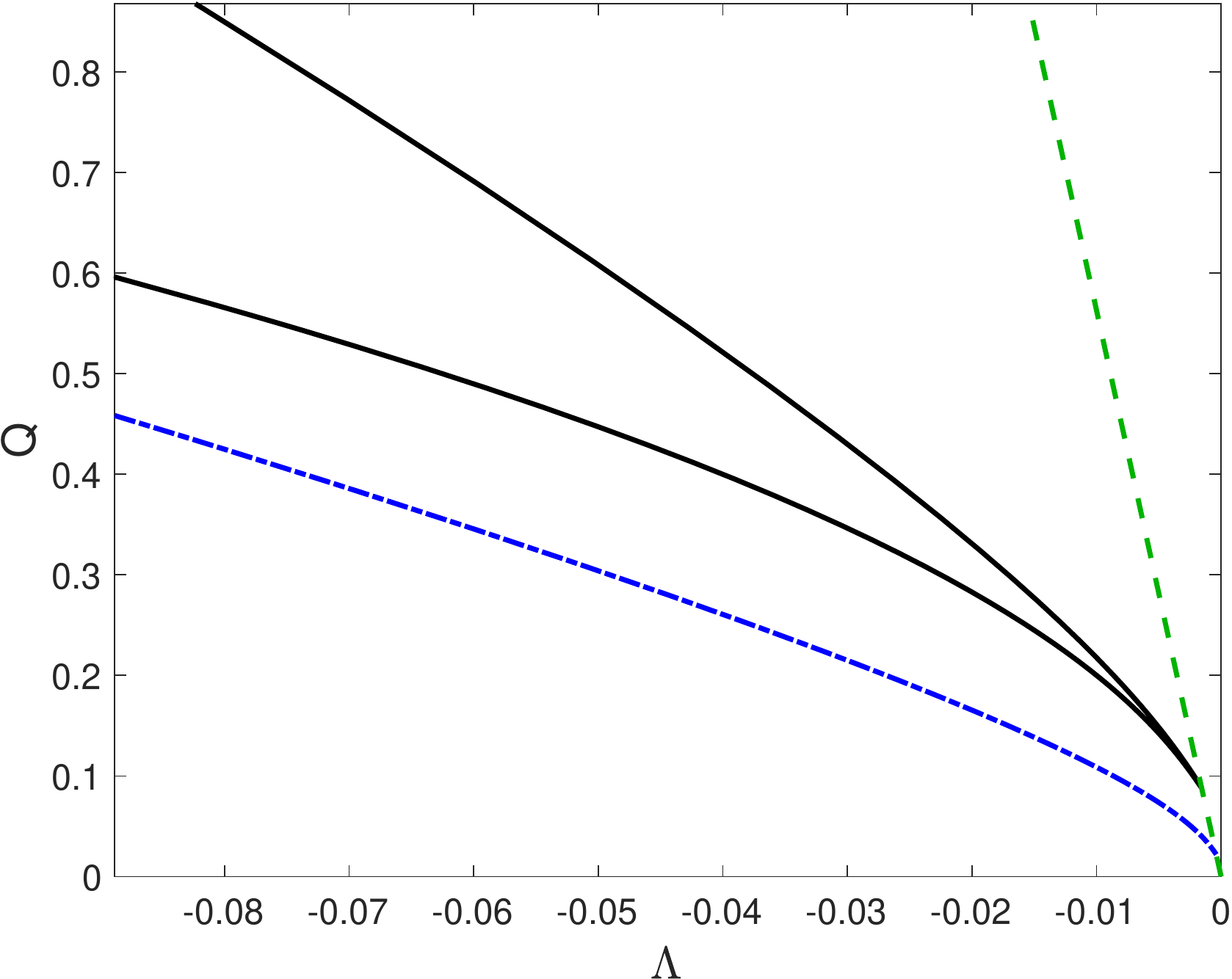} 
   \caption{The analogy of Fig.~\ref{fig:pdeBif} with $L=15$ and $L=50$. As $L$ is increased, the angle with which the two branches of solution approach the transcritical bifurcation decreases, making it appear, locally, more like a pitchfork.}
\label{fig:bigL}
\end{figure}

Therefore, we can expect that as $\epsilon \searrow 0$, each of the transcritical bifurcations on the dumbbell graph becomes increasingly pitchfork-like. In other words, the transcritical bifurcation and the saddle-node nearby merge with each other, and $\Theta_2 \to 0$ at the bifurcation point. While this statement should hold in a neighborhood of the bifurcations, for large values of $\abs{\Lambda}$, the two branches will contain, respectively, one or two pulses, and thus diverge. We investigate this possibility by re-computing the continuation calculation shown in Fig.~\ref{fig:pdeBif} for large values of $L$ in Fig.~\ref{fig:bigL}, which correspond to small values of $\epsilon$. These show the width of the branch narrows greatly for large $L$, and appears more like a pitchfork.

Finally, we note that in the lollipop graph, lacking the symmetry of the dumbbell, the analysis leading to the pitchfork bifurcation in Section~\ref{sec:pitchfork} is not applicable, but that the analysis in Section~\ref{sec:transcritical} would go through without any changes. Therefore we see that eliminating both loops leads to only pitchforks, while eliminating only one leads to only transcriticals. This is an answer to what is special about the dumbbell structure that allows both behaviors. The bifurcation diagram for the lollipop graph will become important in Section~\ref{sec:oneOfEach} when we discuss standing waves on the dumbbell that are built from lollipop standing waves.

\section{Numerical enumeration of stationary solutions}
\label{sec:more}

Given the attention that we have just paid to just two bifurcations, we should point out that the behavior of standing waves is significantly more complex. We have already pointed out that the constant-valued solution undergoes an alternating sequence of pitchfork and transcritical bifurcations, each time its frequency crosses $\Lambda = -\W_j^2/2$ or $\Lambda = -\omega_j^2/2$. In addition, each of the other linear eigenfunctions can be continued into the nonlinear regime in a similar manner. In the remainder of this section, we will enumerate all possible branches of stationary solutions using a combination of numerical continuation, phase planes, and analytical reasoning. A similar enumeration of standing waves was carried out for the tadpole graph by Noja et al.~\cite{Noja:2015hx}. 

Gnutzmann and Waltner have approached this numerical problem differently~\cite{Gnutzmann:2016hma,Gnutzmann:2016ec}. Rather than solve equation~\eqref{stationary}, they use the exact Jacobi elliptic function solutions to define solutions on each edge of the graph. Each of these partial solutions depends on a undetermined parameters: the parameters $\kappa$ or $k$, and $\tau$ in equations~\eqref{cn} and~\eqref{dn}. These then must be chosen to satisfy the vertex conditions~\eqref{continuity} and~\eqref{flux}. Together, these determine a system of equations. The authors then use normal form perturbation theory to simplify this system of equations into a more tractable form, finding good agreement.

\subsection{Two types of loops}

We may classify the different types of stationary solutions by their behavior on the  self-connected edge $\te_1$ (or equivalently $\te_3$).  Since the solution along edge $\te_1$ lies along a level set of the energy, continuity condition~\eqref{continuity} requires $\phi_1(-\pi) = \phi_1(\pi)$ and thus, by the symmetry of the phase planes across the $y$-axis in Fig.~\ref{fig:phaseplane} 
\begin{equation}
\phi_1'(\pi) = \pm \phi_1'(-\pi).
\label{condv1}
\end{equation}

Thus there are two cases. If equation~\eqref{condv1} is satisfied with a minus sign, then condition~\eqref{flux} implies that 
\begin{equation}
\phi_2'(-L) = 2 \phi_1'(\pi). 
\label{incompleteD}
\end{equation}
We will call such a loop \emph{incomplete}. Note that when $L$ is not a multiple of $\pi/2$ both loops of  eigenfunctions from the even and odd families are of this type; compare Fig.~\ref{fig:dumbbellEigs}(a-d). At the vertex $\phi_1 = \phi_2$, by the continuity condition~\eqref{continuity}. Therefore when $\te_1$ is an incomplete loop, any jump in the energy~\eqref{energyODE} at the vertex is determined by the $\frac{1}{2}{(\phi_j')}^2$ term and equation~\eqref{incompleteD} implies $E(\te_1)< E(\te_2)$.

If Equation~\eqref{condv1} is satisfied with the plus sign, then 
\begin{equation}
\phi_2'(-L) = 0
\label{completeD}
\end{equation}
and the solution on the self-directed edge must consist of an integer number of periods of a closed orbit. We will call such a loop \emph{complete}. The nonzero loop of loop-localized eigenfunctions are  of this type; see Fig.~\ref{fig:dumbbellEigs}(e-f). Note that self-directed edges on which the solution is identically zero  can be considered both complete and incomplete. Since $\phi_1'(\pi)$ may be nonzero, condition~\eqref{completeD} implies that when $\te_2$ is a complete loop, then $E(\te_1) \ge E(\te_2)$. 

Therefore, there are three possible types of stationary solution. On the first, both loops are incomplete, on the second, both loops are complete, and the third contains one complete and one incomplete loop.

\subsection{Two incomplete loops}
\label{sec:2incomplete}

We can search for these stationary solutions with two incomplete loops using the following shooting argument. All such stationary solutions must satisfy $\left.\frac{d \phi_j}{d x_j}\right\rvert_{x_j=0}=0 $, $j=1,3$. Now define $f(q,\Lambda,L)=\phi_3'(0)$ found by solving the following three consecutive initial value problems for equation~\eqref{stationary}:
\begin{enumerate}
\item Solve for $\phi_1(x)$ from $x_1 = 0$ to $x_1= \pi$ with initial condition $\phi_1(0)=q$, $\phi_1'(0)=0$.
\item Solve for $\phi_2(x)$ from $x_2 = -L$ to $x_2=L$ with initial condition $\phi_2(-L)=\phi_1(\pi)$, $\phi_2'(-L)=2 \phi_1'(\pi)$.
\item  Solve for $\phi_3(x)$ from $x_3 = -\pi$ to $x_3=0$ with initial condition $\phi_3(-\pi)=\phi_2(L)$, $\phi_3'(-\pi)=\frac{1}{2}\phi_2'(L)$.
\end{enumerate}
Then solutions of $f(q,\Lambda,L)=0$ correspond to solutions of~\eqref{stationary} on $\Gamma$ with the appropriate vertex conditions. This definition is closely related to the map defined by Pelinovsky and Schneider over one cell of an infinite periodic graph~\cite{Pelinovsky:2016fm}. We may then continue these solutions as a function of $\Lambda$ to follow the branches and find bifurcations.

\begin{figure}[htbp] 
   \centering
   \includegraphics[width=4in]{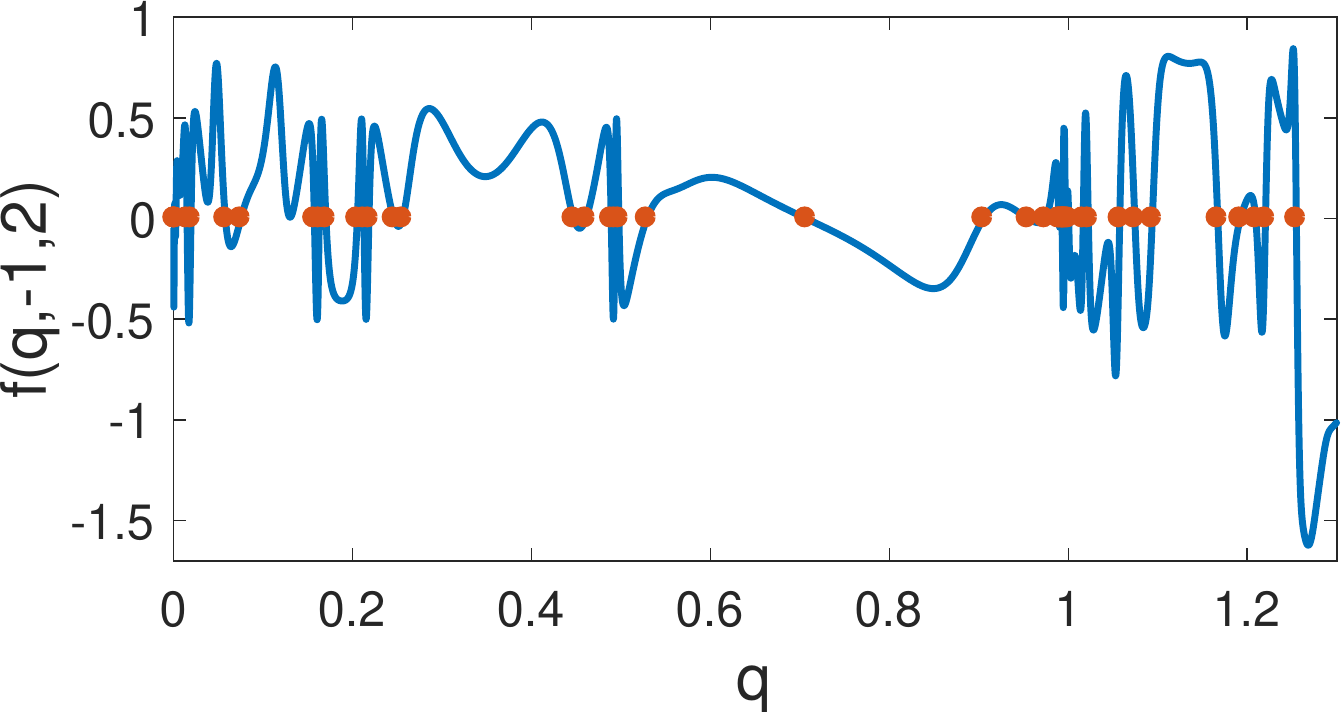} 
   \caption{The shooting function described in the text whose zeros correspond to nonlinear standing waves on the graph $\Gamma$.}
\label{fig:dumbbellShooter}
\end{figure}

\begin{figure}[htbp] 
   \centering
   \includegraphics[width=.4\textwidth]{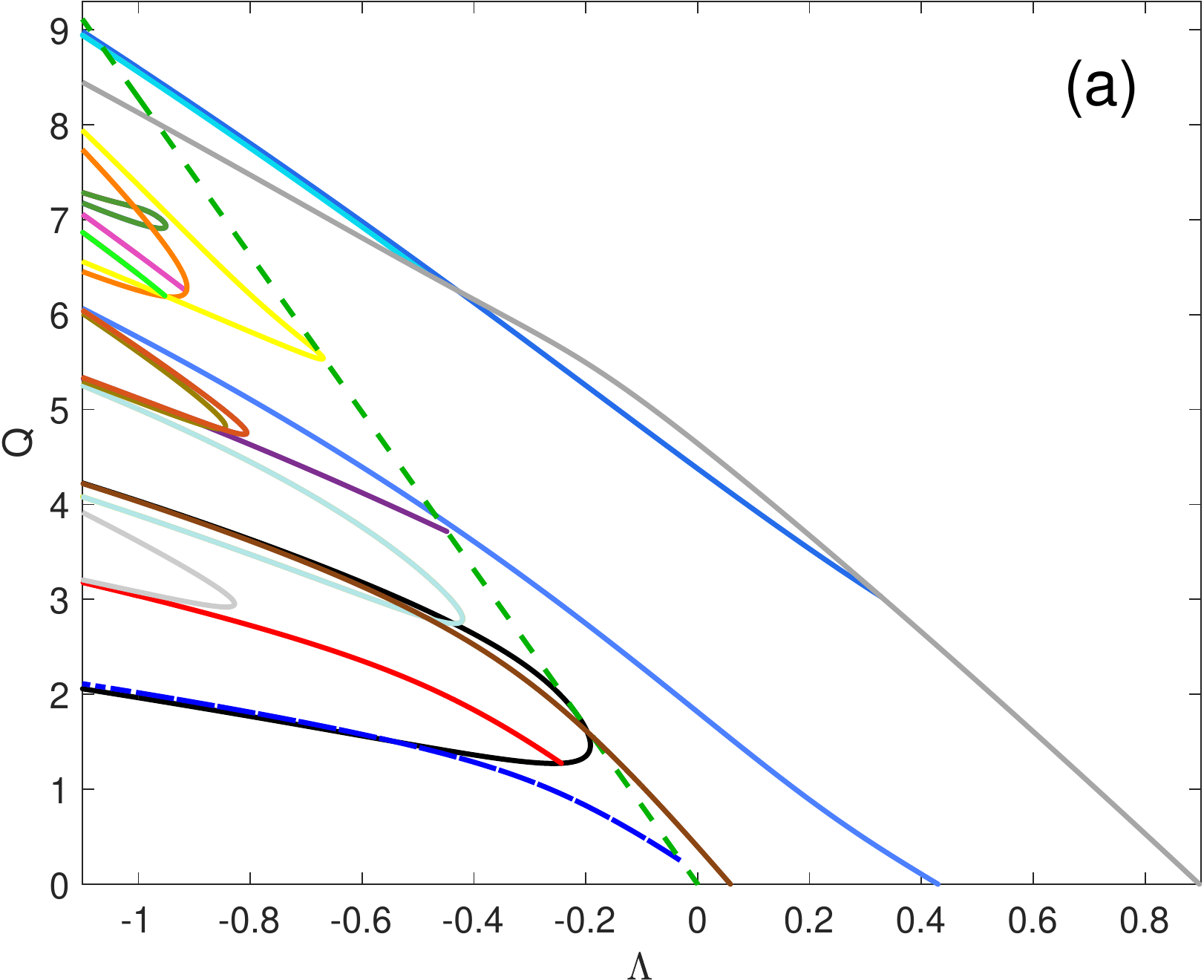} 
   \includegraphics[width=.4\textwidth]{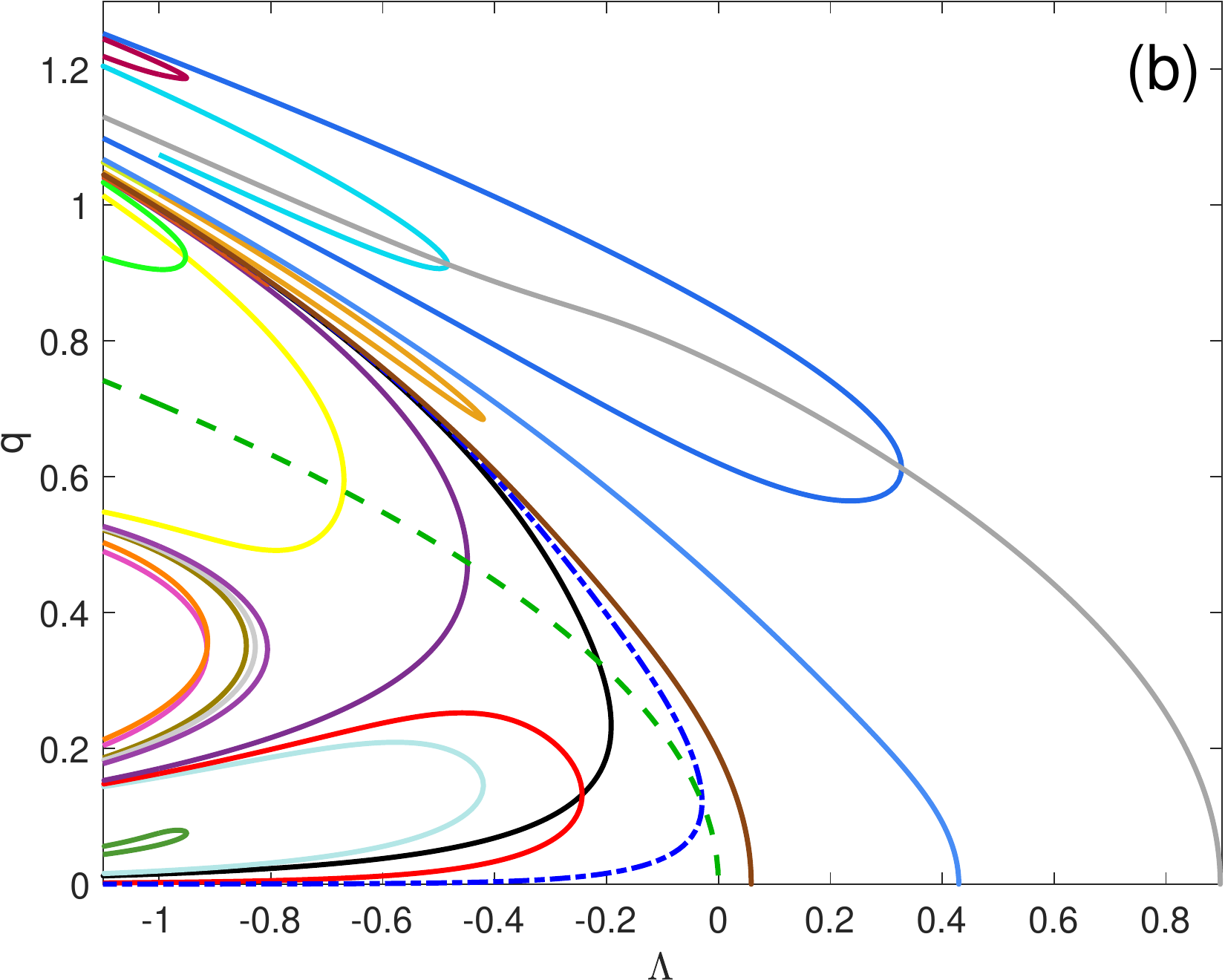} 
   \caption{Two views of a partial bifurcation diagram with $L=2$. (a) Plotting $Q$ the squared $L^2$ norm of the standing wave solutions. (b) Plotting the value $q$ used in the shooting function. Colors of branches are consistent between the two panels and with Fig.~\ref{fig:pdeBif}.}
\label{fig:bigbif}
\end{figure}

To follow this procedure, we first fix $\Lambda=-1$ and $L=2$ and compute $f(q,\Lambda,L)$ over an appropriate interval, which is shown in Fig.~\ref{fig:dumbbellShooter}. By standard arguments about continuous dependence on initial conditions, $f$ is a continuous function of $q$ and we can guarantee finding all such solutions by resolving this graph adequately.	 We then use MATCONT, a MATLAB continuation package, to trace these branches, varying $\Lambda$ and holding $L$ fixed~\cite{Dhooge:2003vy,Dhooge:2008gq}. This is summarized in an incomplete but complex bifurcation diagram in Fig.~\ref{fig:bigbif}, which follows all the solutions to $f(q,-1,2)=0$ with $0<q<1.3$.

First, we find the previously mentioned branches arising alternately in pitchfork and transcritical bifurcations from the constant state. We also note that all other even-symmetric branches undergo symmetry breaking as well, as was previously seen in Fig.~\ref{fig:pdeBif}. This black curve is the centered branch found in Ref.~\cite{Marzuola:2016bl}.  Many additional families bifurcate into existence in saddle-node bifurcations. Some of these are plotted twice in Fig.~\ref{fig:bigbif}(b), as they correspond to asymmetric solutions and their value at the center of either loop may be used as the input parameter $q$ in the shooting function.

An alternative to a numerical shooting method is to use the fact that the solution on each edge is given by a cnoidal or dnoidal function, see Appendix~\ref{sec:cnoidal}, or by a hyperbolic secant and use the boundary conditions~\eqref{continuity} and~\eqref{flux}, together with the complete loop assumption to derive nonlinear equations for the parameters in these solutions. This is the approach taken in reference~\cite{Noja:2015hx}. 

\subsection{Two complete loops}

Solutions of this type can be completely described analytically, including the values at $\Lambda$ for which they bifurcate. Doing so, however, is somewhat unwieldy, as there are many cases to consider. The restriction to each loop must be an integer number of periodic orbits of Equation~\eqref{stationary}, and on the restriction to central edge must be an integer number of half-period orbits, with $\phi_2'(\pm L)=0$. One such solution is shown in Fig.~\ref{fig:oneStationary}.

In Appendix~\ref{sec:cnoidal}, it is shown that when $\Lambda>0$, equation~\eqref{stationaryEdge} has periodic orbits of period $T < \Tcnmax$ given by~\eqref{Tcn}, and that for $\Lambda<0$ it has cnoidal periodic orbits of all periods outside the separatrix, while inside the separatrix, it has dnoidal periodic orbits with period $T > \Tdnmin$ given by~\eqref{Tdn}.

\begin{figure}[htbp] 
   \centering
   \includegraphics[width=2in]{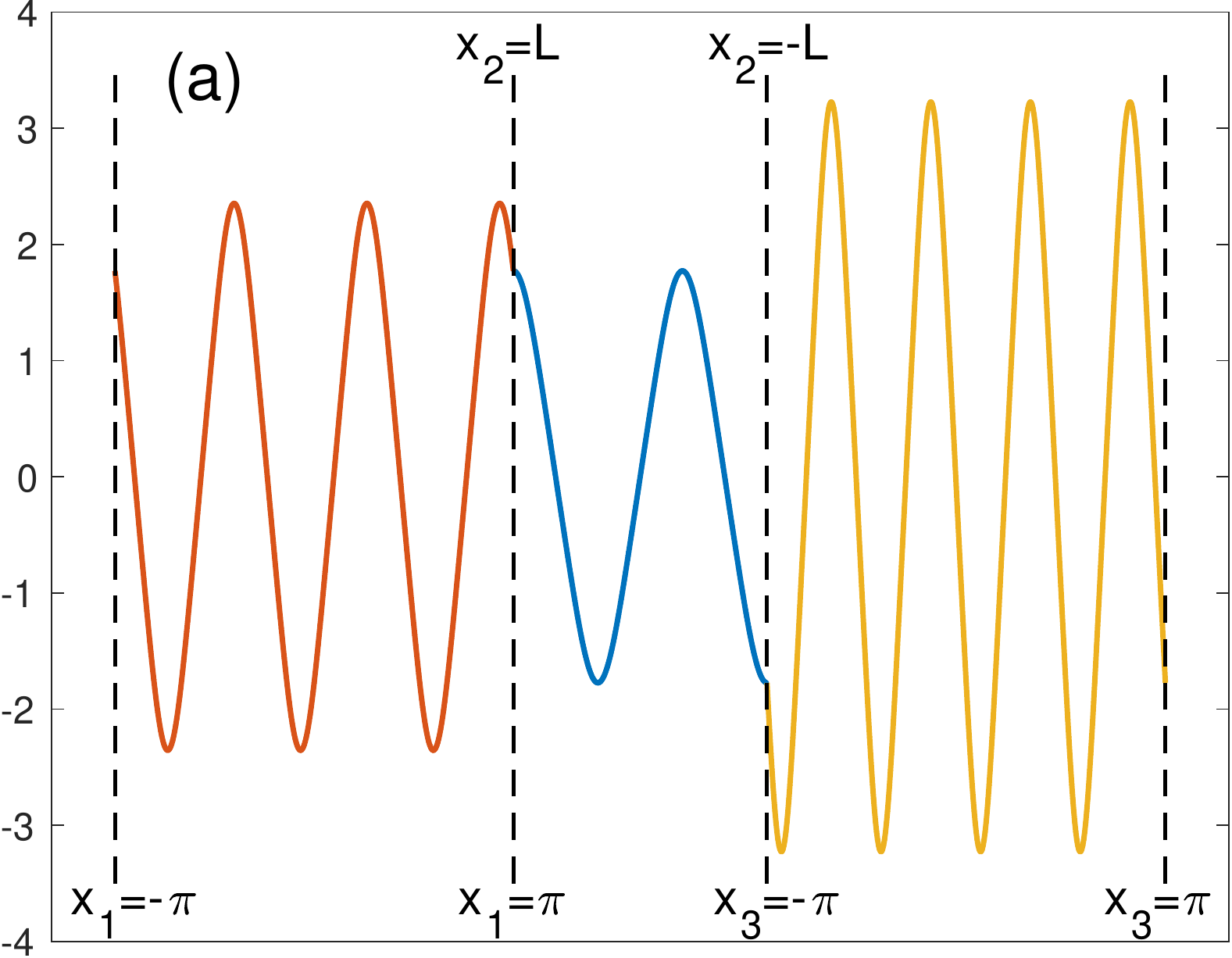} 
   \includegraphics[width=2in]{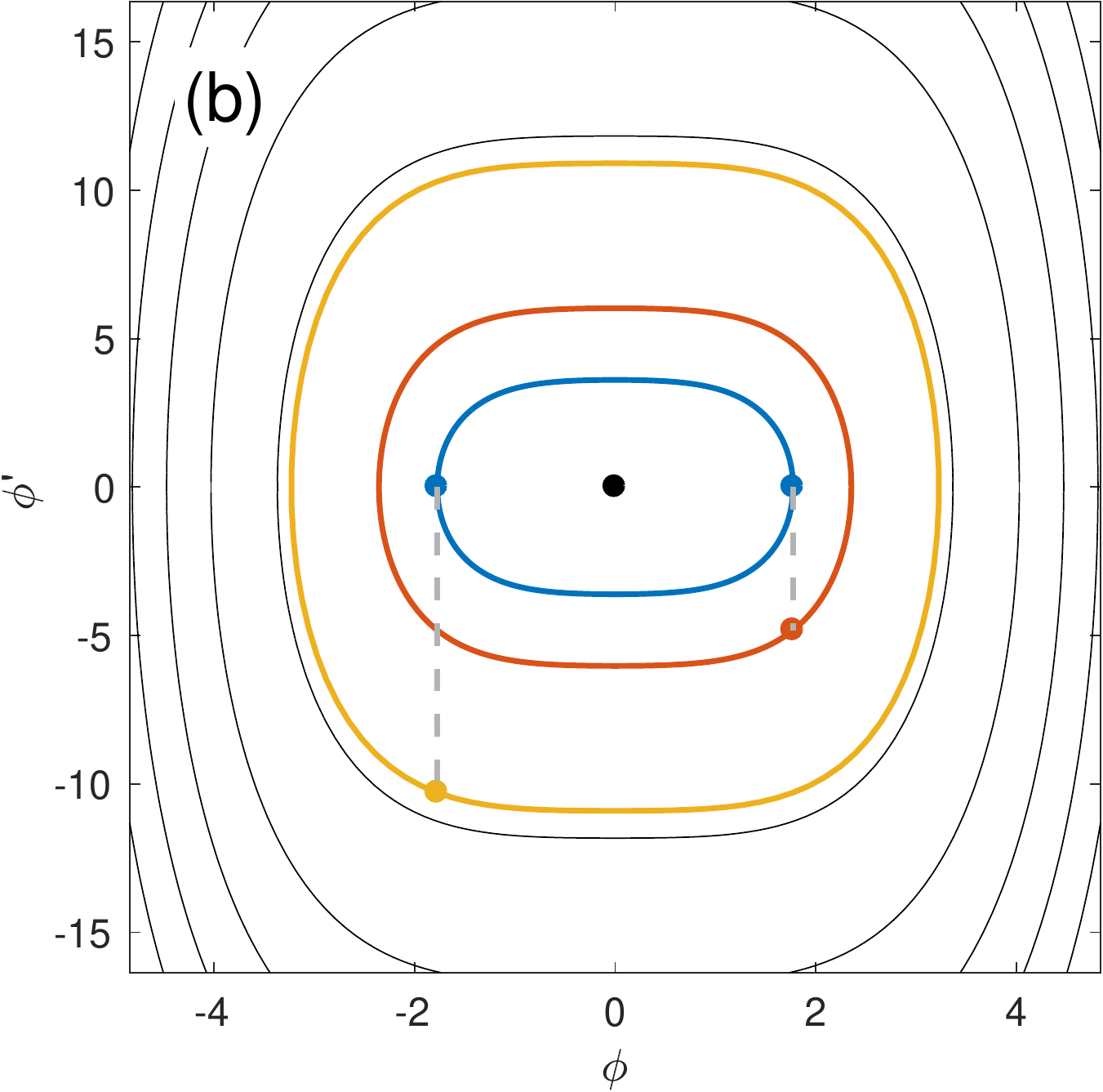} 
   \includegraphics[width=2in]{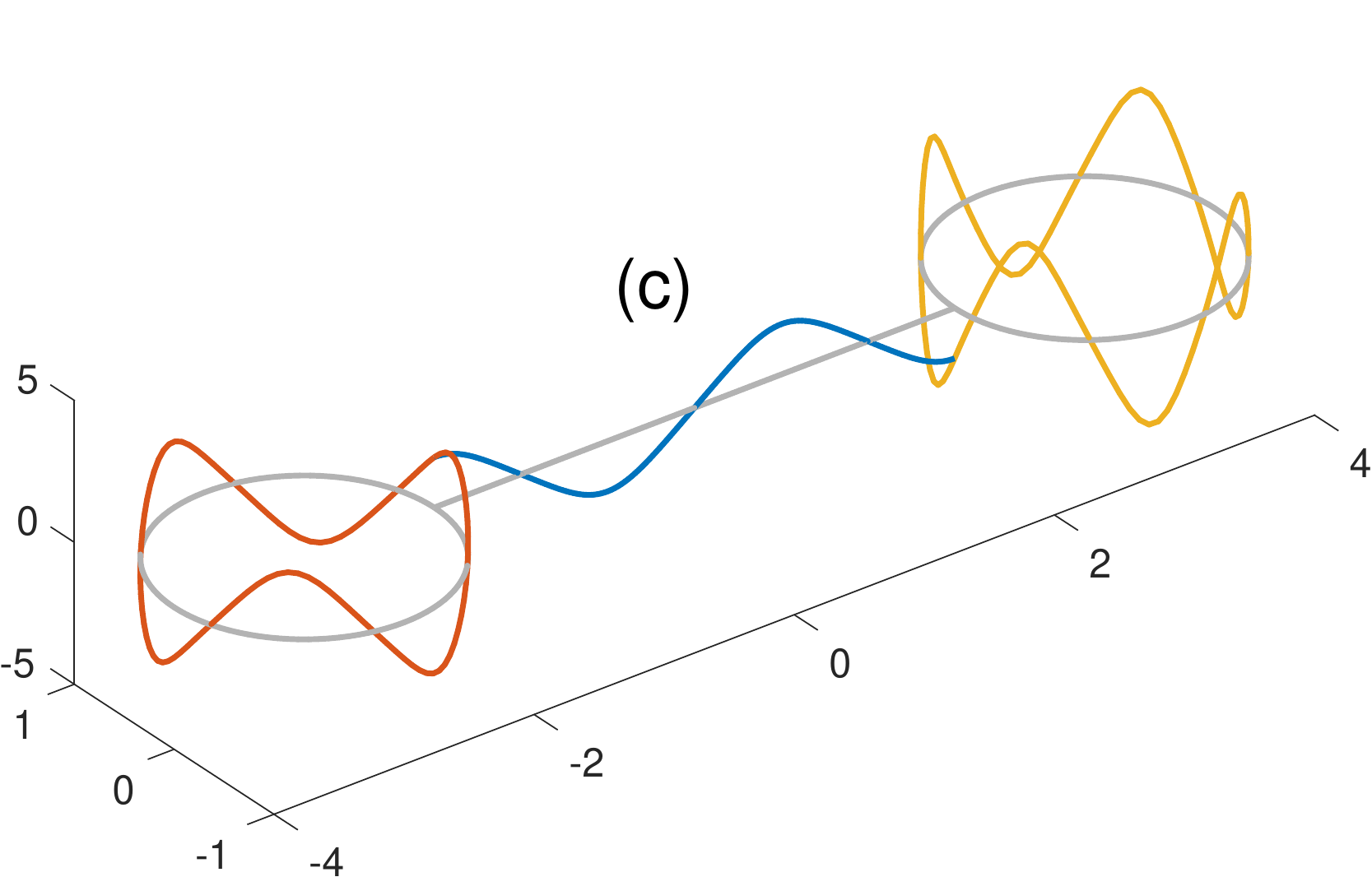} 
   \caption{Three views of a typical solution with two complete loops.}
\label{fig:oneStationary}
\end{figure}

We can classify all such solutions with a triple $(n_1,m,n_3)$ as follows
\begin{itemize}
\item If $n_i=0$ or $m=0$ then the solution vanishes on the indicated edge.
\item We use the label $n_i=\Lambda$ or $m=\Lambda$ if the $\phi= \pm \sqrt{\frac{-\Lambda}2}$ on the indicated edge. This can occur only for $\Lambda<0$.
\item $n_i>0$ if the solution contains exactly $n_i$ periods of the $\cn$ function on loop $i$ and  $n_i<0$ if the solution contains exactly $\abs{n_i}$ periods of the $\dn$ function on loop $i$. The former can only occur for $\Lambda< n_i^2$. and the latter for $\Lambda< -n_i^2/2$. 
\item $m>0$ if the edge $\te_2$ contains exactly $m$ half-periods of the $\cn$ function and and $m<0$ if the central edge contains exactly $\abs{m}$ half-periods of the $\dn$ function. The former can occur only for $\Lambda<m^2 \Lambda^*$ and the latter for $\Lambda<-m^2 \Lambda^*/2$, where $\Lambda*={\left(\frac{\pi}{2L}\right)}^2$. Here we have used the bounds on the periods of cnoidal and dnoidal periods~\eqref{Tcnmax} and~\eqref{Tdnmin}.
\end{itemize}

\begin{figure}[htbp] 
   \centering
   \includegraphics[width=.6\textwidth]{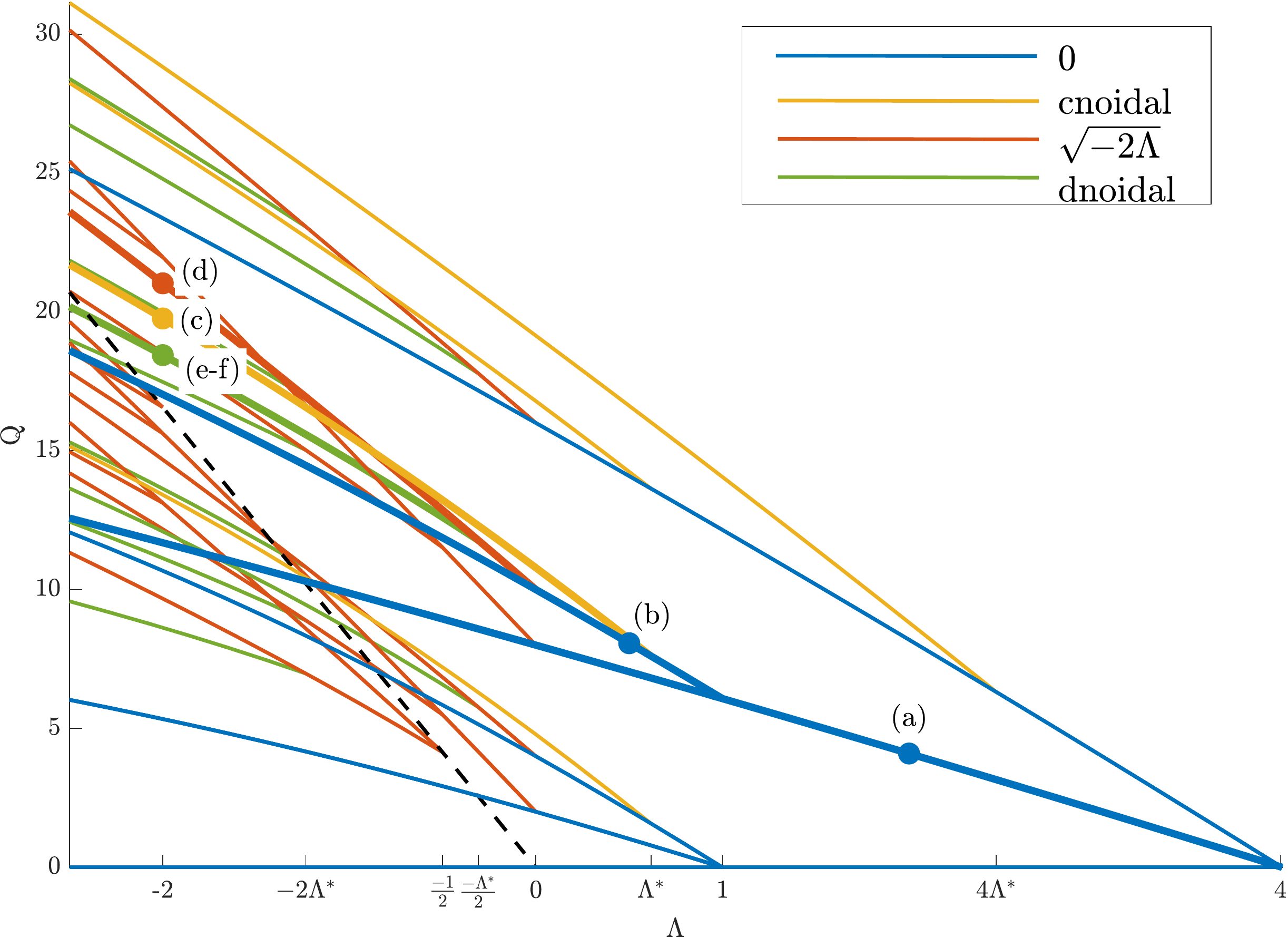} 
   \caption{Bifurcation diagram for solutions with two complete loops. Plotted are solutions with $\abs{n_j}\le 2$ and $\abs{m}\le2$. Color indicates type of solution on the edge $\te_2$. The dashed line shows the nonzero constant solution $\Phi = \sqrt{-\Lambda/2}$.}
\label{fig:ohmygod}
\end{figure}

\begin{figure}[htbp] 
   \centering
   \includegraphics[width=0.32\textwidth]{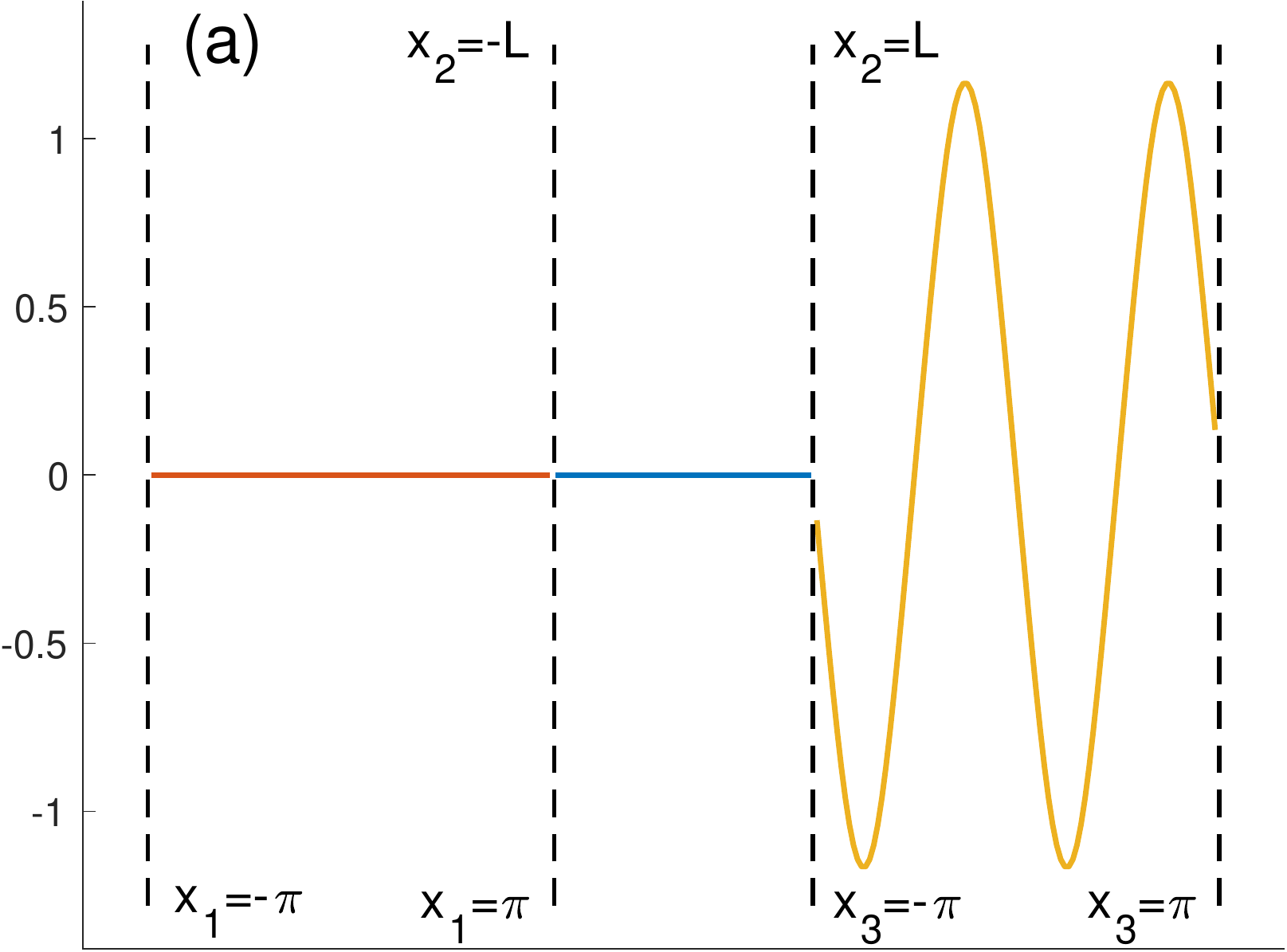}
   \includegraphics[width=0.32\textwidth]{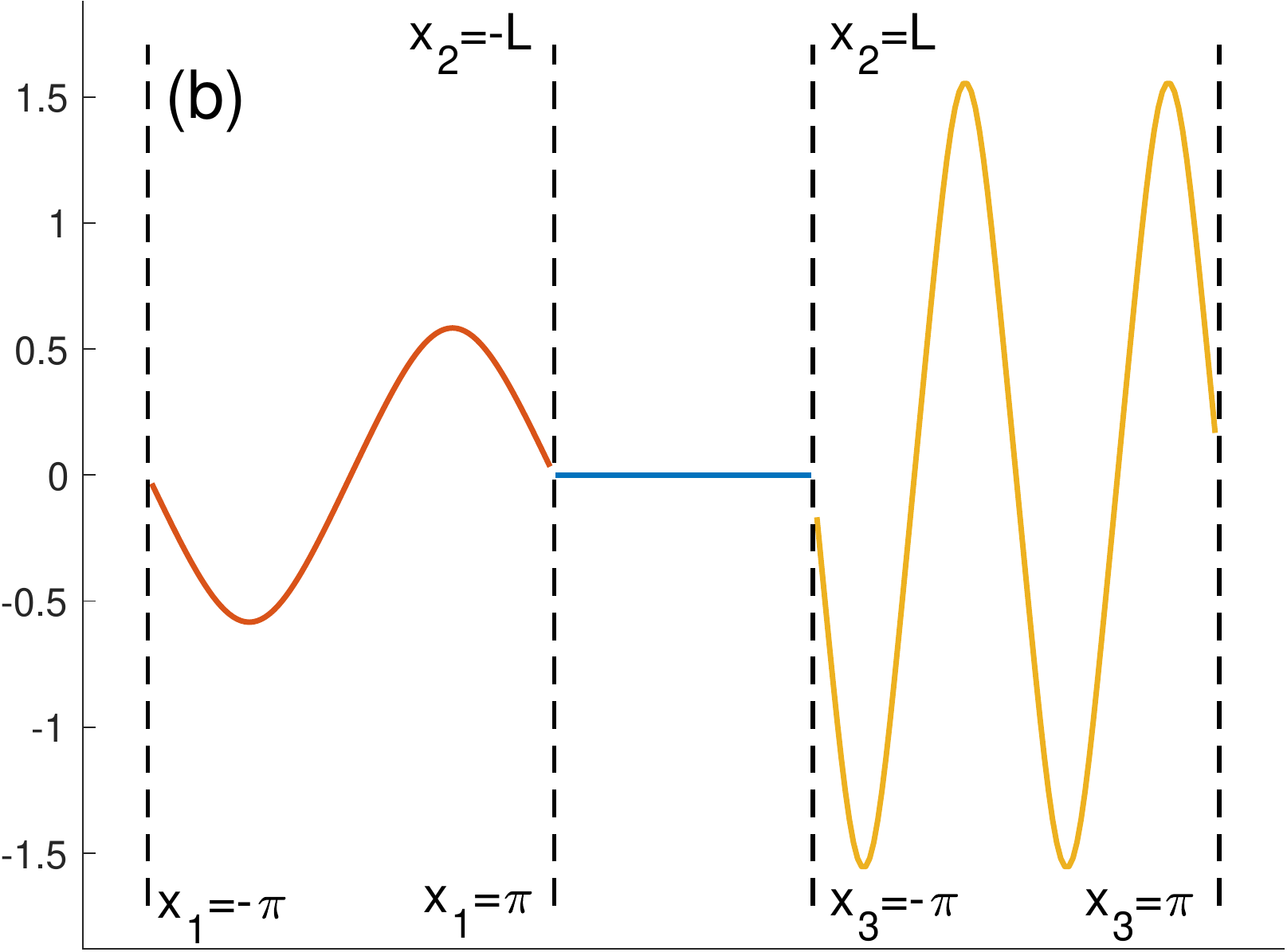}
   \includegraphics[width=0.32\textwidth]{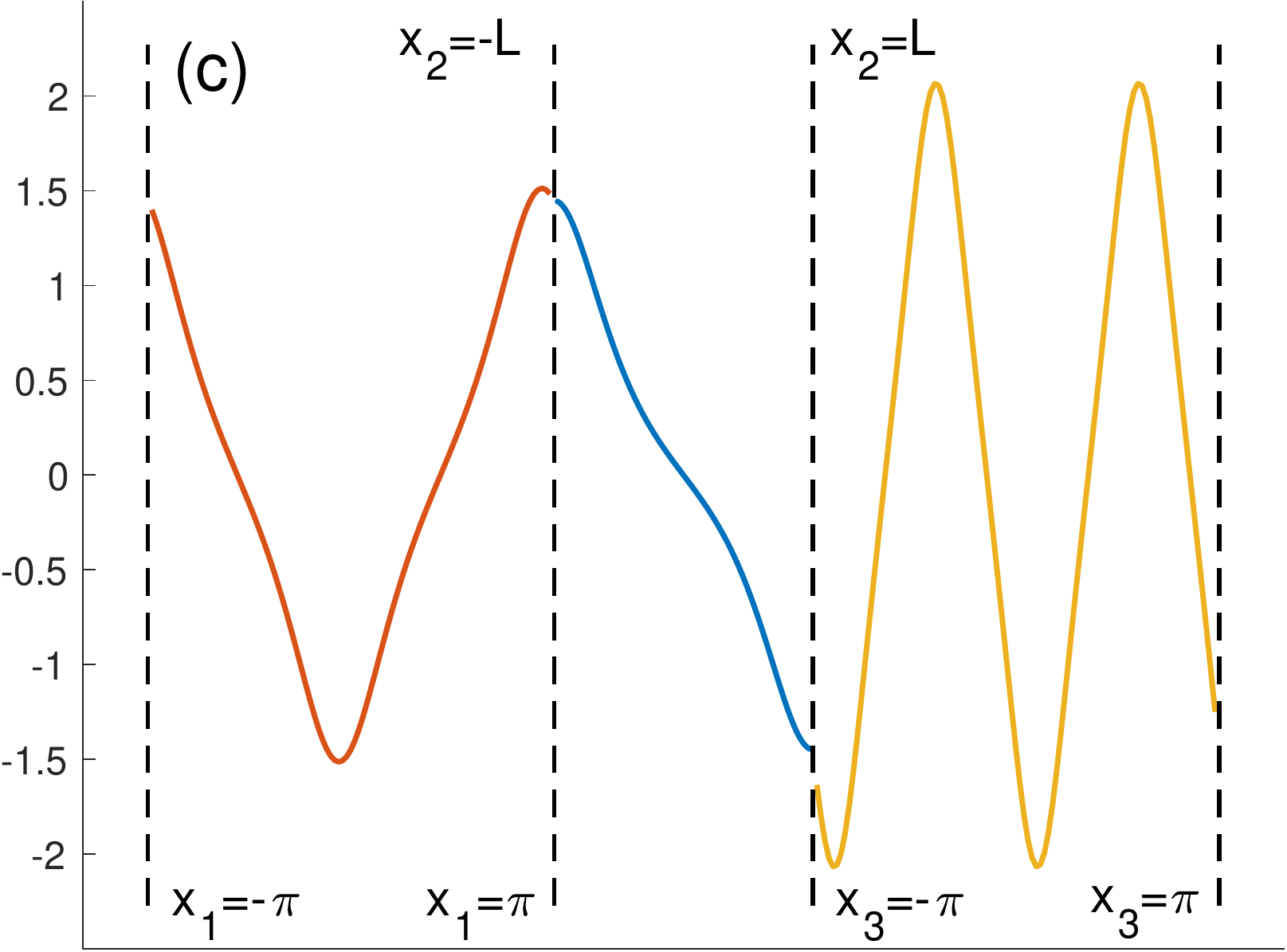}\\
   \includegraphics[width=0.32\textwidth]{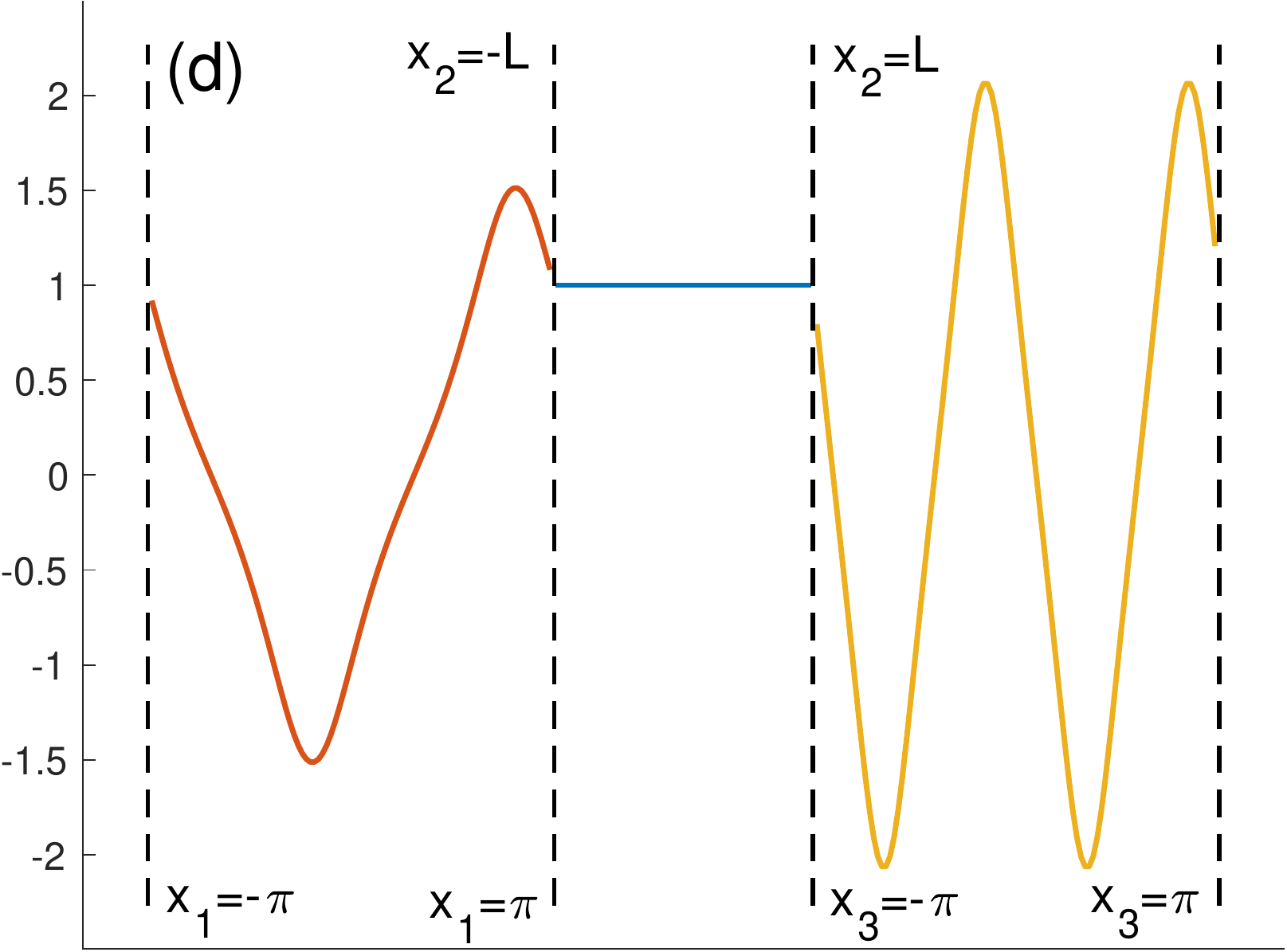}
   \includegraphics[width=0.32\textwidth]{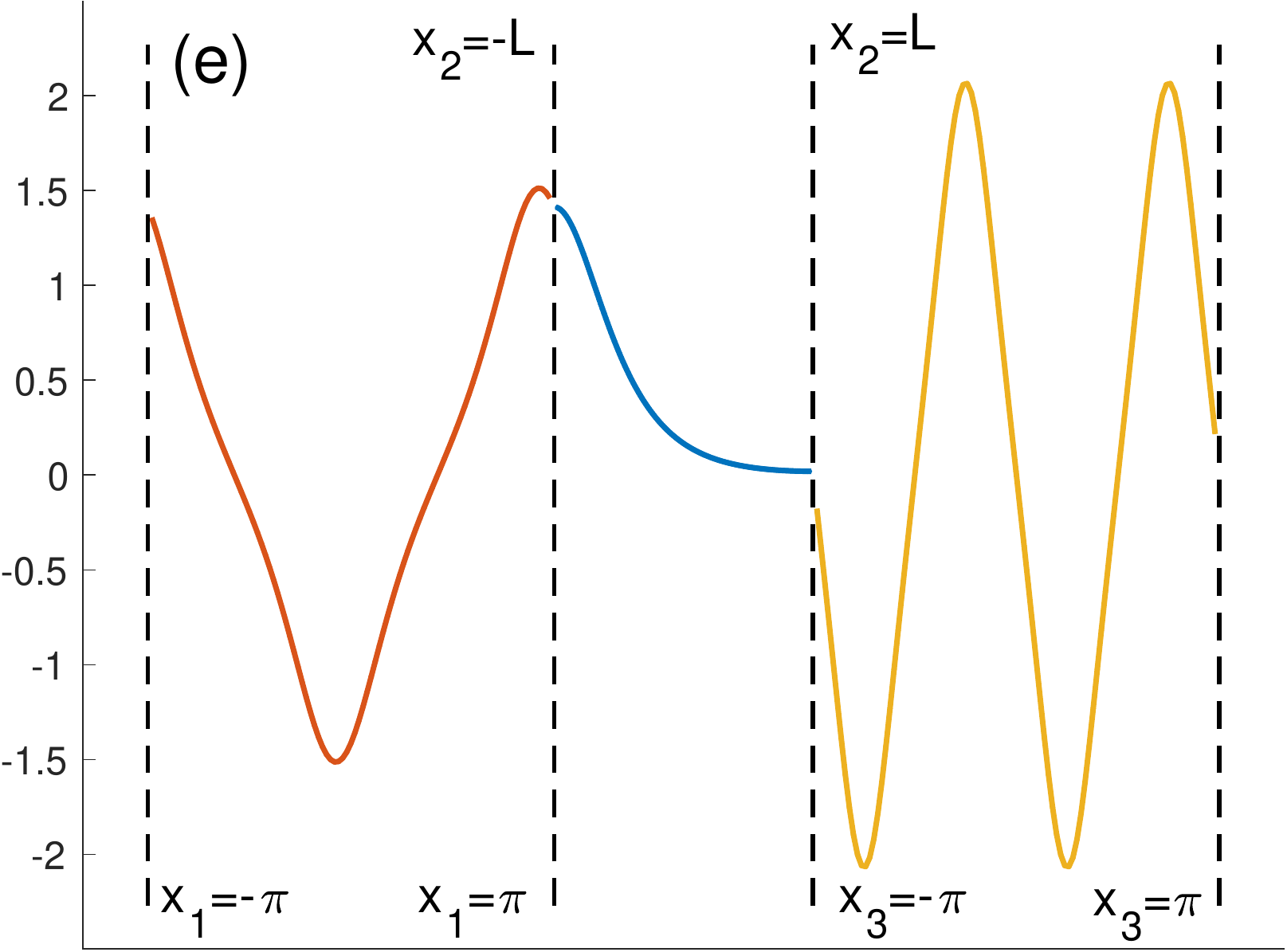}
   \includegraphics[width=0.32\textwidth]{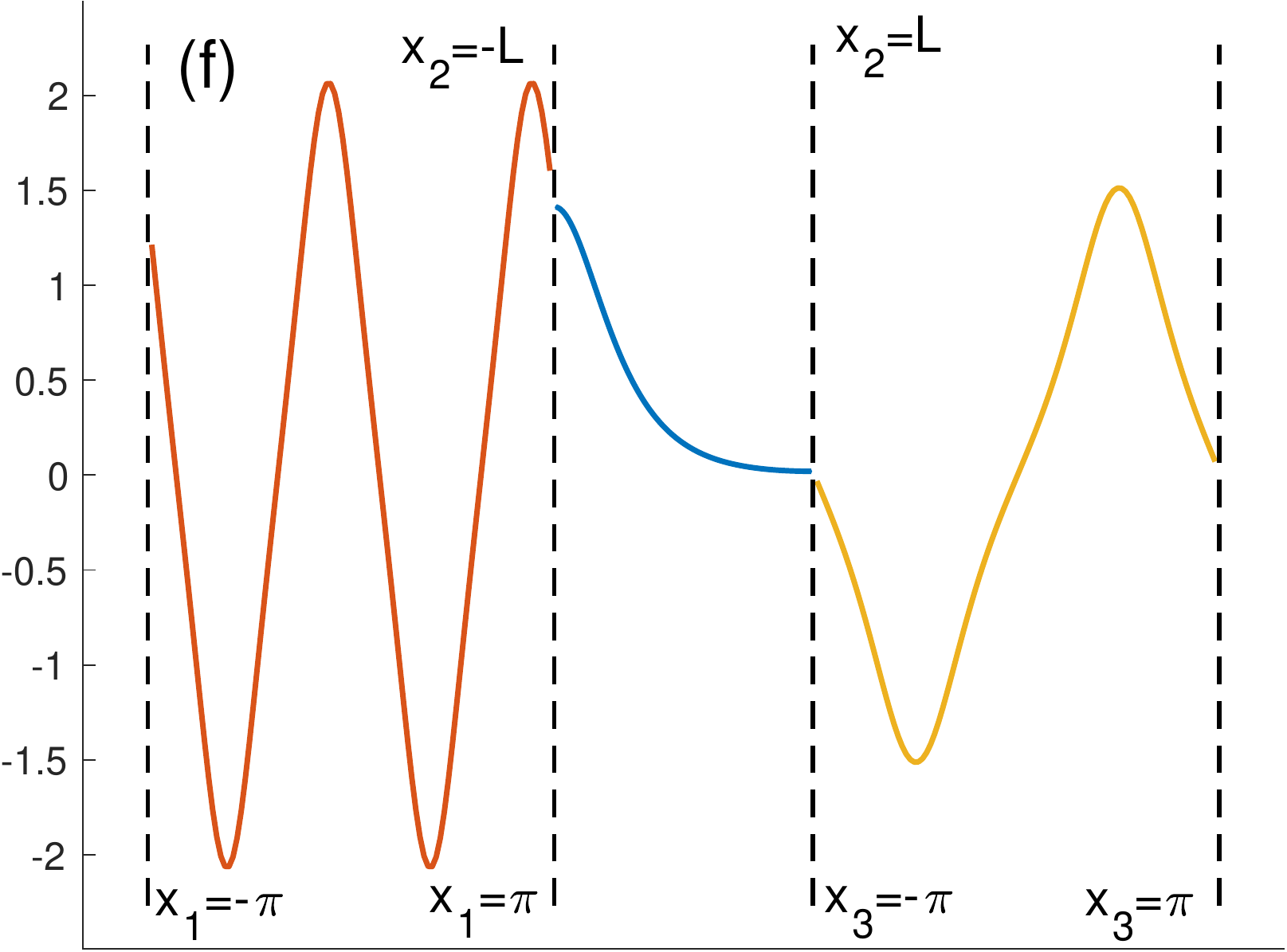}
   \caption{The standing waves at the six marked points in the bifurcation diagram of Fig.~\ref{fig:ohmygod}. \textbf{(a)} $(0,0,2)$, \textbf{(b)} $(1,0,2)$, \textbf{(c)} $(1,1,2)$, \textbf{(d)} $(1,\Lambda,2)$, \textbf{(e)} $(1,-1,2)$, \textbf{(f)} $(2,-1,1)$. Note from \textbf{(e)} and \textbf{(f)} that reversing $n_1$ and $n_3$ is not equivalent to a symmetry operation since a half-period of the $dn$-function has no symmetries. As $\Lambda$ decreases, \tparen{b} bifurcates from \tparen{a}, and then \tparen{c}, \tparen{d}, and \tparen{e-f} bifurcate from (b) in that order.}
\label{fig:completerings}
\end{figure}

Therefore, the trivial solution $(0,0,0)$ exists for all $\Lambda$, and the other solutions emerge as $\Lambda$ is decreased. We present a minimal list, and omit additional solutions obtained from group orbits of these solutions under~\eqref{dumbbellSymmetries}.
\begin{itemize}
\item At $\Lambda = n_1^2$, the solutions $(n_1,0,n_1)$ and $(n_1,0,0)$ bifurcate from $(0,0,0)$. Simultaneously, for all $n_3>n_1$, the solution $(n_1,0,n_3)$ bifurcates from $(0,0,n_3)$.
\item At $\Lambda = m^2 \Lambda^*$, the solution $(n_1,m,n_3)$ bifurcates from $(n_1,0,n_3)$ if $m < 2 L n_{1,3}/\pi$.
\item At $\Lambda = 0$, the solution $(\Lambda,\Lambda,\Lambda)$ bifurcates from $(0,0,0)$. Simultaneously $(n_1,\Lambda,n_3)$ bifurcates from $(n_1,0,n_3)$ for all $n_{1,3}\ge 0$.
\item At $\Lambda = -m^2 \Lambda^*/2$, solutions of the form $(n_1,-\abs{m},n_3)$ bifurcate from those of the form $(n_1,\Lambda,n_3)$. Note that if $m<0$ is odd and $n_1 \neq n_3$ then $(n_1,m,n_3)$ and $(n_3,m,n_1)$ are not related by symmetry.
\item At $\Lambda = -n_1^2/2$, solutions $(-\abs{n_1},\Lambda,-\abs{n_1})$ and $(-\abs{n_1},\Lambda,\Lambda)$ bifurcate from $(\Lambda,\Lambda,\Lambda)$.  Simultaneously $(-\abs{n_1},\Lambda,n_3)$ bifurcates from $(\Lambda,\Lambda,n_3)$ for all $n_3>-\abs{n_1}$.
\end{itemize}
All the resulting branches are displayed in Fig.~\ref{fig:ohmygod}. Solutions at the five indicated points on this bifurcation diagram are displayed in Fig.~\ref{fig:completerings}. 

\subsection{One incomplete and one complete loop}
\label{sec:oneOfEach}

Solutions of this type have two parts. Assume that the incomplete loop lies on edge $\te_1$, then at vertex $\tv_2$, we have $\phi_2'(L) =0$. Therefore the solution on the lollipop-shaped subgraph formed by excluding $\te_3$ is a nonlinear standing wave with Neumann boundary condition at vertex $\tv_2$. We may find all such solutions by the same combination of shooting argument and numerical continuation described in Section~\ref{sec:2incomplete}. For a given value of $\Lambda$, this may be extended to a standing wave on the dumbbell graph if there exists a cnoidal solution of the form~\eqref{cn} with $T = 2\pi/n$ and $\abs{\phi_2(L)} \le \alpha$, or if there exists a dnoidal solution of the form~\eqref{dn} with $T = 2 \pi/n$ and $\sqrt{1-k^2}a \le \abs{\phi_2(L)} < a$. 

\begin{figure}[htbp] 
   \centering
   \includegraphics[width=.52\textwidth]{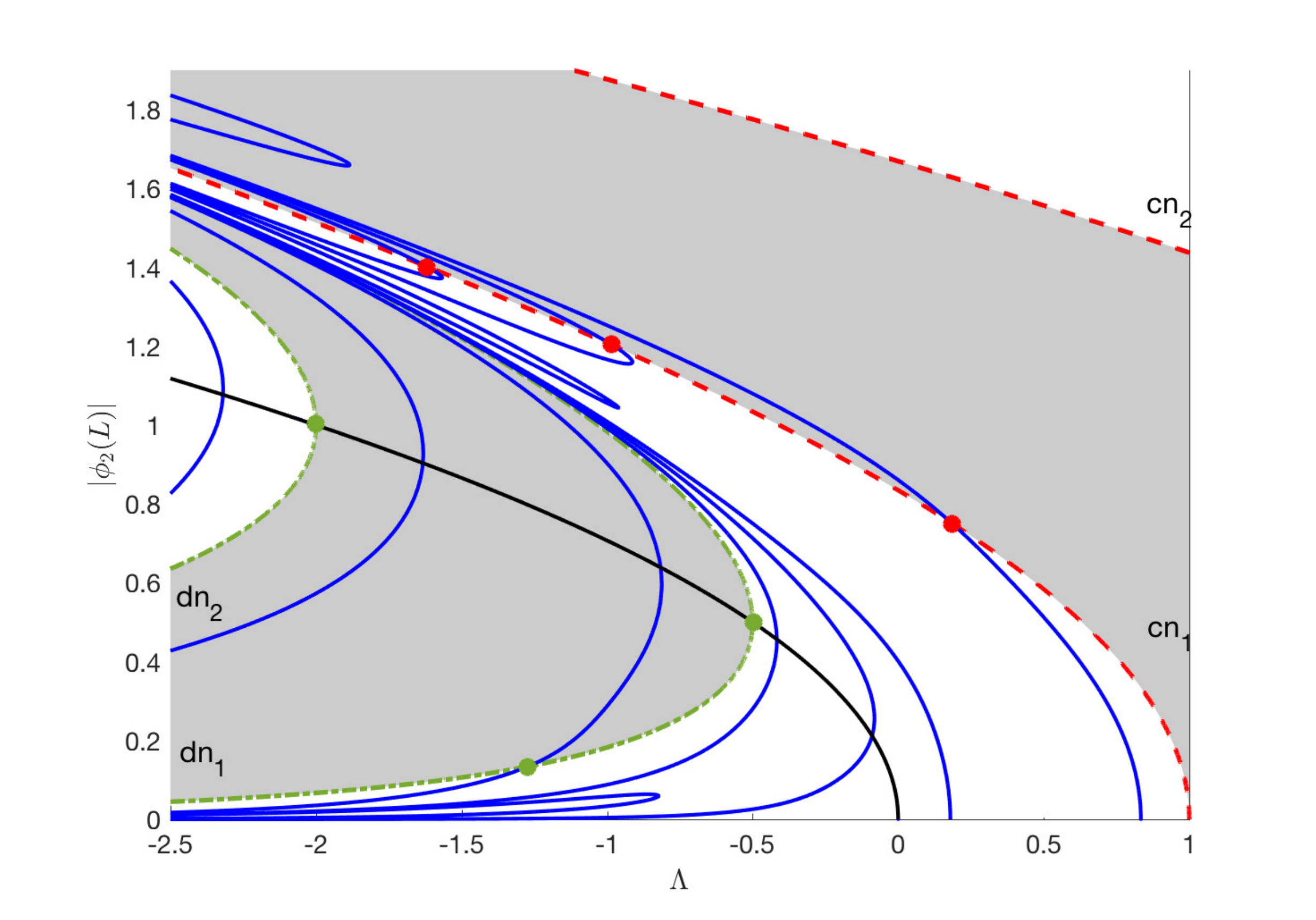} 
   \includegraphics[width=.45\textwidth
   ]{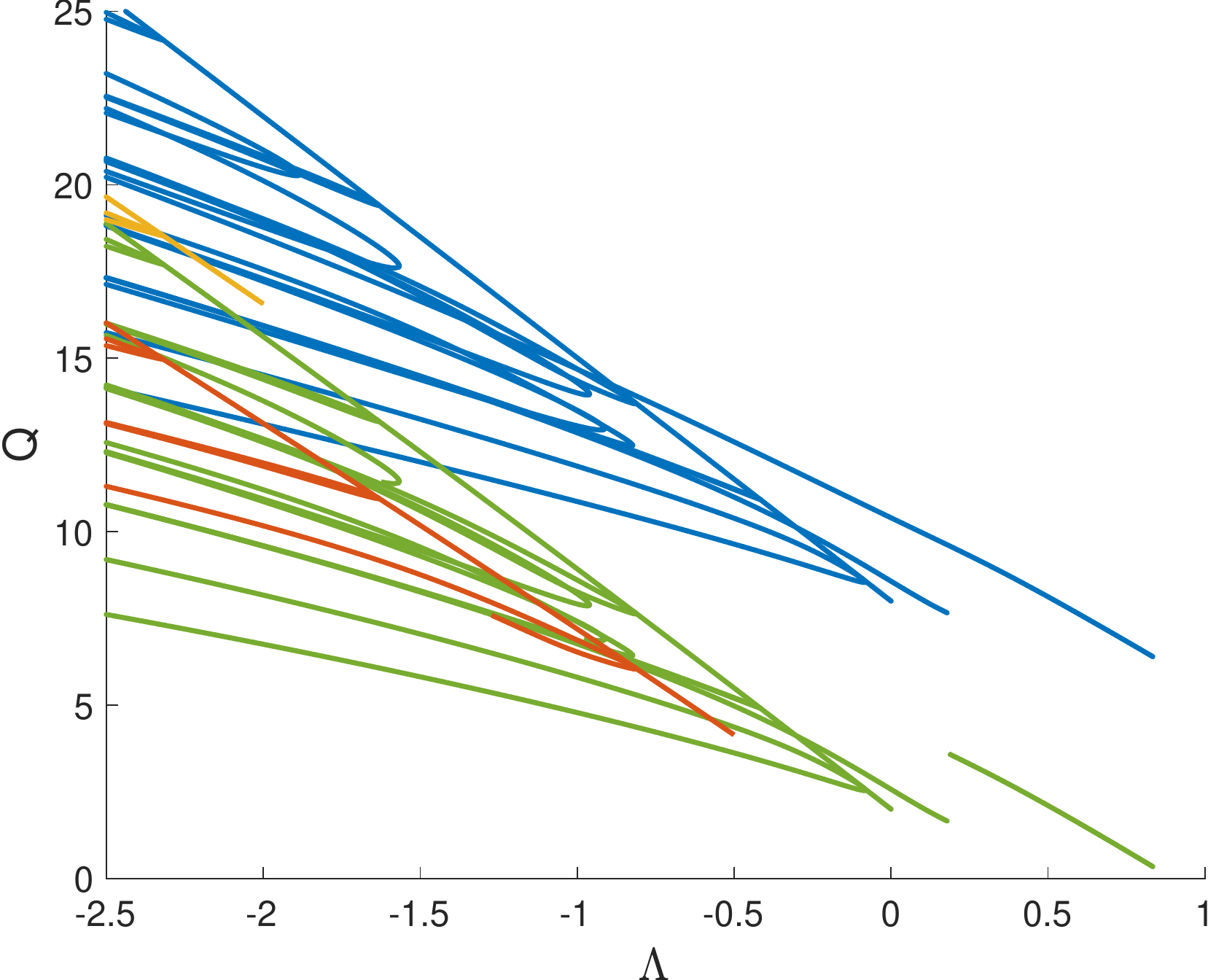} 
   \caption{(a) Solid curves: Partial bifurcation diagram on the lollipop subgraph. Dashed curves indicate the maximum values of the quantized cnoidal solutions and the dash-dot the maximum and minimum values of the quantized conoidal solutions on edge $\te_3$, with the regions between them shaded, alternately, for clarity. The marked points at intersections between the two families of curves indicate saddle node bifurcations of solutions with cnoidal or dnoidal solutions on the edge~$\te_3$. (b) Partial bifurcation diagram on the dumbbell graph.  }
\label{fig:hybridLambda}
\end{figure}

Fig.~\ref{fig:hybridLambda}(a) shows, superimposed, the results of two computations. The solid curves show the solutions of the problem on the lollipop subgraph with Neumann condition at vertex $\tv_2$. We have plotted $\phi_2(\tv_2)$ as a function of $\Lambda$. Also plotted are the maximum value of the cnoidal solutions quantized on the loop, for $n=1,2$ and the minimum and maximum values of the dnoidal solutions quantized on the loop for $n=1,2$. Solutions on the lollipop can be extended to the dumbbell  if $\abs{\phi_2(\tv_2)}$ is below the maximum of the cnoidal function or between the minimum and the maximum of the dnoidal function. In additions to the bifurcations of the lollipop solutions,  saddle-node bifurcations occur where the lollipop solution curves cross the curves of complete loop solutions. 

A bifurcation diagram for these solutions on the dumbbell is shown in Fig.~\ref{fig:hybridLambda}(b). Note that none of the solutions bifurcate from $Q=0$. Points where the curve appears to end are actually saddle-node bifurcations corresponding to the crossings described in the previous paragraph.

\section{Conclusion}
\label{sec:conclusion}

To this author, reference~\cite{Marzuola:2016bl} raised the question of whether the bifurcation scenario seen for NLS on the dumbbell graph is fundamentally different from that seen in other simpler systems. The far simpler bowtie DST system possesses the bifurcation structure identical to what they found, indicating that the phenomenon does not depend on the system being a quantum graph, but suggesting that the topology of the graph is important. This is confirmed by explicit comparison with the analogous problem posed on a line segment. The self-connected edges can be thought of as providing a singular perturbation to this simpler system, one which breaks the symmetry of half the families of solutions and thereby transforms pitchfork bifurcations into transcriticals. By contrast, the lollipop graph has less symmetry than the dumbbell and thus only transcriticals. 

In the opposite direction, conceptually, we have shown how to enumerate the complete set of standing waves for this system, revealing a tremendous amount of complexity. Nonetheless, we have only been able to understand this much because of the relative simplicity of the dumbbell graph. In particular, the loop structure only permits two types of behavior at the vertices, and from this we are able to classify the solutions into three types. Further, the shooting methods described above depend strongly on the simple arrangement of the three edges. On a graph with many branches, or cycles, such shooting methods would be untenable.

We have not discussed stability of these standing waves as was done in~\cite{Noja:2015hx}, other than for the constant valued solution. Of special interest would be the stability of the solutions of the type $(n_1,0,n_3)$ discussed in Section~\ref{sec:2incomplete}. Because the phase of the solution is ill-defined on the edge~$\te_2$, the components on the other two edges can each be multiplied by an arbitrary phase. We may ask whether the stability depends on this phase. Beyond this, we can easily construct solutions which vanish on $\te_2$ and for which the solutions on the other two edges oscillate at different frequencies. Determining the stability of these solutions involves analyzing a genuinely time-dependent problem requiring Floquet theory.

Finally, In addition, recent work by Kirr has made progress toward analysis that identifies all the coherent states of a given nonlinear wave system using global bifurcation theory and perturbations from the large-amplitude limit~\cite{Kirr:2017}. It would be of interest to see if such an approach could be useful in studying the $\Lambda \to -\infty$ limit of the standing wave problem.

\section*{Acknowledgments}
The author thanks Jeremy Marzuola and Dmitry Pelinovsky for many informative discussions and suggestions, and Stephen Shipman and Jianke Yang for reading and commenting on the manuscript. The author gratefully acknowledges the support of the Institute for Mathematics and its Applications (IMA) during the 2016--17 program in Mathematics and Optics, where this research was initiated. Mathematica was used for symbolic calculation and MATLAB for numerical calculation~\cite{Mathematica,matlab}.

\appendix
\section{Jacobi elliptic function solutions}
\label{sec:cnoidal}

This section provides formulas for exact solutions to Equation~\eqref{stationary}, it makes use of basic facts about elliptic integrals and Jacobi elliptic functions that can be found, for example, in Chapters~19 and~22 of the NIST Digital Library of Mathematical Functions~\cite{NIST:DLMF}.

For $\Lambda>0$, the exact solutions are given by cnoidal functions,
\begin{equation}
\Phi = \alpha \cn{(\beta x- \tau,\kappa)},
\label{cn}
\end{equation}
where 
\[
0 \le \kappa < \frac{1}{\sqrt{2}}, \,
\alpha^2 = \frac{\kappa^2 \Lambda}{1-2\kappa^2}, \, \text{and }
\beta^2 = \frac{\Lambda^2}{1-2\kappa^2}.
\]
This has period 
\begin{equation}
\Tcn = 4 K(\kappa) \sqrt{\frac{1-2\kappa^2}{\Lambda}},
\label{Tcn}
\end{equation}
where $K(\kappa)$ is the complete elliptic integral of the first kind, defined by
$$
K(\kappa) = \int_0^{\frac{\pi}{2}} \frac{1}{\sqrt{1-\kappa^2 \sin^2{\theta}}} \ d\theta.
$$
As $\kappa \to 0^+$, the solution trajectories shrink to a point at the origin and
\begin{equation}
T \to \Tcnmax = \frac{2\pi}{\sqrt{\Lambda}},
\label{Tcnmax}
\end{equation}
the period of the linearized solution in a neighborhood of the origin. 
As $\kappa \to {\frac{1}{\sqrt{2}}}^-$, 
\[
T \to 0,
\]
and the diameter of the trajectory diverges.

For $\Lambda<0$, there are two types of periodic orbits. The first lie outside the separatrices shown in Fig.~\ref{fig:phaseplane} and are given by the same formulas as above, only for $\frac{1}{\sqrt{2}}  < \kappa < 1$. As $\kappa \to 1^-$, these orbits approach the separatrices and their period diverges. As $\kappa \to {\frac{1}{\sqrt{2}}}^+$, the diameter of these orbits diverge, and their period approaches zero.

The other types of solution trajectories lie inside one or the other of the separatrices,
\begin{equation}
\Phi = \pm a \dn{(bx-\tau,k)},
\label{dn}
\end{equation}
with 
\[
0\le k < 1, \, \text{and }
a = b = \sqrt{\frac{\Lambda}{k^2-2}}
\]
and period 
\begin{equation}
\Tdn = 2K(k)\sqrt{\frac{k^2-2}{\Lambda}}.
\label{Tdn}
\end{equation}
As $k \to 0^+$, these periodic orbits shrink toward the nonzero fixed points and 
\begin{equation}
T \to \Tdnmin =\frac{\pi \sqrt{2}}{\sqrt{-\Lambda}}.
\label{Tdnmin}
\end{equation}
As $k \to 1^-$, these orbits approach the separatrices and their periods diverge.

\bibliographystyle{habbrv}
\bibliography{bowtieRevised}

\begin{thebibliography}{10}

\bibitem{Adami:2012jb}
R.~Adami, C.~Cacciapuoti, D.~Finco, and D.~Noja.
\newblock {Stationary states of NLS on star graphs}.
\newblock {\em EPL--Europhys. Lett.}, 100:10003, 2012.

\bibitem{Arioli:vf}
M.~Arioli and M.~Benzi.
\newblock {A Finite Element Method for Quantum Graphs}.
\newblock {\em IMA J. Numer. Anal.}, to appear, 2017.

\bibitem{Berkolaiko:2016ue}
G.~Berkolaiko.
\newblock {An elementary introduction to quantum graphs}.
\newblock {\em ArXiv e-prints}, math:arXiv: 1603.07356, 2016.

\bibitem{Berkolaiko:2013}
G.~Berkolaiko and P.~Kuchment.
\newblock {\em Introduction to Quantum Graphs}.
\newblock Mathematical surveys and monographs. American Mathematical Society,
  2013.

\bibitem{Bolte:2014}
J.~Bolte and J.~Kerner.
\newblock Many-particle quantum graphs and {B}ose-{E}instein condensation.
\newblock {\em J. Math. Phys.}, 55:061901, 2014.

\bibitem{Cacciapuoti:2015fe}
C.~Cacciapuoti, D.~Finco, and D.~Noja.
\newblock {Topology-induced bifurcations for the nonlinear Schr{\"o}dinger
  equation on the tadpole graph}.
\newblock {\em Phys. Rev. E}, 91:013206, 2015.

\bibitem{Delourme:2017jd}
B.~Delourme, S.~Fliss, P.~Joly, and E.~Vasilevskaya.
\newblock {Trapped modes in thin and infinite ladder like domains. Part 1:
  Existence results}.
\newblock {\em Asymptotic Anal.}, 103:103--134, 2017.

\bibitem{Dhooge:2003vy}
A.~Dhooge, W.~Govaerts, and Y.~A. Kuznetsov.
\newblock {MATCONT: a MATLAB package for numerical bifurcation analysis of
  ODEs}.
\newblock {\em ACM T. Math. Software}, 29:141--164, 2003.

\bibitem{Dhooge:2008gq}
A.~Dhooge, W.~Govaerts, Y.~A. Kuznetsov, H.~G.~E. Meijer, and B.~Sautois.
\newblock {New features of the software MatCont for bifurcation analysis of
  dynamical systems}.
\newblock {\em Math. Comp. Model. Dyn.}, 14:147--175, 2008.

\bibitem{NIST:DLMF}
{\it NIST Digital Library of Mathematical Functions}.
\newblock http://dlmf.nist.gov/, Release 1.0.15 of 2017-06-01.
\newblock F.~W.~J. Olver, A.~B. {Olde Daalhuis}, D.~W. Lozier, B.~I. Schneider,
  R.~F. Boisvert, C.~W. Clark, B.~R. Miller and B.~V. Saunders, eds.

\bibitem{Eilbeck:2003vl}
J.~C. Eilbeck and M.~Johansson.
\newblock {The discrete nonlinear Schr{\"o}dinger equation--20 years on}.
\newblock In R.~S. MacKay, L.~V\'azquez, and M.~P. Zorzano, editors, {\em
  Proceedings Of The Third Conference On Localization And Energy Transfer In
  Nonlinear Systems}, pages 44--67, Madrid, 2003. World Scientific.

\bibitem{Eilbeck:1985tu}
J.~C. Eilbeck, P.~S. Lomdahl, and A.~C. Scott.
\newblock {The discrete self-trapping equation}.
\newblock {\em Phys. D}, 16:318--338, 1985.

\bibitem{Holm:2011uw}
D.~D. Holm.
\newblock {\em {Geometric Mechanics Part I: Dynamics and Symmetry}}.
\newblock Imperial College Press, 2nd edition, 2011.

\bibitem{Kevrekidis:2009vv}
P.~G. Kevrekidis.
\newblock {\em {The Discrete Nonlinear Schr{\"o}dinger Equation: Mathematical
  Analysis, Numerical Computations and Physical Perspectives}}, volume 232 of
  {\em Springer Tr. Mod. Phys.}
\newblock Springer, Berlin Heidelberg, 2009.

\bibitem{Kirr:2017}
E.-W. Kirr.
\newblock Long time dynamics and coherent states in nonlinear wave equations.
\newblock In R.~Melnik, R.~Makarov, and J.~Belair, editors, {\em Recent
  Progress and Modern Challenges in Applied Mathematics, Modeling and
  Computational Science}, volume~79 of {\em Fields Inst. Commun.} Springer,
  2017.

\bibitem{Kuchment:2007}
P.~Kuchment and O.~Post.
\newblock On the spectra of carbon nano-structures.
\newblock {\em Communications in Mathematical Physics}, 275:805--826, 2007.

\bibitem{Marzuola:2016bl}
J.~L. Marzuola and D.~E. Pelinovsky.
\newblock {Ground State on the Dumbbell Graph}.
\newblock {\em Appl. Math. Res. Express}, 2016:98--145, 2016.

\bibitem{nayfeh2008applied}
A.~Nayfeh and B.~Balachandran.
\newblock {\em Applied Nonlinear Dynamics: Analytical, Computational and
  Experimental Methods}.
\newblock Wiley Series in Nonlinear Science. Wiley, 2008.

\bibitem{Niikuni:2017fu}
H.~Niikuni.
\newblock {Schr{\"o}dinger operators on a periodically broken zigzag carbon
  nanotube}.
\newblock {\em P. Indian Acad. Sci.--Math. Sci.}, 127:471--516, 2017.

\bibitem{Noja:2015hx}
D.~Noja, D.~E. Pelinovsky, and G.~Shaikhova.
\newblock {Bifurcations and stability of standing waves in the nonlinear
  Schr{\"o}dinger equation on the tadpole graph}.
\newblock {\em Nonlinearity}, 28:2343--2378, 2015.

\bibitem{Noja:2017vd}
D.~{Noja}, S.~{Rolando}, and S.~{Secchi}.
\newblock {Standing waves for the NLS on the double-bridge graph and a
  rational-irrational dichotomy}.
\newblock {\em ArXiv e-prints}, math:arXiv:1706.09616, 2017.

\bibitem{Pelinovsky:2016fm}
D.~E. Pelinovsky and G.~Schneider.
\newblock {Bifurcations of Standing Localized Waves on Periodic Graphs}.
\newblock {\em Ann. Henri Poincar{\'e}}, 18:1185--1211, 2016.

\bibitem{Yang:2013hj}
J.~Yang.
\newblock {Stability switching at transcritical bifurcations of solitary waves
  in generalized nonlinear Schr{\"o}dinger equations}.
\newblock {\em Phys. Lett. A}, 377:866--870, 2013.

\end{thebibliography}


\begin{thebibliography}{10}

\bibitem{Adami:2012jb}
R.~Adami, C.~Cacciapuoti, D.~Finco, and D.~Noja.
\newblock {Stationary states of NLS on star graphs}.
\newblock {\em EPL--Europhys. Lett.}, 100:10003, 2012.

\bibitem{Adami:2014hn}
R.~Adami, E.~Serra, and P.~Tilli.
\newblock {NLS ground states on graphs}.
\newblock {\em Calc. Var.}, 54:743--761, 2014.

\bibitem{Adami:2015di}
R.~Adami, E.~Serra, and P.~Tilli.
\newblock {Lack of Ground State for NLSE on Bridge-Type Graphs}.
\newblock In D.~Mugnolo, editor, {\em Mathematical Technology of Networks},
  volume 128 of {\em Springer Proc. in Math. and Stat.}, pages 1--11. Springer,
  2015.

\bibitem{Adami:2016hx}
R.~Adami, E.~Serra, and P.~Tilli.
\newblock {Negative Energy Ground States for the $L^2$-Critical NLSE on Metric
  Graphs}.
\newblock {\em Commun. Math. Phys.}, 352(1):387--406, 2016.

\bibitem{Adami:2016do}
R.~Adami, E.~Serra, and P.~Tilli.
\newblock {Threshold phenomena and existence results for NLS ground states on
  metric graphs}.
\newblock {\em Journal of Functional Analysis}, 271:201--223, 2016.

\bibitem{Adami:2017te}
R.~Adami, E.~Serra, and P.~Tilli.
\newblock {Nonlinear dynamics on branched structures and networks}, 2017,
  arXiv: 1705.00529v1.

\bibitem{Berkolaiko:2017}
G.~Berkolaiko.
\newblock An elementary introduction to quantum graphs.
\newblock In {\em Geometric and computational spectral theory}, volume 700 of
  {\em Contemp. Math.}, pages 41--72. Amer. Math. Soc., Providence, RI, 2017.

\bibitem{Berkolaiko:2013}
G.~Berkolaiko and P.~Kuchment.
\newblock {\em Introduction to Quantum Graphs}.
\newblock Mathematical surveys and monographs. Amer. Math. Soc., 2013.

\bibitem{Berkolaiko:2018}
G.~Berkolaiko, Y.~Latushkin, and S.~Sukhtaiev.
\newblock Limits of quantum graph operators with shrinking edges, 2018,
  arXiv:1806.00561.

\bibitem{Bolte:2014}
J.~Bolte and J.~Kerner.
\newblock Many-particle quantum graphs and {B}ose-{E}instein condensation.
\newblock {\em J. Math. Phys.}, 55:061901, 2014.

\bibitem{Cacciapuoti:2015fe}
C.~Cacciapuoti, D.~Finco, and D.~Noja.
\newblock {Topology-induced bifurcations for the nonlinear Schr{\"o}dinger
  equation on the tadpole graph}.
\newblock {\em Phys. Rev. E}, 91:013206, 2015.

\bibitem{Cacciapuoti:2017hr}
C.~Cacciapuoti, D.~Finco, and D.~Noja.
\newblock {Ground state and orbital stability for the NLS equation on a general
  starlike graph with potentials}.
\newblock {\em Nonlinearity}, 30:3271--3303, 2017.

\bibitem{Delourme:2017jd}
B.~Delourme, S.~Fliss, P.~Joly, and E.~Vasilevskaya.
\newblock {Trapped modes in thin and infinite ladder like domains. Part 1:
  Existence results}.
\newblock {\em Asymptotic Anal.}, 103:103--134, 2017.

\bibitem{Dhooge:2003vy}
A.~Dhooge, W.~Govaerts, and Y.~A. Kuznetsov.
\newblock {MATCONT: a MATLAB package for numerical bifurcation analysis of
  ODEs}.
\newblock {\em ACM T. Math. Software}, 29:141--164, 2003.

\bibitem{Dhooge:2008gq}
A.~Dhooge, W.~Govaerts, Y.~A. Kuznetsov, H.~G.~E. Meijer, and B.~Sautois.
\newblock {New features of the software MatCont for bifurcation analysis of
  dynamical systems}.
\newblock {\em Math. Comp. Model. Dyn.}, 14:147--175, 2008.

\bibitem{NIST:DLMF}
{NIST Digital Library of Mathematical Functions}.
\newblock http://dlmf.nist.gov/, Release 1.0.15 of 2017-06-01.
\newblock F.~W.~J. Olver, A.~B. {Olde Daalhuis}, D.~W. Lozier, B.~I. Schneider,
  R.~F. Boisvert, C.~W. Clark, B.~R. Miller and B.~V. Saunders, eds.

\bibitem{Eilbeck:2003vl}
J.~C. Eilbeck and M.~Johansson.
\newblock {The discrete nonlinear Schr{\"o}dinger equation--20 years on}.
\newblock In R.~S. MacKay, L.~V\'azquez, and M.~P. Zorzano, editors, {\em
  Proceedings Of The Third Conference On Localization And Energy Transfer In
  Nonlinear Systems}, pages 44--67, Madrid, 2003. World Scientific.

\bibitem{Eilbeck:1985tu}
J.~C. Eilbeck, P.~S. Lomdahl, and A.~C. Scott.
\newblock {The discrete self-trapping equation}.
\newblock {\em Phys. D}, 16:318--338, 1985.

\bibitem{Glendinning:1994ub}
P.~Glendinning.
\newblock {\em {Stability, Instability and Chaos}}.
\newblock An Introduction to the Theory of Nonlinear Differential Equations.
  Cambridge University Press, 1994.

\bibitem{Gnutzmann:2016hma}
S.~Gnutzmann and D.~Waltner.
\newblock {Stationary waves on nonlinear quantum graphs: General framework and
  canonical perturbation theory}.
\newblock {\em Phys. Rev. E}, 93:032204, 2016.

\bibitem{Gnutzmann:2016ec}
S.~Gnutzmann and D.~Waltner.
\newblock {Stationary waves on nonlinear quantum graphs. II. Application of
  canonical perturbation theory in basic graph structures}.
\newblock {\em Phys. Rev. E}, 94:062216, 2016.

\bibitem{Golubitsky:2013aa}
M.~Golubitsky and D.~Schaeffer.
\newblock {\em Singularities and Groups in Bifurcation Theory vol. {I}}.
\newblock Springer New York, 2013.

\bibitem{Govaerts:2000vv}
W.~J.~F. Govaerts.
\newblock {\em {Numerical Methods for Bifurcations of Dynamical Equilibria}}.
\newblock SIAM, Jan. 2000.

\bibitem{Holm:2011uw}
D.~D. Holm.
\newblock {\em {Geometric Mechanics Part I: Dynamics and Symmetry}}.
\newblock Imperial College Press, 2nd edition, 2011.

\bibitem{Kevrekidis:2009vv}
P.~G. Kevrekidis.
\newblock {\em {The Discrete Nonlinear Schr{\"o}dinger Equation: Mathematical
  Analysis, Numerical Computations and Physical Perspectives}}, volume 232 of
  {\em Springer Tr. Mod. Phys.}
\newblock Springer, Berlin Heidelberg, 2009.

\bibitem{Kirr:2017}
E.-W. Kirr.
\newblock Long time dynamics and coherent states in nonlinear wave equations.
\newblock In R.~Melnik, R.~Makarov, and J.~Belair, editors, {\em Recent
  Progress and Modern Challenges in Applied Mathematics, Modeling and
  Computational Science}, volume~79 of {\em Fields Inst. Commun.}, pages
  59--88. Springer, 2017.

\bibitem{Kuchment:2007}
P.~Kuchment and O.~Post.
\newblock On the spectra of carbon nano-structures.
\newblock {\em Communications in Mathematical Physics}, 275:805--826, 2007.

\bibitem{Marzuola:2016bl}
J.~L. Marzuola and D.~E. Pelinovsky.
\newblock Ground state on the dumbbell graph.
\newblock {\em Appl. Math. Res. Express}, 2016:98--145, 2016.

\bibitem{Marzuola:2017}
J.~L. Marzuola and D.~E. Pelinovsky.
\newblock Ground state on the dumbbell graph (v4), 2017, arXiv:1509.04721.

\bibitem{Marzuola:2010fl}
J.~L. Marzuola and M.~I. Weinstein.
\newblock {Long time dynamics near the symmetry breaking bifurcation for
  nonlinear Schr{\"o}dinger/Gross-Pitaevskii equations}.
\newblock {\em Discrete Contin. Dyn. Syst.}, 28:1505--1554, 2010.

\bibitem{nayfeh2008applied}
A.~Nayfeh and B.~Balachandran.
\newblock {\em Applied Nonlinear Dynamics: Analytical, Computational and
  Experimental Methods}.
\newblock Wiley Series in Nonlinear Science. Wiley, 2008.

\bibitem{Niikuni:2017fu}
H.~Niikuni.
\newblock {Schr{\"o}dinger operators on a periodically broken zigzag carbon
  nanotube}.
\newblock {\em P. Indian Acad. Sci.--Math. Sci.}, 127:471--516, 2017.

\bibitem{Noja:2015hx}
D.~Noja, D.~E. Pelinovsky, and G.~Shaikhova.
\newblock {Bifurcations and stability of standing waves in the nonlinear
  Schr{\"o}dinger equation on the tadpole graph}.
\newblock {\em Nonlinearity}, 28:2343--2378, 2015.

\bibitem{Noja:2017vd}
D.~{Noja}, S.~{Rolando}, and S.~{Secchi}.
\newblock {Standing waves for the NLS on the double-bridge graph and a
  rational-irrational dichotomy}, 2017, arXiv:1706.09616.

\bibitem{Pelinovsky:2012hb}
D.~E. Pelinovsky and T.~V. Phan.
\newblock {Normal form for the symmetry-breaking bifurcation in the nonlinear
  Schr{\"o}dinger equation}.
\newblock {\em J. Diff. Eq.}, 253:2796--2824, 2012.

\bibitem{Pelinovsky:2016fm}
D.~E. Pelinovsky and G.~Schneider.
\newblock {Bifurcations of Standing Localized Waves on Periodic Graphs}.
\newblock {\em Ann. Henri Poincar{\'e}}, 18:1185--1211, 2016.

\bibitem{matlab}
{The Mathworks{,} Inc.}
\newblock {MATLAB Release 2018a}.
\newblock Natick, Massachusetts, United States.

\bibitem{Mathematica}
{Wolfram Research{,} Inc.}
\newblock Mathematica, {V}ersion 11.3.
\newblock Champaign, IL, 2018.

\bibitem{Yang:2009}
J.~Yang.
\newblock {Newton-conjugate-gradient methods for solitary wave computations}.
\newblock {\em J. Comput. Phys.}, 228:7007--7024, 2009.

\bibitem{Yang:2012ux}
J.~Yang.
\newblock {Classification of solitary wave bifurcations in generalized
  nonlinear Schr{\"o}dinger equations}.
\newblock {\em Stud. Appl. Math.}, 129:133--162, 2012.

\bibitem{Yang:2018}
J.~Yang.
\newblock Personal communication, 2018.

\end{thebibliography}
\end{document}